\documentclass[10pt]{article}

\usepackage{array}
\usepackage{color} 
\usepackage{mathrsfs}
\usepackage{graphicx}
\usepackage{amsmath}
\usepackage{bbm}
\usepackage{amsfonts}
\usepackage{amssymb}
\usepackage{amsbsy}
\usepackage[T1]{fontenc}
\usepackage{theorem}
\usepackage{stmaryrd}
\usepackage{tipa}
\usepackage{textcomp}
\usepackage{subfigure}
\usepackage{epsfig}
\usepackage[sort, colon]{natbib}
\usepackage{lmodern}
\usepackage{framed}
\usepackage[subfigure]{tocloft}
\usepackage{subeqnarray}
\usepackage{alphalph}
\usepackage[refpage,cfg,prefix]{nomencl}
\usepackage{multirow}
\usepackage{xifthen}
\usepackage{enumerate}
\usepackage{color}
\usepackage{wasysym}
\usepackage{bibentry}
\usepackage{todonotes}
\nobibliography*
\usepackage{verbatim}
\usepackage{booktabs}
\usepackage[hyperindex,breaklinks]{hyperref}
\hypersetup{
colorlinks=true,
urlcolor=blue,
linkcolor=black,
citecolor=black,
}
\usepackage{breakurl}
\usepackage{url}
\usepackage{setspace}
\usepackage{xr}
\usepackage{wrapfig}
\usepackage{mathtools}
\usepackage{dsfont}
\usepackage[toc,page]{appendix}
\usepackage{filecontents}

\makenomenclature

\usepackage[labelfont=bf,labelsep=period,justification=raggedright]{caption}

\makeatletter
\renewcommand{\@biblabel}[1]{\quad#1.}
\makeatother

\allowdisplaybreaks

\date{}

\newcommand{\bs}[1]{\boldsymbol{#1}}

\newcommand{\partder}[2]{\frac{\partial #1}{\partial #2}}

\newcommand{\secpartder}[2]{\frac{\partial^2 #1}{\partial #2 ^2}}

\newcommand{\set}[2]{\left\{#1:#2\right\}}

\newcommand{\argmin}{\operatornamewithlimits{argmin}}

\newcommand{\smallsub}[2]{#1_{\text{{\tiny #2}}}}

\newcommand{\listofalgorithms}{\textbf{\Huge{List of Algorithms}}}
\newlistof{Algorithm}{exp}{\listofalgorithms}
\newcounter{instructioncounter}
 
\newenvironment{Algorithm}[2][]
{\refstepcounter{Algorithm}
\begin{framed}\addcontentsline{exp}{Algorithm}{\protect\numberline{\theAlgorithm} #1}\par\begin{center}\textbf{Algorithm \theAlgorithm : #2}\end{center}\begin{list}
{\bf{(\arabic{Algorithm}\alph{instructioncounter}})}{\usecounter{instructioncounter}}}{\end{list}\end{framed}}

\graphicspath{{./Figures/}}

\begin{document}

\begin{flushleft}
{\Large
\textbf{Incorporating domain growth into hybrid methods for reaction-diffusion systems}
}
\\
Cameron A. Smith$^{1,\ast}$, 
Christian A. Yates$^{1}$
\\
$^1$\textbf{Centre for Mathematical Biology, Department of Mathematical Sciences, University of Bath, Claverton Down, Bath, BA2 7AY, United Kingdom}\\
\end{flushleft}

Key words: Reaction--diffusion, domain growth, hybrid methods.

\begin{abstract}
Reaction--diffusion mechanism are a robust paradigm that can be used to represent many biological and physical phenomena over multiple spatial scales. Applications include intracellular dynamics, the migration of cells and the patterns formed by vegetation in semi-arid landscapes. Moreover, domain growth is an important process for embryonic growth and wound healing. There are many numerical modelling frameworks capable of simulating such systems on growing domains, however each of these may be well suited to different spatial scales and particle numbers. Recently, spatially extended hybrid methods on static domains have been produced in order to bridge the gap between these different modelling paradigms in order to represent multiscale phenomena. However, such methods have not been developed with domain growth in mind. In this paper, we develop three hybrid methods on growing domains, extending three of the prominent static domain hybrid methods. We also provide detailed algorithms to allow others to employ them. We demonstrate that the methods are able to accurately model three representative reaction-diffusion systems accurately and without bias.
\end{abstract}

\section{Introduction} \label{sect:Intro}

The reaction--diffusion paradigm can be employed to model a range of biological and physical scenarios over multiple length scales, from representing vegetation patterns in semi-arid landscapes \citep{sherratt2005avs} and the study of epidemics \citep{volpert2009rdw}, to intracellular dynamics \citep{khan2011scd, andasari2012iid, zhuge2000dsb}. These systems couple the random movement of particles (which when considered at a continuum level manifests as the movement of particle density down the concentration gradient) and the interaction of particles with each other and potentially with the domain boundaries.

Domain growth is a process which underpins many biological processes, and it is therefore important that we have accurate and efficient modelling methods to represent it. Examples span many biological applications, including the growth and shrinkage of tissue \citep{yates2014dcm, wolpert2015pod} and neural crest cell migration \citep{mort2016rdm, kulesa2010cnc, mclennan2012mmc}. Domain growth has also been shown to play an important role in theoretical studies of pattern formation in reaction-diffusion systems \citep{crampin1999rdg, taylor2015dab, woolley2011srd}.

Reaction--diffusion systems can be modelled in several different ways, each of which has different suitabilities depending on the scale of the system being modelled. The coarsest of the three methods that we focus on is the macroscale, which uses partial differential equations (PDEs) in order to represent the system. PDEs model how the continuum \textit{density} of particles evolve in time, and are only suitable if the number of particles is high enough to consider a continuum limit. The reaction--diffusion PDE on a growing domain consists of four components (see equation \eqref{eqn:PDE-EC}): a first order differential of density with respect to time which describes the change of concentration in time; a second-order differential of density with respect to space which represents diffusion; a first-order differential of density with respect to space representing domain growth, and finally a term which represents reactions, if any are present. The macroscale is generally quick to implement using a number of established techniques (for example, see \citet{brenner2004fem, eymard2000fvm, morton2005nsp, smith1985nsp}), and there are often analytical approaches that can be employed to investigate such systems. However, if particle numbers are too low, the assumption that the continuum limit holds may break down, and stochastic fluctuations may be found to play a more pivotal role. Moreover, the deterministic mean-field PDE may not fully agree with its stochastic counterparts described below. Attempts to derive a deterministic equivalent to a non-linear stochastic model may result in an infinite hierarchy of interrelated equations. This results in the need for moment closure, which necessarily leads to the loss of some of the information encapsulated in the higher-order moments. 

The second modelling paradigm that we consider is the mesoscale, where we split the spatial domain into a series of compartments, within which particles reside. These particles are able to jump between neighbouring compartments, and are able to interact/react with others within their own compartment. These events are given exponential waiting times whose rates dictate the evolution of the system. There are a large number of algorithms to simulate such systems, both exact \citep{gillespie1977ess, gibson2000ees, anderson2007mnr} and approximate \citep{gillespie2001aas, auger2006rla}. This middle scale is generally slower than the macroscale, but provides more fine-grained accuracy when particle numbers are lower. In order to incorporate domain growth at least two methods that have been proposed. The first is a local method proposed by \citet{baker2010fmm}, which chooses a compartment uniformly at random to instantaneously double in size and then divide. The particles in the parent compartment are distributed into the two daughter compartments according to some symmetric probability distribution. This method causes a build-up of particles at the boundaries when growth occurs on a faster time-scale than diffusion \citep{smith2019uol}. The second method (and the method we will employ in this work), is a global method introduced by \citet{smith2019uol}. Compartments grow uniformly until a growth event is due to occur. At this point, the boundaries between compartments are redrawn to include one extra compartment, and the particles are redistributed appropriately. This method works well in both high and low diffusion regimes.

Finally, at the microscale, we investigate Brownian-based dynamics. At this, the smallest scale that we employ, individual particles are tracked and updated in continuous space. There are several techniques that can be employed in order to simulate a system at this scale, including the time-driven mechanisms of Brownian motion for purely diffusive processes and Smoluchowski dynamics \citep{smoluchowski1917vem} when reactions are involved, or the event-driven Green's function reaction dynamics (GFRD) \citep{van2005gfr}. Under time-driven algorithms, particles diffuse and are diluted according to an appropriate stochastic differential equation (SDE). As well as each particle's location, if the system requires second- or higher-order reactions, we also need to calculate the pairwise distances between all particles at every time-step, which means that while this modelling paradigm is the most accurate, it is also the slowest. Added to this, if the system in question is diffusion limited, a very small time-step is required to accurately resolve the dynamics. More efficient time-stepping is employed by the event-driven GFRD. This sets a maximum time-step and the solution to the Smoluchowski equation in order to combine diffusion and interactions, whilst accounting for the additional error which is introduced in doing so. 

There is another, even finer, scale of spatially resolved model known as molecular dynamics, which we do not consider in this paper. We direct the interested reader to \citep{holley1971mhp, durr1981mmb} for more information. 

Many biological problems of interest are genuinely multiscale \citep{markevich2004ssb, black2012sfe, gillespie2013psa, robinson2014atr}. Consequently we require methods that are able to resolve the dynamics at the appropriate scale. Spatially extended hybrid methods \citep{smith2018seh} are one such class of techniques that are able to do this. These methods employ two or more reaction-diffusion modelling paradigms (described above) to represent the dynamics in different areas of the domain in the most appropriate way. In regions with low particle numbers, for example, one of the finer-grained stochastic methods might be employed at the cost of reduced simulation efficiency. However, in regions of high particle numbers --- enough to consider a continuum limit --- the more computationally efficient PDE may be used. There are many examples of spatially extended hybrid methods \citep{yates2015pcm, moro2004hms, spill2015ham, flegg2015cmc, flegg2012trm, robinson2014atr, smith2018arm, franz2013mrd, alexander2002ars, erban2014fmd, erban2016caa}, all of which focus on a static (non-growing) domain. This paper extends three of these methods onto uniformly growing domains. The pseudo-compartment method (PCM) \citep{yates2015pcm} is a macroscopic-to-mesoscopic method which uses an interface to divide the domain into two subdomains. Particles are able to jump between the two subdomains via a `pseudo-compartment' adjacent to the interface within the PDE subdomain. The ghost cell method (GCM) proposed by \citet{flegg2015cmc} is a mesoscopic-to-microscopic method that makes similar use of an extra compartment, coined the ghost cell, adjacent to the interface in the microscopic subdomain in order to couple compartment-based and Brownian dynamics. Finally, the auxiliary region method (ARM) \citep{smith2018arm} is a macroscopic-to-microscopic method which, similar to the PCM and GCM, which employs a mesoscale auxiliary region at the interface in order to allow particles to jump between the two subdomians.

The rest of this paper will be set out as follows. In Section \ref{sect:Equiv} we briefly explain how the macroscopic, mesoscopic and microscopic can be considered equivalent to each other. In Sections \ref{sect:PCM}--\ref{sect:ARM} we introduce the PCM, GCM and ARM, explain the key differences between the algorithms for a static and growing domain, and present the algorithms in full. We present representative results for multiple test problems in Section \ref{sect:Results}, and discuss our findings in Section \ref{sect:Discussion}.

\section{Equivalence Framework} \label{sect:Equiv}

In this section we present each of the three modelling paradigms that are the constituent parts of our three hybrid methods. In Section \ref{sect:Equiv-Macro} we introduce the PDE approach. We present the mesoscopic approach in Section \ref{sect:Equiv-Meso} and briefly demonstrate how we can consider it to be equivalent to the PDE in the appropriate limit. Finally, in Section \ref{sect:Equiv-Micro} we do the same for the individual, particle-based dynamics. All numerical algorithms can be found in the Supplementary Material.

\subsection{Macroscopic modelling} \label{sect:Equiv-Macro}

Firstly we consider the macroscale PDE representation. Consider a population with density $u(x,t)$ undergoing diffusion at a position $x$ on an exponentially growing one-dimensional domain $(0,L(t))\subseteq\mathbb{R}$ at time $t>0$. We consider just a single dimension and exponential growth (with rate $\rho$) for ease of description here, but the methods outlined can be extended to higher dimensions, and will work for any form of uniform growth. Under these assumptions, the length of the domain is $L(t) = L_0\exp\{\rho t\}$, where $L_0$ is the initial length of the domain, and the concentration evolves according to the following PDE:
\begin{equation}
\partder{u}{t}(x,t) = D\secpartder{u}{x}(x,t) - \rho\partder{(xu(x,t))}{x},
\label{eqn:PDE-EC}
\end{equation} which holds for $x\in(0,L(t))$ and for $t>0$. This description of domain growth is known as the Eulerian representation in Eulerian coordinates, $x$ and $t$. The first term on the right-hand side represents the spread of particles due to diffusion (with Fickian diffusion coefficient $D$) and the second term is the dilution and spread of concentration caused by the stretching of the spatial domain. We will employ this PDE in order to demonstrate the equivalence of the two finer-scale methods to this first one.

In order to simulate this PDE, we need to switch from the Eulerian coordinates above, in which the domain grows in time, to Lagrangian coordinates, in which the domain remains static in time. We do this through the following change of coordinates:
\begin{align}
x &= X\exp\{\rho t\}, \label{eqn:EC_to_LC_x}\\
t &= \tau. \label{eqn:EC_to_LC_t}
\end{align}
Note now that the length of the domain in the Lagrangian spatial variable $X$ ranges between $0$ and the fixed value $L_0$. Further, the Lagrangian and Eulerian temporal variables, $\tau$ and $t$ respectively, coincide here. There are times when it is useful to rescale time, such as when finding analytical solutions to the diffusion equation on uniformly growing domains \citep{simpson2015esl}. For more details on the Lagrangian PDE and how we solve it numerically, please see Section \ref{SUPP-sect:Macro_Num} of the supplementary material (SM).

\subsection{Mesoscopic modelling} \label{sect:Equiv-Meso}

In this section, we describe the mesoscale representation, which is the ``middle'' scale that we will consider. In order to model at this level, we will divide the spatial domain $(0,L(t))$ into a number of compartments labelled $C_i(t)$ for $i\in\{1,...,K(t)\}$, where $K(t)$ is the (time-dependent) number of compartments at time $t$. Particles lie within these compartments and are able to jump between neighbouring compartments mimicking diffusion, and can interact with one another within the same compartment through $R$ reaction channels. We will define the state of the system at time $t$ to be $\bs{N}(t)$, where $N_i(t)$ is the number of particles at time $t$ in compartment $i$. Throughout this paper, we will implement the modified next reaction method \citep{anderson2007mnr} in order to advance the system forwards in time. This method is used for explicit time-varying propensity functions (the propensity function is a proxy for the rate of that event occurring), which are important due to the domain growth. For an explanation of the method and the algorithm, see the Supplementary Material, Section \ref{SUPP-sect:Meso_MNRM}.

To extend the domain, we utilise the stretching method of \citet{smith2019uol}. As the domain grows through a series of discrete fixed size extensions, the compartments grow in length uniformly. Once a stretch event has been determined to occur, the compartment boundaries are redrawn, making each compartment smaller and making way for a new compartment. The particles are then appropriately redistributed assuming a uniform distribution of particles across each compartment. This algorithm can be included in the main mesoscale algorithm in two ways. Firstly, propensity functions for the splitting events (events where a new compartment is added) can be added to the list of propensity functions, making this stochastic in time and space. Alternatively (and the method we utilise here), we calculate the deterministic time at which new compartments should be added to effect a particular domain growth pattern, and add them at this time. In  this way, the growth events are still stochastic in space, but are now time-deterministic.

It can be shown, in the limit of small compartment size and fast inter-compartment jumping rate, that the average behaviour of this mesoscopic model can be described by the Eulerian PDE \eqref{eqn:PDE-EC}. We briefly present the main steps of the calculation in the Supplementary Material (Section \ref{SUPP-sect:Meso_Equiv}). For a more complete calculation please see \citep{smith2019uol} for the mean equations and Section \ref{SUPP-sect:Meso_Equiv} in the Supplementary Material for the small compartment limit. 

\subsection{Microscopic modelling} \label{sect:Equiv-Micro}

Within this section, we introduce the finest scale modelling paradigm that we will employ for the hybrid methods presented in this paper. The microscale tracks the individual locations of particles which diffuse and are repositioned (due to domain growth) according to a stochastic differential equation (SDE) as well as interacting depending on proximity to one another. Particles diffuse through a Brownian motion, and the growth is implemented using a deterministic drift term. In practice, this means that each particles is ``pulled along'' as the domain stretches.

Let $X_t\in(0,L(t))$ be the position of a particle at time $t$. Then this evolves according to: 
\begin{equation}
\text{d}X_t = \rho X_t\ \text{d}t + \sqrt{2D}\ \text{d}W_t. \label{eqn:SDE-EC}
\end{equation}
Here, the term on the left hand side denotes the change in position, the first term on the right hand side is the drift term representing the repositioning of the particles caused by the stretching of the domain, and the second term on the right hand side is the diffusion. Note also that $\text{d}W_t$ is a standard Weiner process. 

In order to show that the density of particles evolving according to this SDE is described by the Eulerian PDE \eqref{eqn:PDE-EC}, we look to the Fokker-Planck equation (FPE) (otherwise known as the Kolmogorov forward equation (KFE)) \citep{risken1996fpe}. As with the mesoscopic case, we briefly outline the derivation in the SM (Section \ref{SUPP-sect:Micro_Num}), and refer the interested reader to the aforementioned reference. 

\section{The pseudo-compartment method} \label{sect:PCM}

The first hybrid method we will adapt to incorporate uniform domain growth is the macroscopic-to-mesoscopic pseudo-compartment method (PCM) \citep{yates2015pcm}. At time $t$ we decompose the domain as follows. The PDE subdomain occupies the region $\smallsub{\Omega}{P}(t) = (0,I(t))$, where $I(t)$ is the location of the interface at time $t$, and the mesoscopic subdomain is $\smallsub{\Omega}{C}(t) = (I(t),L(t))$, where $L(t)$ is the total length of the domain. The values of $I(t)$ and $L(t)$ will be calculated deterministically from the initial position of these boundaries, and the growth process, via the following pair of equations
\begin{align*}
I(t) &= I(0)\exp\{\rho t\},\\
L(t) &= L(0)\exp\{\rho t\}.
\end{align*}

\begin{figure}[ht!!!!!!!!!!!!!!]
\centering
\includegraphics[width=0.8\textwidth]{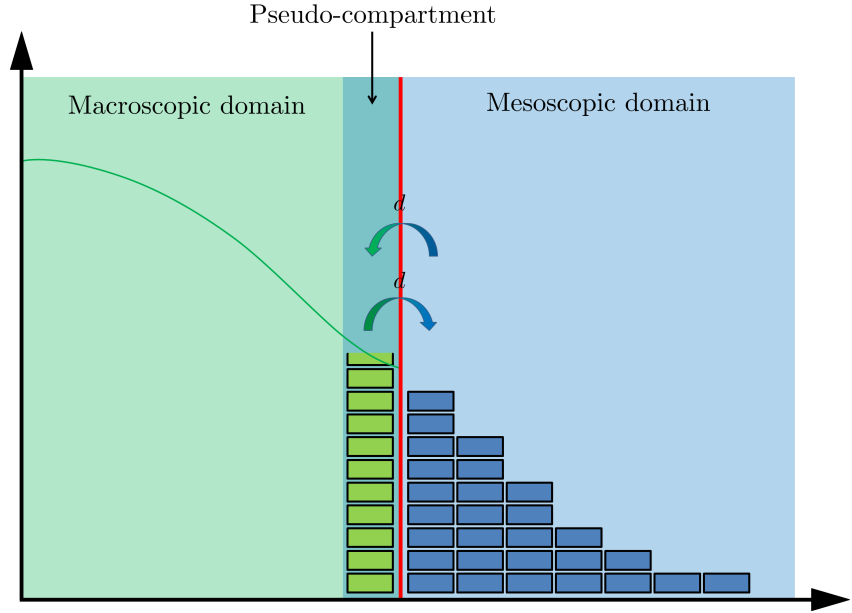}
\caption{A schematic of the static PCM \citep{yates2015pcm}. The green line denotes the density of particles in the macroscopic domain, while the blue rectangles represent particles within each compartment. The red line is the interface, and the green rectangles represent the number of ``pseudo-particles'', obtained by direct integration of the PDE solution over the pseudo-compartment. The arrows crossing the interface denote the movement of the pseudo-particles between the two subdomains.}
\label{fig:PCM_Schematic}
\end{figure}

In order to couple the macroscale and mesoscale, we allow particles to jump between the first mesoscopic compartment (next to the interface (vertical red line) in Figure \ref{fig:PCM_Schematic}) and the so-called ``pseudo-compartment'', which is a region of space the width of one compartment next to the interface on the PDE side of the domain. Mass may only cross the interface through this jumping mechanism and not though continuum PDE mass flowing over the boundary. As such, a zero-flux boundary condition is imposed on the PDE at the interface.

Suppose that the number of particles in compartment $C_1(t)$ at time $t$ is $n_1(t)$, and that it is of width $h_c(t)$. Then the region of space occupied by the pseudo-compartment of length $h_{PC}(t)$ is denoted $C_{-1}(t)$ and contains $n_{-1}(t)$ particles, where 
\begin{equation}
n_{-1}(t) = \int_{C_{-1}(t)}{u(x,t)\ dx}.
\label{eqn:Pseudo_Particles}
\end{equation}
In order to incorporate jumps into and out of the pseudo-compartment, we introduce two events into the list of mesoscopic events, one for a jump into the pseudo-compartment from compartment 1, and one for a jump in the other direction. Since, on the growing domain these compartments are potentially of different widths, the appropriate jumping rates would be $D/(h_c(t)^2)$ and $D/(h_{PC}(t)h_c(t))$ respectively. The second of these takes into account the differing compartment sizes. In the case that $h_c(t)$ and $h_{PC}(t)$ are the same, this collapses to the usual diffusive jump rate.

The reasons that the length of a ``regular'' compartment $h_c(t)$ is typically different from the length of the pseudo-compartment $h_{PC}(t)$ are subtle and due to the differences between the Lagrangian (static) and Eulerian (growing) coordinates. The PDE is being solved using a fixed mesh-width in Lagrangian coordinates, and as such, the mesh grows when considered in Eulerian coordinates. On the other hand, the compartment size is set in Eulerian coordinates, and as a result, the number of Lagrangian PDE points per Eulerian compartment decreases in time.

On the growing domain, the lengths of these compartments are calculated using other simulation parameters. The length of the compartments $C_i(t)$ for $i\in\{1,...,K(t)\}$ is given by the length of the mesoscopic part of the domain, divided by the number of compartments $K(t)$. Therefore 
\begin{equation}
h_c(t) = \frac{L(t)-I(t)}{K(t)}.
\label{eqn:Compartment_Size}
\end{equation}
The pseudo-compartment size is calculated from the current compartment size as follows. As well as having compartment properties we also need to solve the PDE in the region occupied by the pseudo-compartment. As such, we will set its length to be an integer number of PDE mesh points. The PDE mesh, when considered from the Eulerian perspective, has width $h_p(t) = h_p(0)\exp\{\rho t\}$, and we calculate the number of PDE mesh points that would be in the pseudo-compartment, if it was of length $h_c(t)$, to be 
\begin{equation*}
p_c(t) = \frac{h_c(t)}{h_p(t)}.
\end{equation*}
The value of $p_c(t)$ is generally not an integer, so in order to obtain an integer we round (up or down or to the nearest integer) the value of $p_c(t)$ and multiply this by $h_p(t)$ in order to find $h_{PC}(t)$.

If a particle is chosen to jump out of $C_1(t)$ and into $C_{-1}(t)$, we firstly reduce $n_1(t)$ by one, and then add a particle's worth of mass to the pseudo-compartment according to: 
\begin{equation}
\bs{U}_{PC} = \bs{U}_{PC} + \frac{1}{h_{PC}(t)}\bs{1},
\end{equation}
 where the vector $\bs{U}_{PC}$ is the numerical approximation to the solution $u(x,t)$ at the PDE nodes contained within the pseudo-compartment, and $\bs{1}$ is a vector of ones of the appropriate size. If a particle jumps out of the pseudo-compartment and into $C_1(t)$, we add one particle to $n_1(t)$ and remove a particle's worth of mass uniformly across the pseudo-compartment 
 \begin{equation}
\bs{U}_{PC} = \bs{U}_{PC} - \frac{1}{h_{PC}(t)}\bs{1}.
\end{equation}
 The algorithm for the implementation of the growing pseudo-compartment method (gPCM) can be found in Algorithm \ref{alg:PCM}, and a schematic for the static case is given in in Figure \ref{fig:PCM_Schematic}. The algorithm is given for diffusion only, however reactions may be incorporated through any appropriate method (see for example \citep{erban2009smr}).

\begin{Algorithm}{The growing pseudo-compartment method (Diffusion only)} \label{alg:PCM}
\item[] \textbf{Initialise:} Initial time --- $t=0$; Final time --- $t_f$; Initial compartment size --- $h_c$; Initial pseudo-compartment size --- $h_{pc}$; PDE solution --- $\bs{U}$; Number of pseudo-particles --- $n_{-1}$; Number of compartments --- $K$; Compartment particle numbers --- $\bs{n}$; Propensity functions --- $a_i$ for $i\in\{1,...,2K+1\}$; Internal clock times --- $T_i$ for $i\in\{1,...,2K+1\}$; Next firing times --- $P_i$ for $i\in\{1,...,2K+1\}$; Times until next event --- $\Delta t_i$ for $i\in\{1,...,2K+1\}$; PDE update step --- $\Delta t$; Time until next PDE update --- $t_p$; Time until next split event --- $t_s$; Time until next PDE re-mesh event --- $t_r$.

\item At time $t > 0$: \label{alg_step:PCM_return}
\begin{enumerate}
	\item Calculate $\Delta = \min\set{i\in\{1,...,2K+1\}}{\Delta t_i}$ and $\beta = \argmin\set{i\in\{1,...,2K+1\}}{\Delta t_i}$. Set $t_\Delta = t + \Delta$.
	\item If $\min\{t_\Delta,t_p,t_s,t_r\} = t_\Delta $:
	\begin{enumerate}
		\item For every $i\in\{1,...,2K+1\}$, update $T_i$ according to $$T_i \leftarrow T_i + \frac{1}{2\rho}a_i\left(1-\exp\{-2\rho\Delta\}\right).$$
		\item For event $\beta$, set $$P_\beta \leftarrow P_\beta + \ln\left(\frac{1}{u_1}\right) \text{, where }u_1\sim\text{Unif}(0,1).$$
		\item Enact the event $\beta$:
		\begin{itemize}
			\item If the event $\beta$ corresponds to a jump from the pseudo-compartment to the first compartment, set $n_1 \leftarrow n_1 + 1$ and set $\bs{U}_{pc} \leftarrow \bs{U}_{pc} - 1/h_{pc}\bs{1}$, where $\bs{U}_{pc}$ are the pseudo-compartment nodes of the PDE solution, and $\bs{1}$ is a vector of ones of the appropriate size.
			\item If the event $\beta$ corresponds to a jump from the first compartment to the pseudo-compartment, set $n_1 \leftarrow n_1 - 1$ and set $\bs{U}_{pc} \leftarrow \bs{U}_{pc} + 1/h_{pc}\bs{1}$.
			\item Otherwise, set $\bs{n} \leftarrow \bs{n} + \bs{\nu}_\beta$, where $\bs{\nu}_\beta$ is the stoichiometric vector for the event $\beta$.
		\end{itemize}
		\item Set $t=t_\Delta$.
	\end{enumerate}
	\item Else if $ \min\{t_\Delta,t_p,t_s,t_r\} = t_s$:
	\begin{enumerate}
		\item Enact a growth event according to \citet{smith2019uol} (see Algorithm \ref{SUPP-alg:Stretch}). Set $K' = K+1$. 
		\item For every $i\in\{1,...,2K+1\}$, update $T_i$ according to $$T_i \leftarrow T_i + \frac{1}{2\rho}a_i\left(1-\exp\{-2\rho(t_s - t)\}\right).$$
		\item For $i\in\{2K+2,2K+3\}$, set $T_i = 0$ and $P_i = \ln\left(1/u_{2,i}\right)$, where $u_{2,i}\sim\text{Unif}(0,1)$.
		\item Set $t = t_s$. Update $t_s$.
		\item Set $K \leftarrow K'$.
	\end{enumerate}
	\item Else if $\min\{t_\Delta,t_p,t_s,t_r\} = t_p$:
	\begin{enumerate}
		\item For every $i\in\{1,...,2K+1\}$, update $T_i$ according to $$T_i \leftarrow T_i + \frac{1}{2\rho}a_i\left(1-\exp\{-2\rho(t_p - t)\}\right).$$
		\item Enact a PDE update step using Algorithm \ref{SUPP-alg:PDE}.
		\item Set $t \leftarrow t_p$. Set $t_p \leftarrow t_p + \Delta t$.
	\end{enumerate}
	\item Else:
	\begin{enumerate}
		\item For every $i\in\{1,...,2K+1\}$, update $T_i$ according to $$T_i \leftarrow T_i + \frac{1}{2\rho}a_i\left(1-\exp\{-2\rho(t_r - t)\}\right).$$
		\item Re-mesh the PDE solution according to Algorithm \ref{SUPP-alg:Remesh}
		\item Set $h_p \leftarrow h_p/2$.
		\item Set $t \leftarrow t_r$. Update $t_r$ according to Algorithm \ref{SUPP-alg:Remesh}.
	\end{enumerate}
	\item Update $h_c$ and $h_{pc}$.
	\item Update all propensity functions $a_i$, for $i\in\{1,...,2K+1\}$.
	\item Update $\Delta t_i$ according to $$\Delta t_i = -\frac{1}{2\rho}\ln\left(1-\frac{2\rho(P_i-T_i)}{a_i}\right),$$for $i\in\{1,...,2K+1\}$.
\end{enumerate}
\item If $t < t_f$, return to \ref{alg_step:PCM_return}. Otherwise, end.
\end{Algorithm}

\section{The ghost-cell method} \label{sect:GCM}

Within this section, we describe the growing ghost cell method (gGCM), the static counterpart of which was proposed by \citet{flegg2015cmc}. This hybrid method couples the mesoscopic and microscopic descriptions of reaction-diffusion systems. We define the two spatial domains to be $\smallsub{\Omega}{B}(t) = (0,I(t))$ for the microscopic, Brownian-based dynamics, and $\smallsub{\Omega}{C}(t) = (I(t),L(t))$ for the compartment-based subdomain. The definitions of $I(t)$ and $L(t)$ are the same as for gPCM.

The coupling is implemented in a similar way to the PCM. A ghost cell (which is analogous to the pseudo-compartment for the PCM) is created within the microscopic subdomain, adjacent to the interface. Transport of mass over the interface is implemented using the mesoscopic approach. As such, the microscopic subdomain has a reflective boundary at the interface to ensure that no particles are able to move across through that medium.

\begin{figure}[ht!!!!!!!!!!!!!!]
\centering
\includegraphics[width=0.8\textwidth]{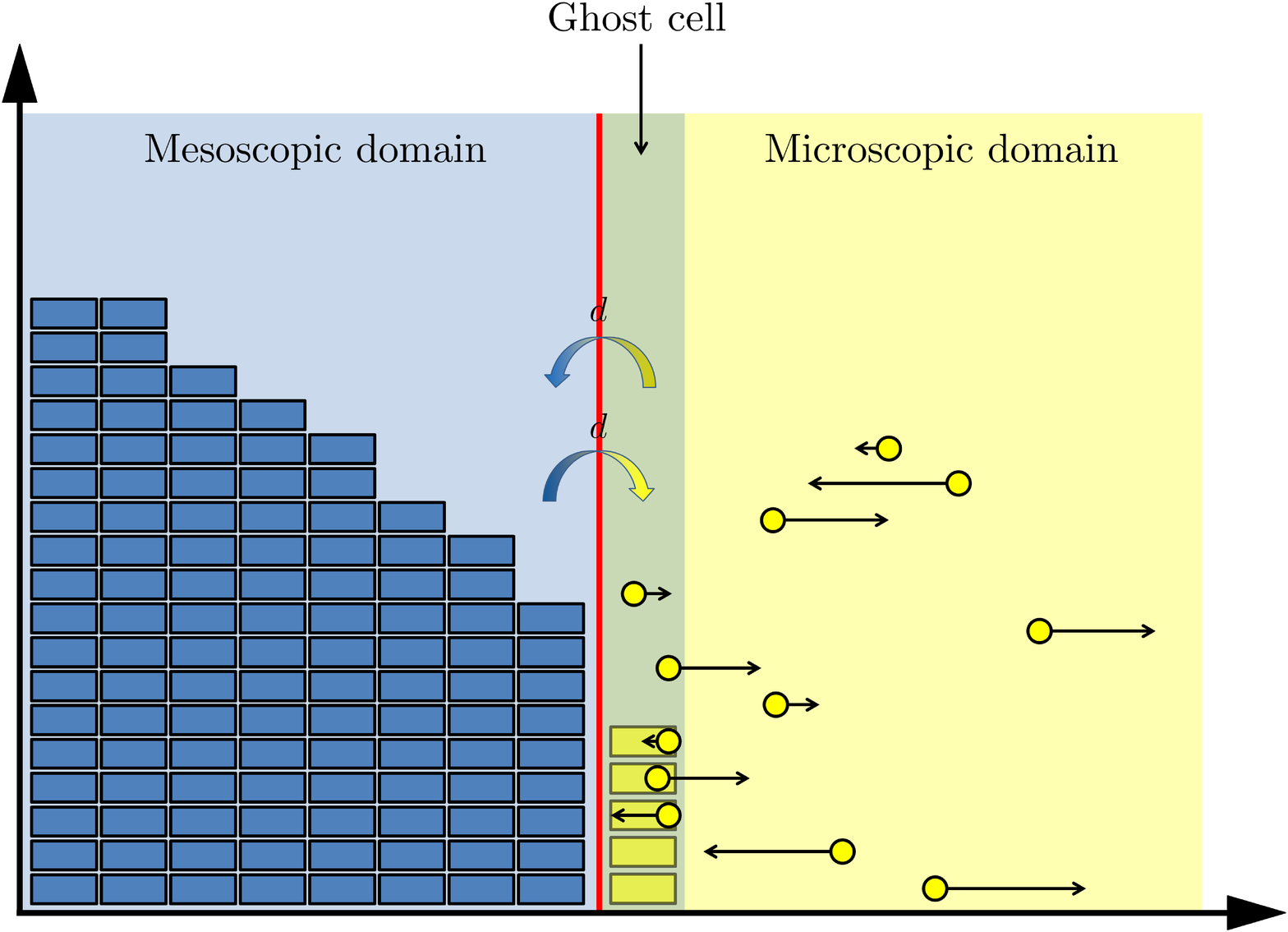}
\caption{A schematic for the static GCM \citep{flegg2015cmc}. The blue rectangles and red line are the same as in Figure \ref{fig:PCM_Schematic}. The yellow dots, denote the positions of the individual particles, with arrows denoting the next jump size and direction. Note that we have given each particle a different height to aid clarity, but all particles lie on the axis in reality. The yellow rectangles are the number of ghost cell particles, while the arrows over the interface denote the direction of travel for the ghost cell particles. We further note that this diagram has the two subdomains in the opposite order compared to the description in the text.}
\label{fig:GCM_Schematic}
\end{figure}

Unlike in the PCM, since we don't have two different discretisation lengths being used in the numerical realisation of the algorithm (PDE and compartment meshes), we are free to choose the ghost-cell to be the same size as the other compartments in the simulation (as they change in size), and we set it to $h_c(t)$ as defined in equation \eqref{eqn:Compartment_Size}. The ghost cell therefore occupies the region $\smallsub{C}{GC}(t) = (I(t)-h_c(t),I(t))$. The propensity function for particles to jump from the ghost cell into the first compartment in the compartment-based regime is $D/(h_{PC}(t)^2)$, multiplied by the number of particles $\smallsub{n}{GC}(t)$ in the ghost cell. $\smallsub{n}{GC}(t)$ is calculated simply by counting the number of particles within the ghost cell, so 
\begin{equation*}
\smallsub{n}{GC}(t) = \sum_{i=1}^{\smallsub{n}{B}(t)}{\mathbbm{1}_{[y_i(t)\in\smallsub{C}{GC}(t)]}},
\label{eqn:GhostCell_Particles}
\end{equation*}
where $\mathbbm{1}_{[A]}$ is the indicator function that is one if $A$ is true and 0 otherwise, $y_i(t)$ is the location of Brownian particle $i$ at time $t$, and $\smallsub{n}{B}(t)$ is the total number of particles in the Brownian-based subdomain at time $t$.

When a particle jumps from the ghost cell to the first compartment, we increase $n_1(t)$ by one and remove one of the ghost cell particles uniformly at random whilst simultaneously reducing $\smallsub{n}{GC}(t)$ by 1. When a jump occurs from the first compartment into the ghost cell, we reduce $n_1(t)$ by one and add a new particle to the ghost cell by sampling a uniform position within the ghost cell (that is, add a new particle at a position $y^* \sim \text{Unif}(I(t)-h_c(t),I(t))$) and subsequently increase  $\smallsub{n}{GC}(t)$ by 1. Domain growth is implemented using the stretching method \citep{smith2019uol} for the mesoscale, and via the SDE for the microscale. The algorithm for the growing ghost-cell method can be found in Algorithm \ref{alg:GCM}, and a schematic for the static case is given in Figure \ref{fig:GCM_Schematic}.

\begin{Algorithm}{The growing ghost cell method (Diffusion only)} \label{alg:GCM}
\item[] \textbf{Initialise:} Initial time --- $t=0$; Final time --- $t_f$; Compartment size --- $h_c$; Positions of particles --- $\bs{y}$; Number of compartments --- $K$; Compartment particle numbers --- $\bs{n}$; Propensity functions --- $a_i$ for $i\in\{1,...,2K+1\}$; Internal clock times --- $T_i$ for $i\in\{1,...,2K+1\}$; Next firing times --- $P_i$ for $i\in\{1,...,2K+1\}$; Times until next event --- $\Delta t_i$ for $i\in\{1,...,2K+1\}$; Brownian update step --- $\Delta t$; Time until next Brownian update --- $t_b$; Time until next splitting event --- $t_s$.
\item At time $t > 0$: \label{alg_step:GCM_return}
\begin{enumerate}
	\item Calculate $\Delta = \min\set{i\in\{1,...,2K+1\}}{\Delta t_i}$ and $\beta = \argmin\set{i\in\{1,...,2K+1\}}{\Delta t_i}$. Set $t_\Delta = t + \Delta$.
	\item If $\min\{t_\Delta,t_b,t_s\}= t_\Delta $:
	\begin{enumerate}
		\item For every $i\in\{1,...,2K+1\}$, update $T_i$ according to $$T_i \leftarrow T_i + \frac{1}{2\rho}a_i\left(1-\exp\{-2\rho\Delta\}\right).$$
		\item For event $\beta$, set $$P_\beta \leftarrow P_\beta + \ln\left(\frac{1}{u_1}\right) \text{, where }u_1\sim\text{Unif}(0,1).$$
		\item Enact the event $\beta$:
		\begin{itemize}
			\item If the event $\beta$ corresponds to a jump from the ghost cell to the first compartment, set $n_1 \leftarrow n_1 + 1$ and remove a particle from the ghost cell uniformly at random.
			\item If the event $\beta$ corresponds to a jump from the first compartment to the ghost cell, set $n_1 \leftarrow n_1 - 1$ and add a new particle to the ghost cell by drawing $u_2\sim\text{Unif}(0,1)$ and set the new particle's position $y^*$ to be $y^*=I(t) - u_2h_c(t)$.
		\end{itemize}
		\item Set $t=t_\Delta$.
	\end{enumerate}
	\item Else if $\min\{t_\Delta,t_b,t_s\} = t_s$:
	\begin{enumerate}
		\item Enact a growth event according to \citet{smith2019uol} (see Algorithm \ref{SUPP-alg:Stretch} in the SM). Set $K' = K+1$.
		\item For every $i\in\{1,...,2K+1\}$, update $T_i$ according to $$T_i \leftarrow T_i + \frac{1}{2\rho}a_i\left(1-\exp\{-2\rho(t_s - t)\}\right).$$
		\item For $i\in\{2K+2,2K+3\}$, set $T_i = 0$ and $P_i = \ln\left(1/u_{3,i}\right)$, where $u_{3,i}\sim\text{Unif}(0,1)$.
		\item Set $t = t_s$. Update $t_s$.
		\item Set $K \leftarrow K'$.
	\end{enumerate}
	\item Else:
	\begin{enumerate}
		\item For every $i\in\{1,...,2K+1\}$, update $T_i$ according to $$T_i \leftarrow T_i + \frac{1}{2\rho}a_i\left(1-\exp\{-2\rho(t_b - t)\}\right).$$
		\item Enact a Brownian update step using algorithm \ref{SUPP-alg:Ind-Update}.
		\item Set $t \leftarrow t_b$. Set $t_b \leftarrow t_b + \Delta t$.
	\end{enumerate}
	\item Update $h_c(t)$.
	\item Update all propensity functions $a_i$, for $i\in\{1,...,2K+1\}$.
	\item Update $\Delta t_i$ according to $$\Delta t_i = -\frac{1}{2\rho}\ln\left(1-\frac{2\rho(P_i-T_i)}{a_i}\right),$$for $i\in\{1,...,2K+1\}$.
\end{enumerate}
\item If $t<t_f$, return to \ref{alg_step:GCM_return}. Otherwise, end.
\end{Algorithm}

\section{The auxiliary region method} \label{sect:ARM}

\begin{figure}[ht!!!!!!!!!!!]
\centering
\includegraphics[width=0.8\textwidth]{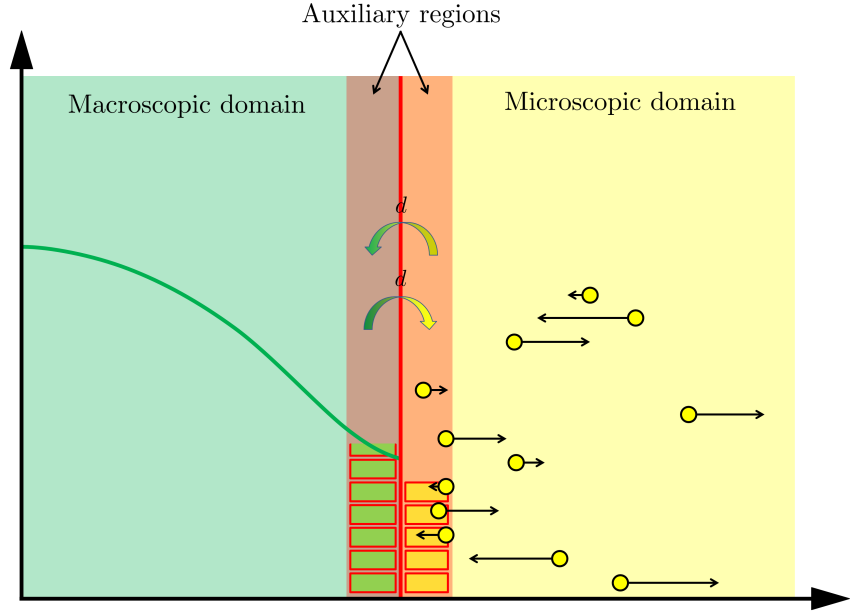}
\caption{A schematic for the static ARM \citep{smith2018arm}. The green and red lines are the same as in Figure \ref{fig:PCM_Schematic}, while the yellow dots and arrows are the same as in Figure \ref{fig:GCM_Schematic}. The green and yellow rectangles denote the number of PDE and Brownian auxiliary particles respectively. The arrows over the interface denote the movement of these auxiliary particles.}
\label{fig:ARM_Schematic}
\end{figure}

Within this section, we describe the third and final of our growing hybrid methods, the growing auxiliary region method (gARM) \citep{smith2018arm}. This method couples the macro and microscales, using a similar methodology to both the PCM and the GCM. Auxiliary regions are set up on either side of the interface, which act as compartments for the purpose of allowing particles to move between the two subdomains. A schematic for the static version of the method is in Figure \ref{fig:ARM_Schematic}.

The two auxiliary regions are equally sized compartments (although it is possible to have them at different sizes), where particle numbers in the PDE auxiliary region are calculated as in the pseudo-compartment of the gPCM for the macroscopic subdomain, and particles in the Brownian auxiliary region are calculated as in the ghost cell of the gGCM for the microscopic domain. As is the case in each of those other hybrid methods, compartment-based jumping via the auxiliary regions is the only mechanism by which particles may pass between the subdomains. As such, the PDE requires a zero-flux boundary condition at the interface, and the Brownian-based dynamics need an equivalent reflective boundary.

In order to calculate the size of the auxiliary regions, we use a similar idea to that employed in the gPCM. We set the size of a compartment to be $h$ initially, and use this to find the actual auxiliary region size, $\smallsub{h}{AR}(t)$ by firstly calculating how many PDE mesh points lie within this initial size, via: 
\begin{equation*}
p_c(t) = \frac{h}{h_p(t)},
\end{equation*}
where $h_p(t)$ is as in Section \ref{sect:PCM}, the mesh-width in Eulerian coordinates. We convert this to an integer number of PDE mesh points by rounding it to the nearest integer, and convert it back to a length by multiplying by $h_p(t)$. Using this, we can write the regions occupied by the auxiliary regions to be $\smallsub{\Omega}{PA}(t) = (I(t)-\smallsub{h}{AR}(t),I(t))$ for the PDE auxiliary region and $\smallsub{\Omega}{BA}(t) = (I(t),I(t)+\smallsub{h}{AR}(t))$ for the Brownian-based auxiliary region.

Particle numbers for the PDE auxiliary region, $\smallsub{n}{PA}(t)$, and the Brownian auxiliary region, $\smallsub{n}{BA}(t)$, are calculated as in the PCM and GCM respectively, and are given by 
\begin{align}
\smallsub{n}{PA}(t) = \int_{\smallsub{\Omega}{PA}(t)}{u(x,t)\ dx}, \label{eqn:PDE_AR_Numbers}\\
\smallsub{n}{BA}(t) = \sum_{i=1}^{\smallsub{N}{B}(t)}{\mathbbm{1}_{[y_i(t)\in\smallsub{\Omega}{BA}(t)]}}. \label{eqn:Bro_AR_Numbers}
\end{align}
The implementation of the gARM is described in Algorithm \ref{alg:ARM} below.

\begin{Algorithm}{The growing auxiliary region method (Diffusion only)} \label{alg:ARM}
\item[] \textbf{Initialise:} Initial time --- $t=0$; Final time --- $t_f$; PDE solution --- $\bs{U}$; Positions of particles --- $\bs{y}$; Propensity functions --- $a_i$ for $i = 1,2$; Internal clock times --- $T_i$ for $i=1,2$; Next firing times --- $P_i$ for $i=1,2$; Times until next event --- $\Delta t_i$ for $i=1,2$; PDE/Brownian update step --- $\Delta t$; Time until next PDE/Brownian update --- $t_b$; Time until next re-mesh event --- $t_r$; Auxiliary region size $\smallsub{h}{AR}$.
\item At time $t > 0$: \label{alg_step:ARM_return}
\begin{enumerate}
	\item Calculate $\Delta = \min\set{i\in\{1,2\}}{\Delta t_i}$ and $\beta = \argmin\set{i\in\{1,2\}}{\Delta t_i}$. Set $t_\Delta = t + \Delta$.
	\item If $ \min\{t_\Delta,t_b,t_r\} = t_\Delta$:
	\begin{enumerate}
		\item For every $i\in\{1,2\}$, update $T_i$ according to $$T_i \leftarrow T_i + \frac{1}{2\rho}a_i\left(1-\exp\{-2\rho\Delta\}\right).$$
		\item For event $\beta$, set $$P_\beta \leftarrow P_\beta + \ln\left(\frac{1}{u_1}\right) \text{, where }u_1\sim\text{Unif}(0,1).$$
		\item Enact the event $\beta$:
		\begin{itemize}
			\item If the event $\beta$ corresponds to a jump from the PDE auxiliary region to the microscopic auxiliary region, draw $u_2\sim\text{Unif}(I(t)-\smallsub{h}{AR},I(t))$ and place a new particle at that position, and set $\bs{U}_{pc} \leftarrow \bs{U}_{pc} - 1/h_{pc}\bs{1}$, where $\bs{U}_{pc}$ are the pseudo-compartment nodes of the PDE solution, and $\bs{1}$ is a vector of ones of the appropriate size.
			\item If the event $\beta$ corresponds to a jump from the microscopic auxiliary region to the PDE auxiliary region, choose one of the particles contained in the microscopic auxiliary region uniformly at random and remove it, and set $\bs{U}_{pc} \leftarrow \bs{U}_{pc} + 1/h_{pc}\bs{1}$.
		\end{itemize}
		\item Set $t=t_\Delta$.
	\end{enumerate}
	\item Else if $ \min\{t_\Delta, t_b, t_r\} = t_b$:
	\begin{enumerate}
		\item For $i\in\{1,2\}$, update $T_i$ according to $$T_i \leftarrow T_i + \frac{1}{2\rho}a_i\left(1-\exp\{-2\rho(t_b - t)\}\right).$$
		\item Enact a Brownian update step using algorithm \ref{SUPP-alg:Ind-Update}.
		\item Enact a PDE update step using algorithm \ref{SUPP-alg:PDE}.
		\item Set $t \leftarrow t_b$. Set $t_b \leftarrow t_b + \Delta t$.
	\end{enumerate}
	\item Else:
	\begin{enumerate}
		\item For $i\in\{1,2\}$, update $T_i$ according to $$T_i \leftarrow T_i + \frac{1}{2\rho}a_i\left(1-\exp\{-2\rho(t_r - t)\}\right).$$
		\item Re-mesh the PDE solution according to Algorithm \ref{SUPP-alg:Remesh}
		\item Set $h_p \leftarrow h_p/2$.
		\item Set $t \leftarrow t_r$. Update $t_r$ according to Algorithm \ref{SUPP-alg:Remesh}.
	\end{enumerate}
	\item Update $\smallsub{h}{AR}$.
	\item Update all propensity functions $a_i$, for $i\in\{1,2\}$.
	\item Update $\Delta t_i$ according to $$\Delta t_i = -\frac{1}{2\rho}\ln\left(1-\frac{2\rho(P_i-T_i)}{a_i}\right),$$for $i\in\{1,2\}$.
\end{enumerate}
\item If $t<t_f$, return to \ref{alg_step:ARM_return}. Otherwise, end.
\end{Algorithm}

\section{Results} \label{sect:Results}

Within this section, we present results from three test problems, for all three of the methods described in Sections \ref{sect:PCM}-\ref{sect:ARM}. The three test problems are designed to evaluate the performance of the algorithms in comparison to the corresponding PDE solutions. As such, the examples are relatively simple so that the PDE is in exact correspondence with the expected behaviour of the individual-based methods (i.e. no reactions of order higher than one). These choices mean that discrepancies between the mean behaviour of the hybrid methods and the PDE solutions can be be attributed directly to the hybridisation. All examples will be on a one-dimensional, exponentially growing domain, but can be straightforwardly extended to higher dimensions on Cartesian domains, and to other forms of uniform domain growth.

The next three subsections will be devoted to the three test problems which will each assess a different aspect of the performance of the three hybrid algorithms.

\subsection{Test problem 1: Maintaining uniformity}

The first test problem verifies	 that the algorithms are able to maintain a uniform particle distribution under pure diffusion. For this, we will use the growing domain diffusion equation: \begin{align}
\partder{u}{t}(x,t) &= D\secpartder{u}{x}(x,t) - \rho\partder{(xu(x,t))}{x} && x\in(0,2\exp\{\rho t\}),\ t>0,\\
\partder{u}{x}(0,t) &= 0 && t > 0,\\
\partder{u}{x}(2\exp\{\rho t\},t) &= 0 && t>0\\
u(x,0) &= \frac{M}{2} && x \in [0,2].
\end{align}Here, $M$ is the number of particles in the system, and all other parameters are as in Sections \ref{sect:PCM}---\ref{sect:ARM}. This PDE system has an analytical solution of \begin{equation}
u(x,t) = \frac{M}{2}\exp\{-\rho t\}
\end{equation} We run this example with a diffusion coefficient of $D=0.0025$, an exponential growth rate of $\rho=0.001$ and $M=500$ particles. Each hybrid simulation is averaged over 1000 independent repeats for comparison and error plotting purposes.
 
\begin{figure}
\begin{center}
\vspace{-20pt}
\subfigure[][]{\includegraphics[width=0.31\textwidth]{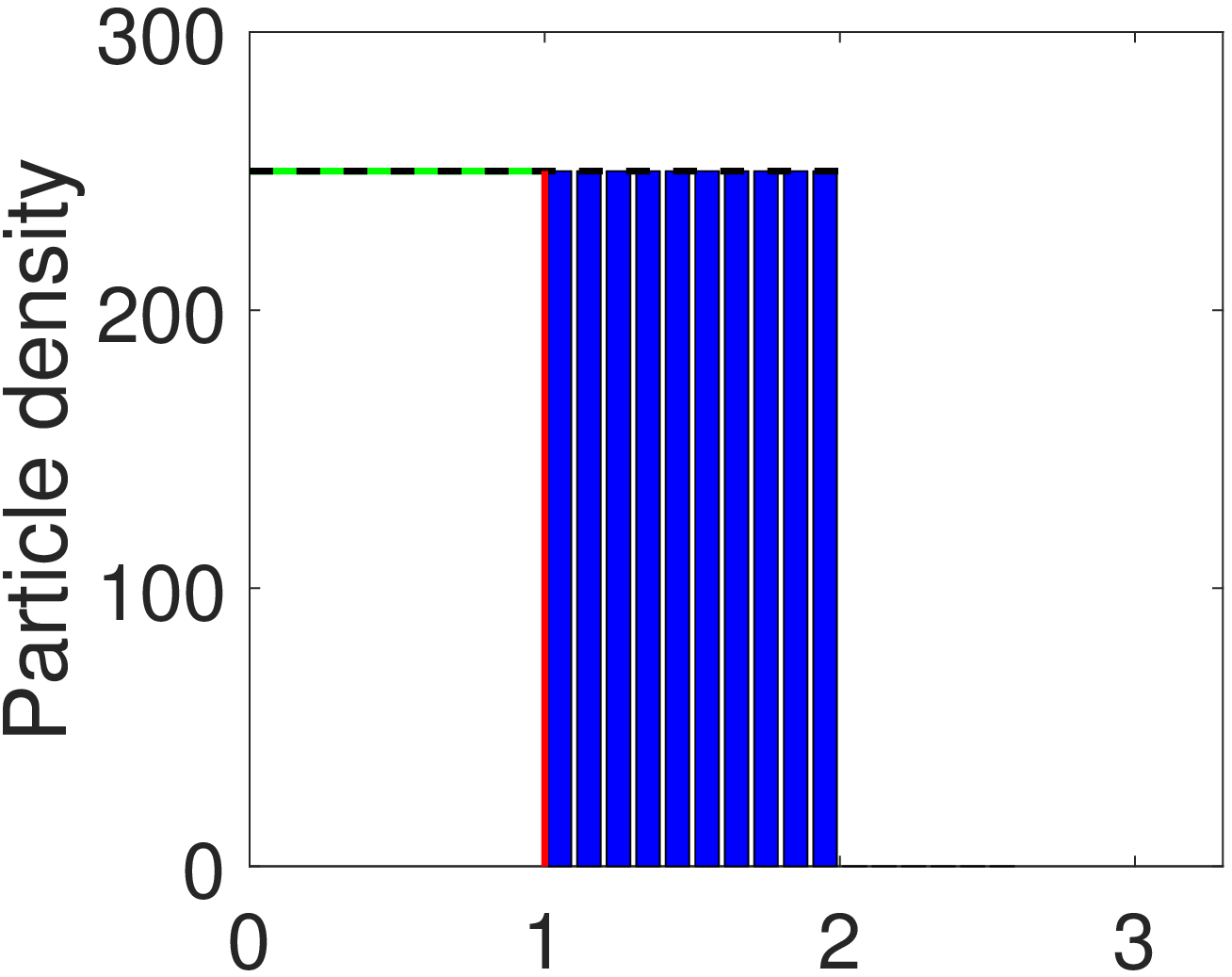}
\label{fig:Uniform_Results_PCM_0}}
\subfigure[][]{\includegraphics[width=0.31\textwidth]{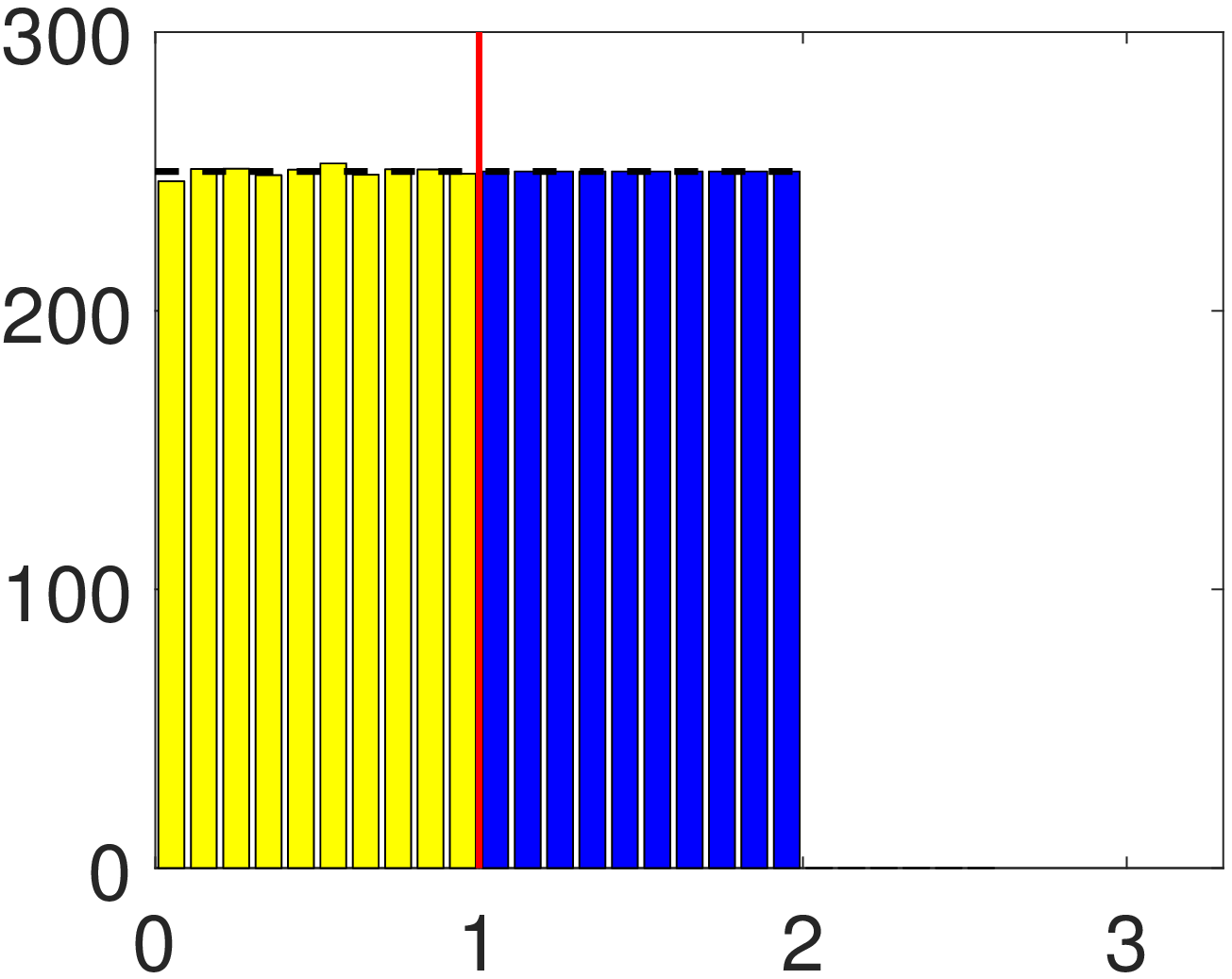}
\label{fig:Uniform_Results_GCM_0}}
\subfigure[][]{\includegraphics[width=0.31\textwidth]{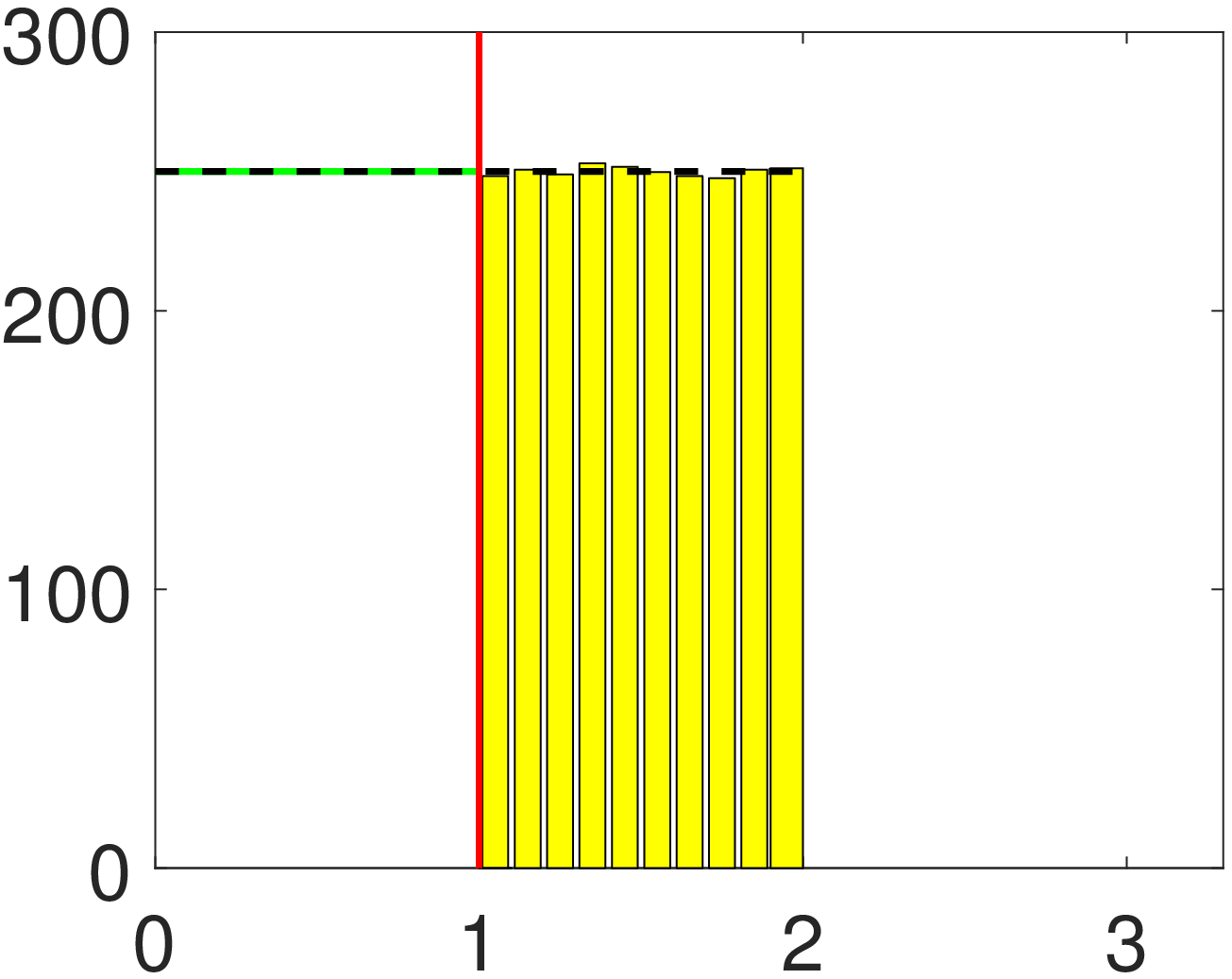}
\label{fig:Uniform_Results_ARM_0}}
\subfigure[][]{\includegraphics[width=0.31\textwidth]{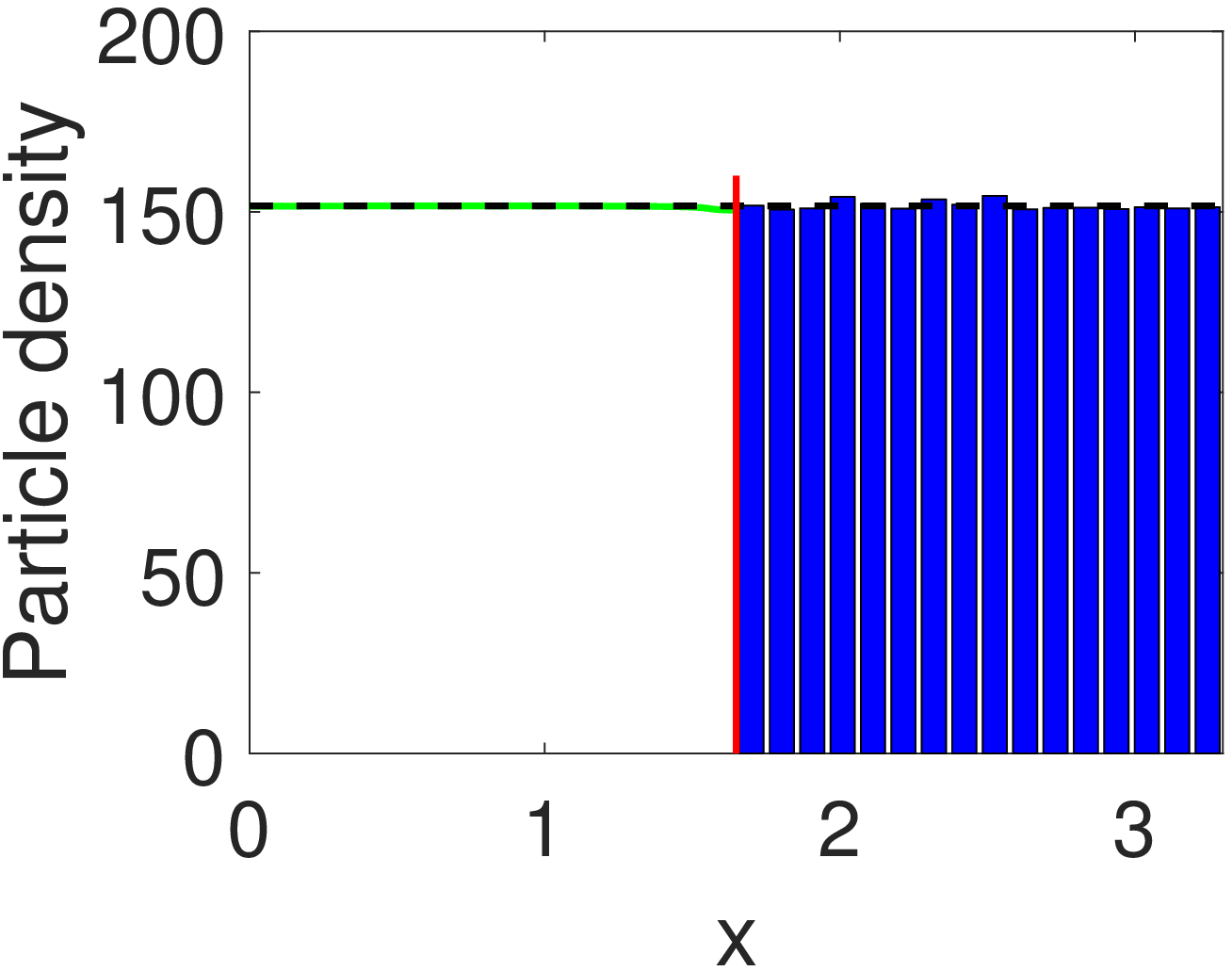}
\label{fig:Uniform_Results_PCM_T}}
\subfigure[][]{\includegraphics[width=0.31\textwidth]{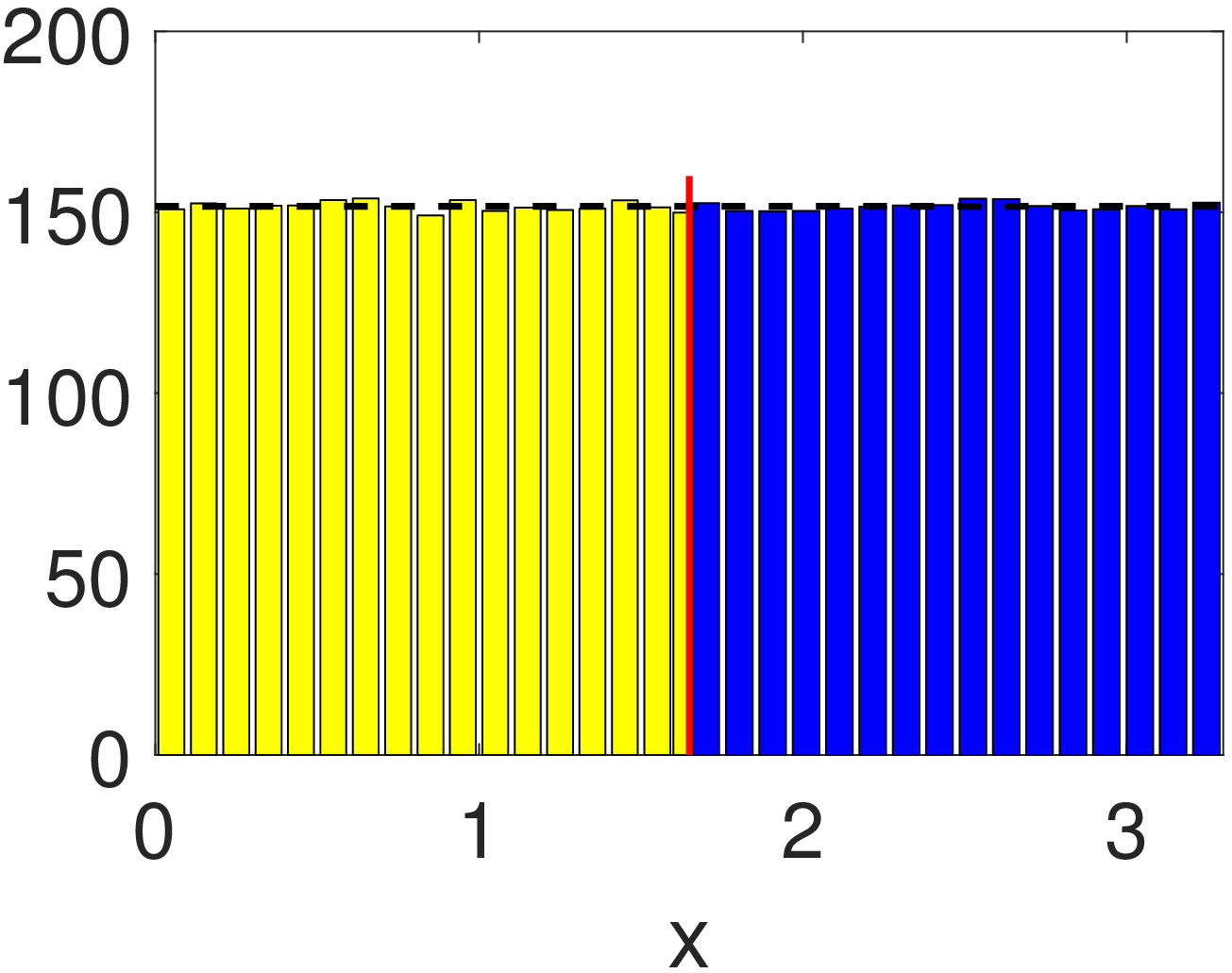}
\label{fig:Uniform_Results_GCM_T}}
\subfigure[][]{\includegraphics[width=0.31\textwidth]{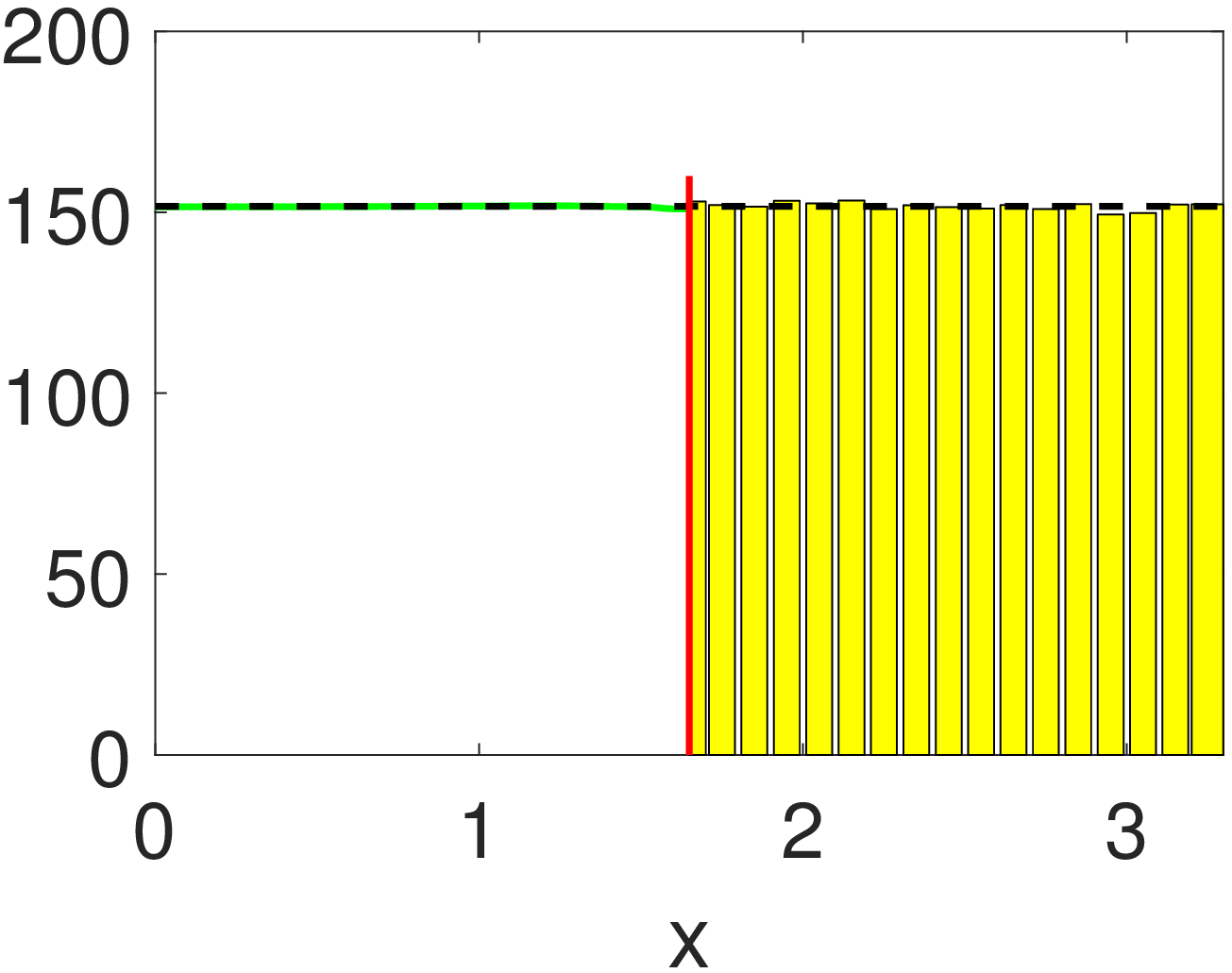}
\label{fig:Uniform_Results_ARM_T}}
\subfigure[][]{\includegraphics[width=0.31\textwidth]{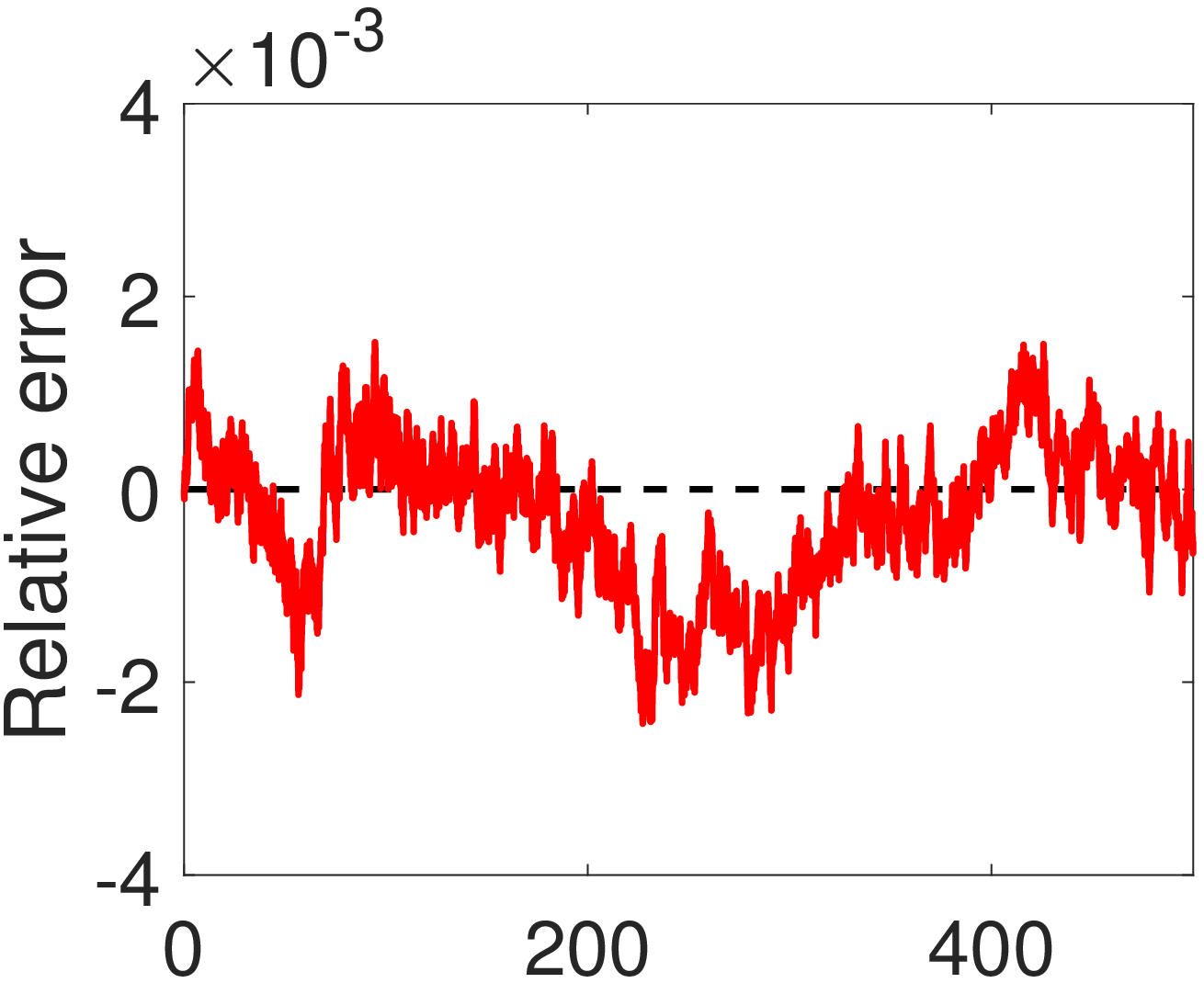}
\label{fig:Uniform_Results_PCM_L}}
\subfigure[][]{\includegraphics[width=0.31\textwidth]{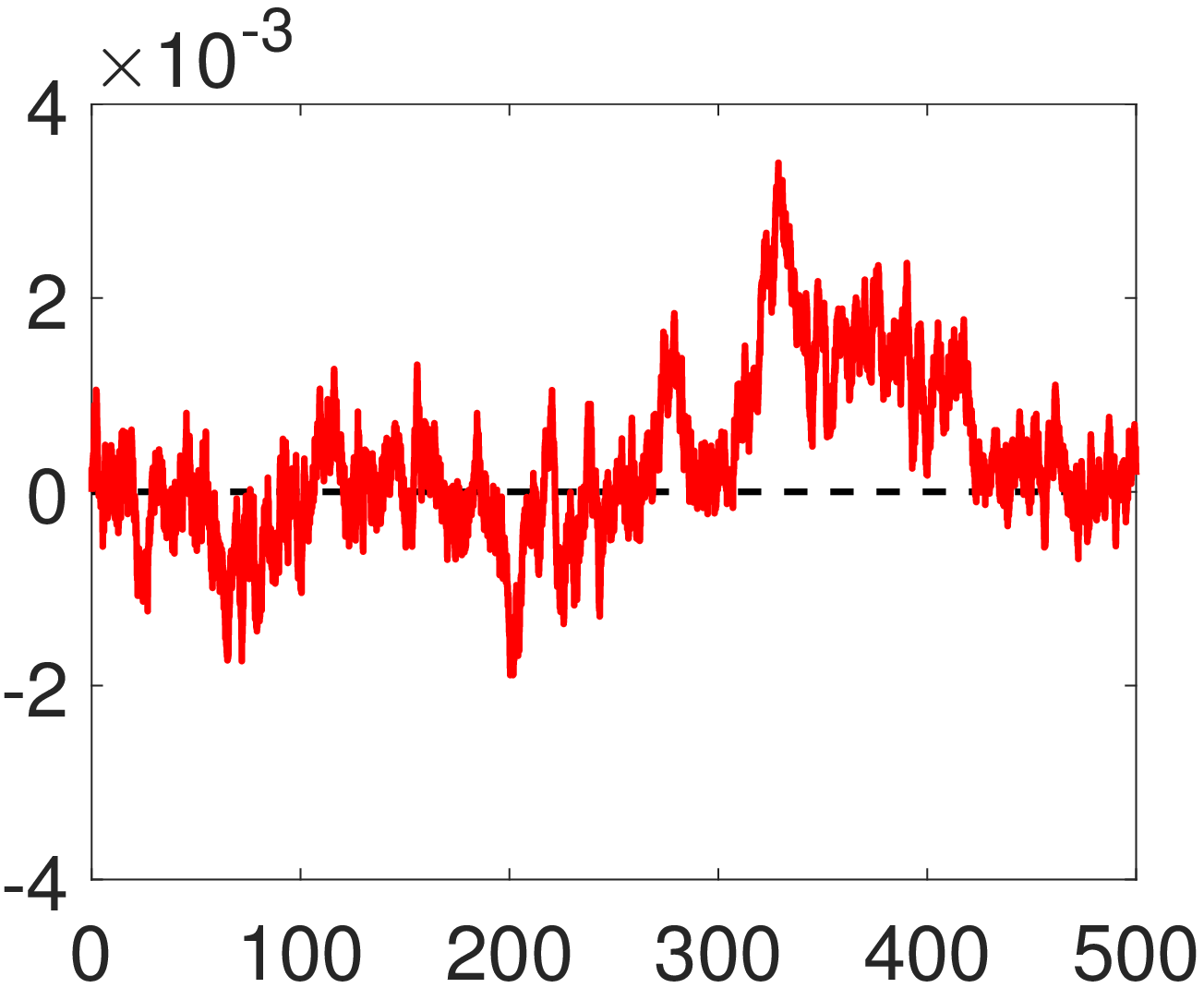}
\label{fig:Uniform_Results_GCM_L}}
\subfigure[][]{\includegraphics[width=0.31\textwidth]{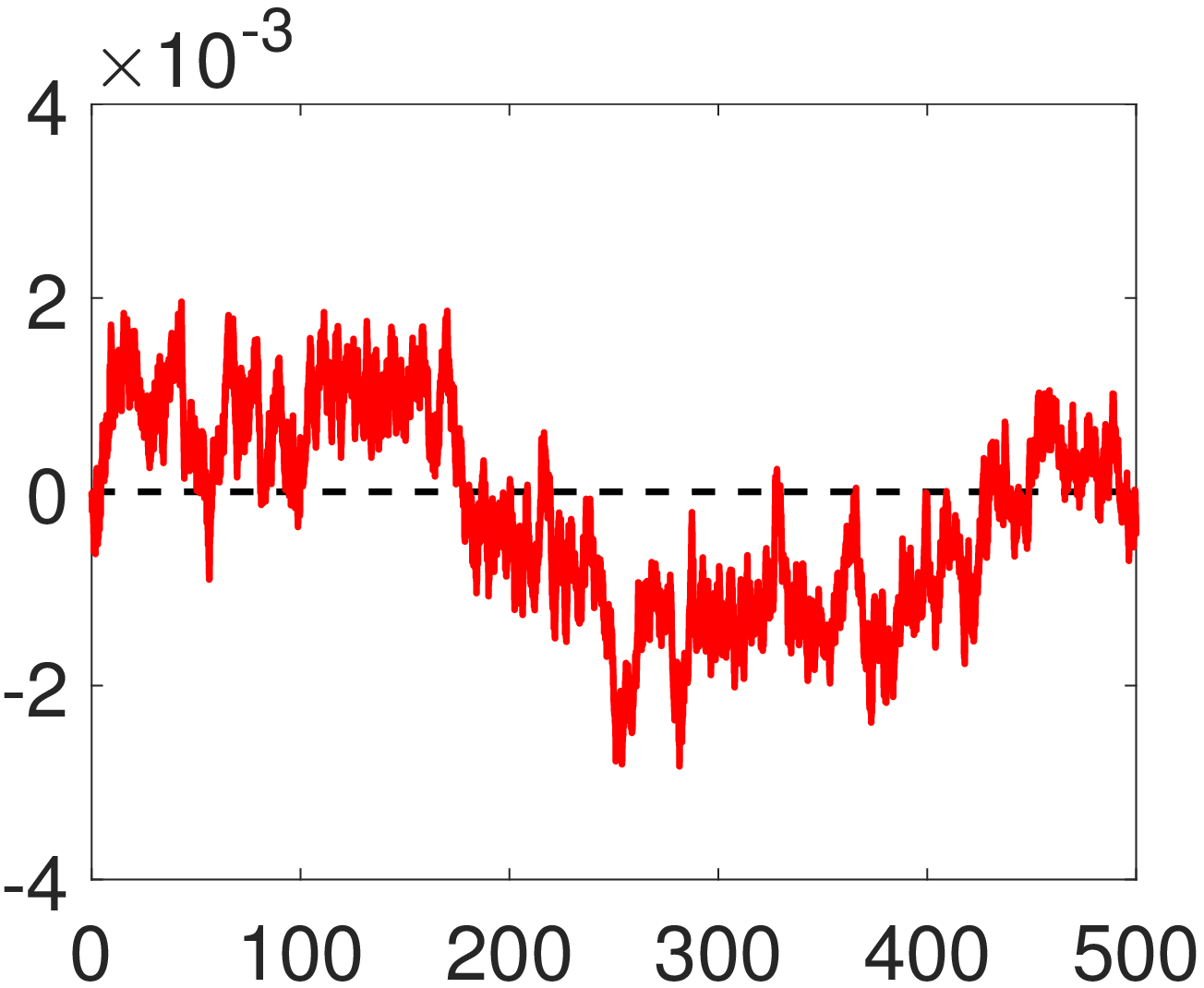}
\label{fig:Uniform_Results_ARM_L}}
\subfigure[][]{\includegraphics[width=0.31\textwidth]{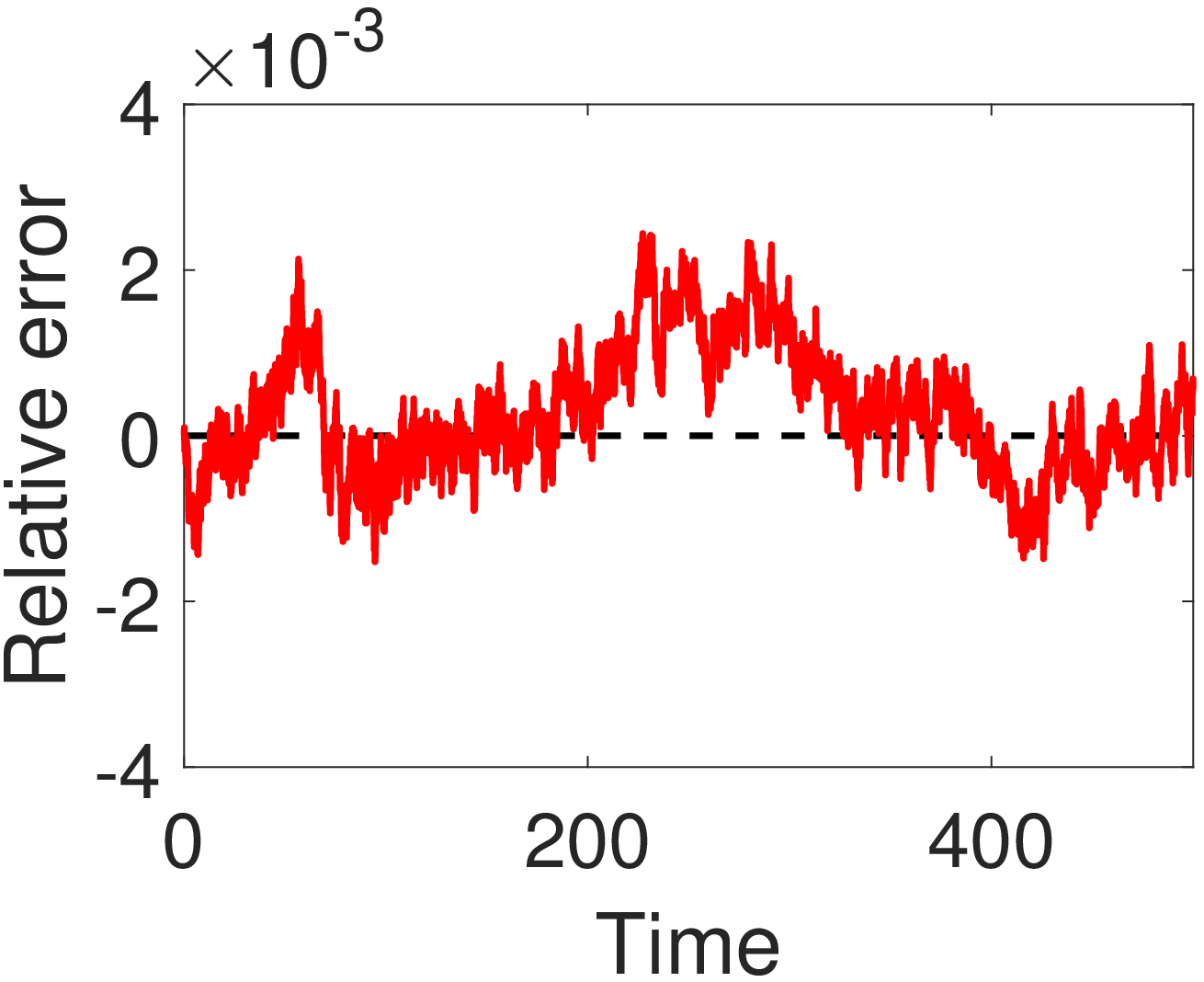}
\label{fig:Uniform_Results_PCM_R}}
\subfigure[][]{\includegraphics[width=0.31\textwidth]{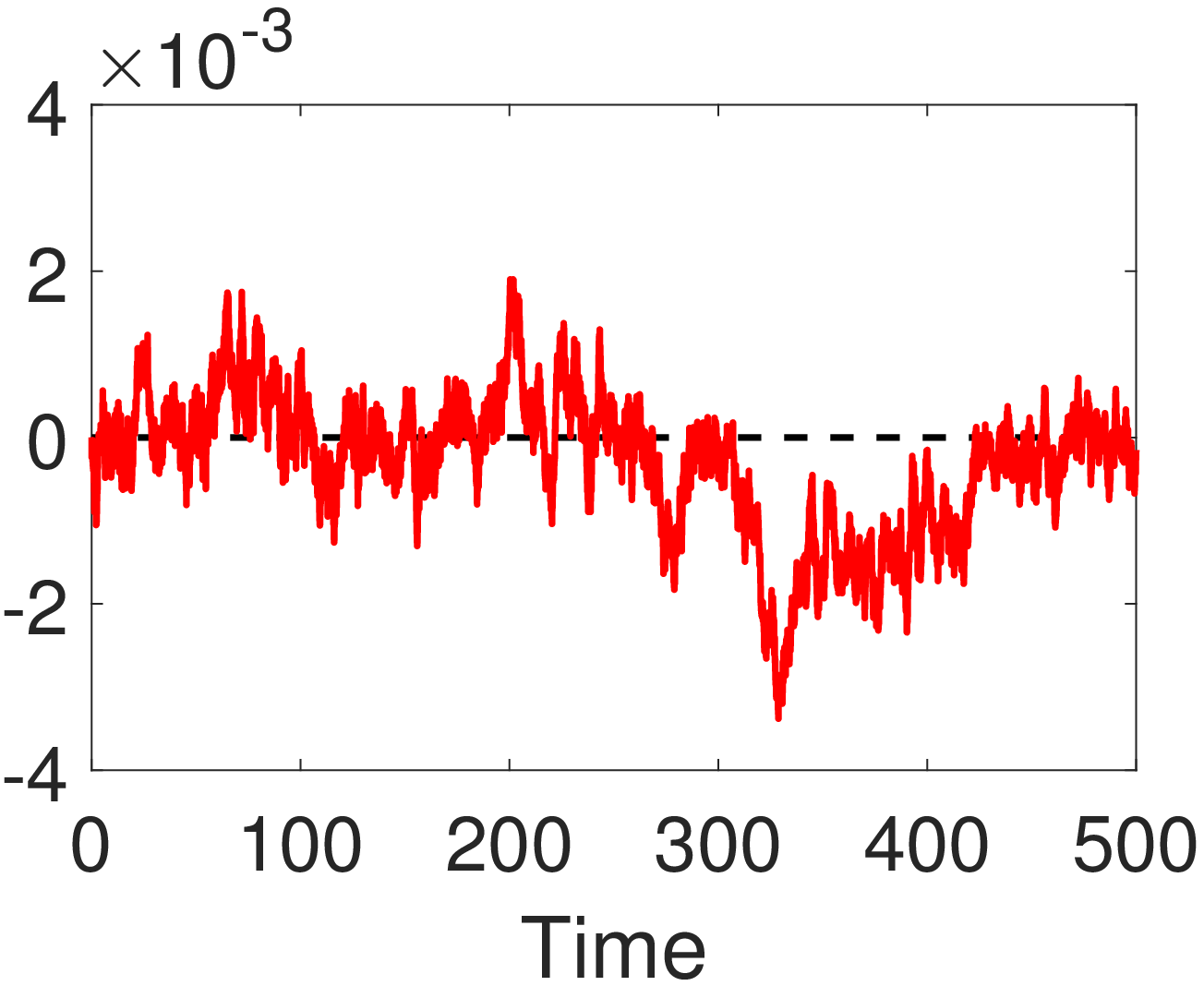}
\label{fig:Uniform_Results_GCM_R}}
\subfigure[][]{\includegraphics[width=0.31\textwidth]{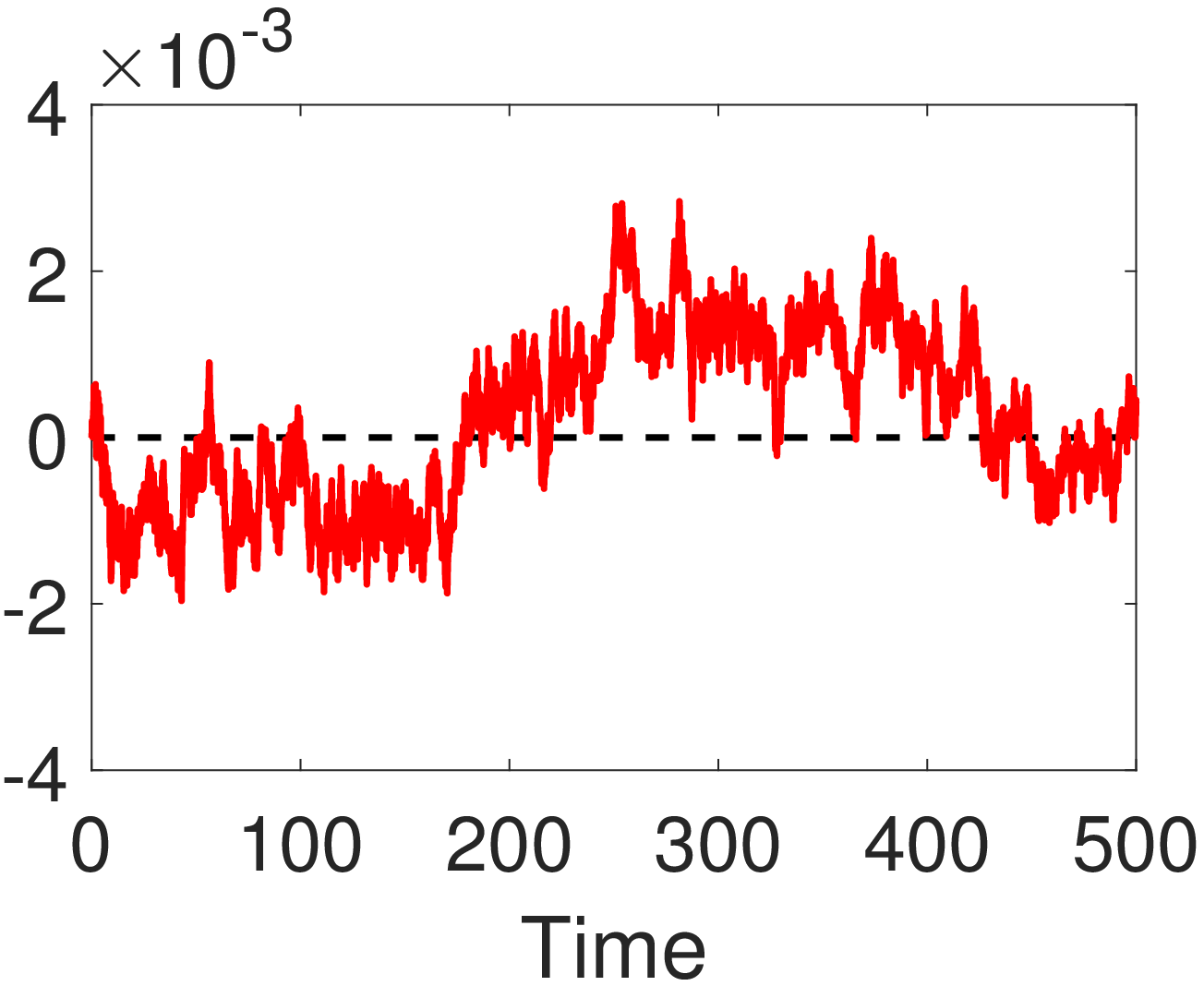}
\label{fig:Uniform_Results_ARM_R}}
\end{center}
\vspace{-10pt}
\caption{Results for Test Problem 1. We present results for the gPCM (Section \ref{sect:PCM}) in column 1, the gGCM (Section \ref{sect:GCM}) in column 2 and the gARM (Section \ref{sect:ARM}) in column 3. Figures \subref{fig:Uniform_Results_PCM_0}-\subref{fig:Uniform_Results_ARM_T} are snapshots of each method at the initial (row 1) and final ($t=500$, row 2) times of the simulation for the three methods. Green lines denote the PDE solution of the hybrid methods, blue bars are the particle densities for the mesoscale within the hybrid methods, and the microscale densities are denoted by the yellow bars, where we have binned particles onto the same mesh as the compartments. The red vertical lines on each plot denote the position of the interface at that time, and the black dashed line is the solution of the PDE across the whole domain, which we consider as our ground truth. Figures \subref{fig:Uniform_Results_PCM_L}-\subref{fig:Uniform_Results_ARM_R} display the relative errors in the left subdomain (row 3) and right subdomain (row 4) corresponding to the different modelling paradigms in the hybrid methods. The red curves denoting the relative error are given by the formula \eqref{eqn:Left_Error} for the left subdomain, with an analogous formula for the right side. The black dashed line corresponds to an error of 0. There is no bias in the error in the positive or negative direction for any of the methods.}
\label{fig:Uniform_Results}
\end{figure}

In Figure \ref{fig:Uniform_Results}, we present the results for each of the hybrid methods (gPCM in column 1, gGCM in column 2 and gARM in column 3), with snapshots of the solution at the initial time (row 1) and final time $t_f = 500$ (row 2), and then the relative errors in the left (row 3) and right (row 4) halves of the domain. The relative error is calculated as \begin{equation}
\smallsub{E}{L}(t) = \frac{\smallsub{n}{L}^H(t) - \smallsub{n}{L}^P(t)}{\smallsub{n}{L}^P(t)},
\label{eqn:Left_Error}
\end{equation} for the left side of the domain. Here, $\smallsub{E}{L}(t)$ is the relative error in the left side of the domain at time $t$, $\smallsub{N}{L}^H(t)$ is the number of particles in the left subdomain of the hybrid method at time $t$ and $\smallsub{n}{L}^P(t)$ is the number of particles calculated in the left-hand side of the PDE solution at time $t$. As can be seen from the plots, each hybrid method is able to correctly maintain uniformity with no bias in particle numbers to either subdomain.

\subsection{Test problem 2: Testing flux}

The second test problem is designed to assess the ability of each of the hybrid methods to cope with a non-zero flux acrosss the interface. For this, we use the same PDE from test problem 1, but change the boundary conditions to ensure a flux over the entire domain. The PDE with its boundary and initial conditions are below:\begin{align}
\partder{u}{t}(x,t) &= D\secpartder{u}{x}(x,t) - \rho\partder{(xu(x,t))}{x} && x\in(0,2\exp\{\rho t\}),\ t>0,\\
-D\partder{u}{x}(0,t) &= Ru(2\exp\{\rho t\},t) && t > 0,\label{eqn:periodic_left}\\
-D\partder{u}{x}(2\exp\{\rho t\},t) &= Ru(2\exp\{\rho t\},t) && t>0,\label{eqn:robin_right}\\
u(x,0) &=  \frac{M}{2} && x \in [0,2].
\end{align} 
Equation \eqref{eqn:robin_right} specifies a Robin boundary condition and equation \eqref{eqn:periodic_left} is the corresponding condition which ensures periodic boundaries: particles that exit the domain at the right-hand end will re-enter the domain at the left-hand end. The resultant steady state (when the domain is not growing) is a linear gradient in density. 

\begin{figure}
\begin{center}
\subfigure[][]{\includegraphics[width=0.31\textwidth]{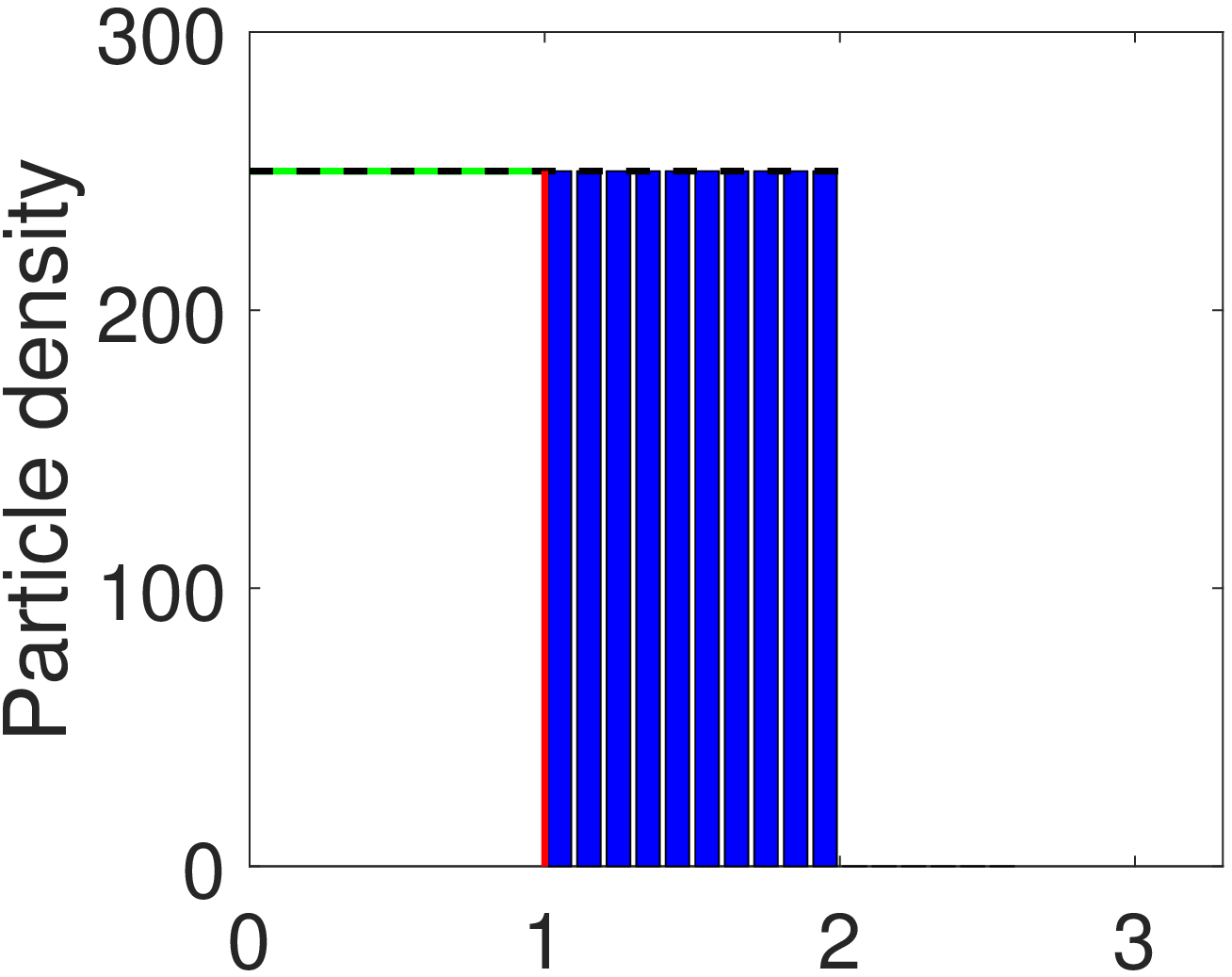}
\label{fig:Gradient_Results_PCM_0}}
\subfigure[][]{\includegraphics[width=0.31\textwidth]{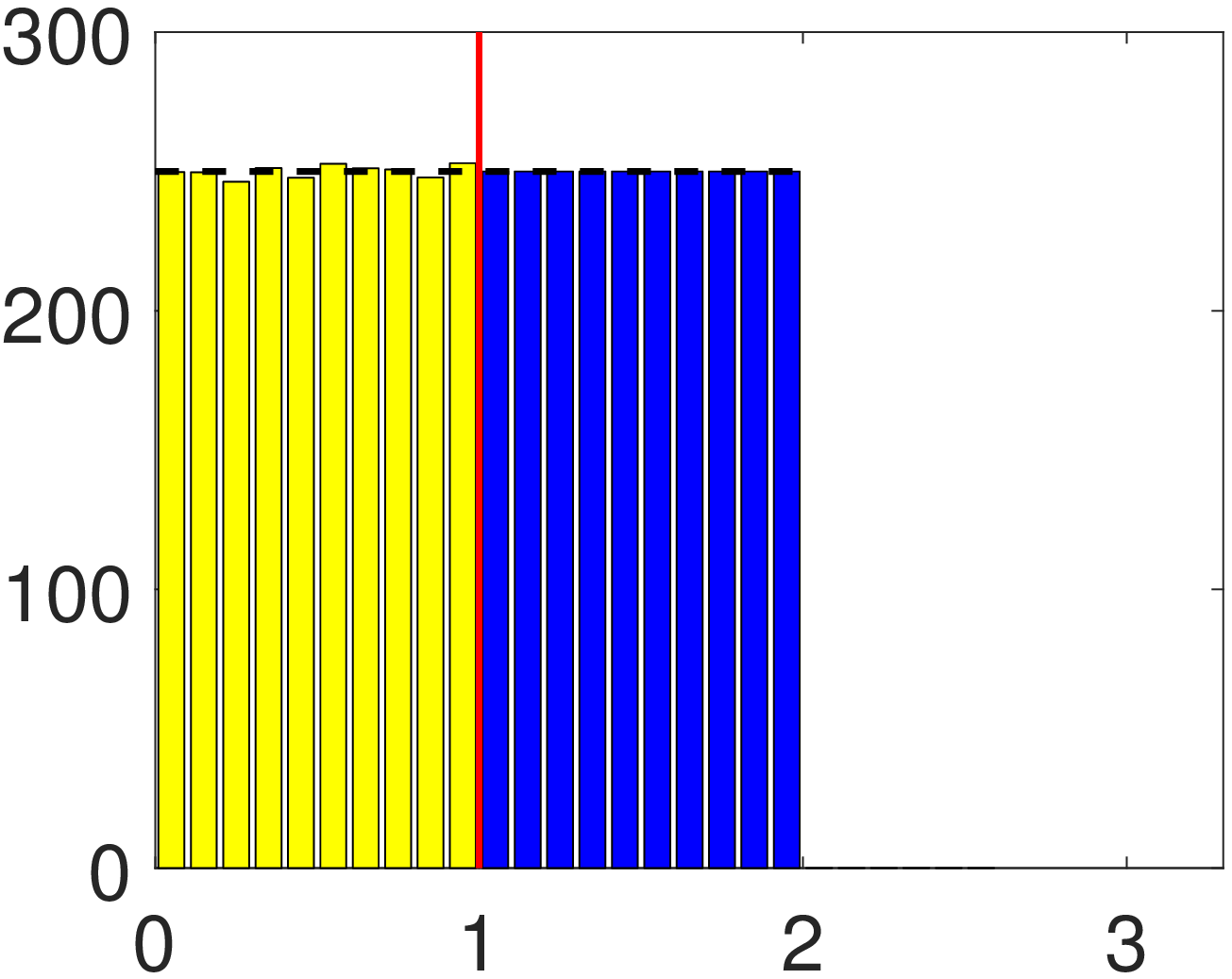}
\label{fig:Gradient_Results_GCM_0}}
\subfigure[][]{\includegraphics[width=0.31\textwidth]{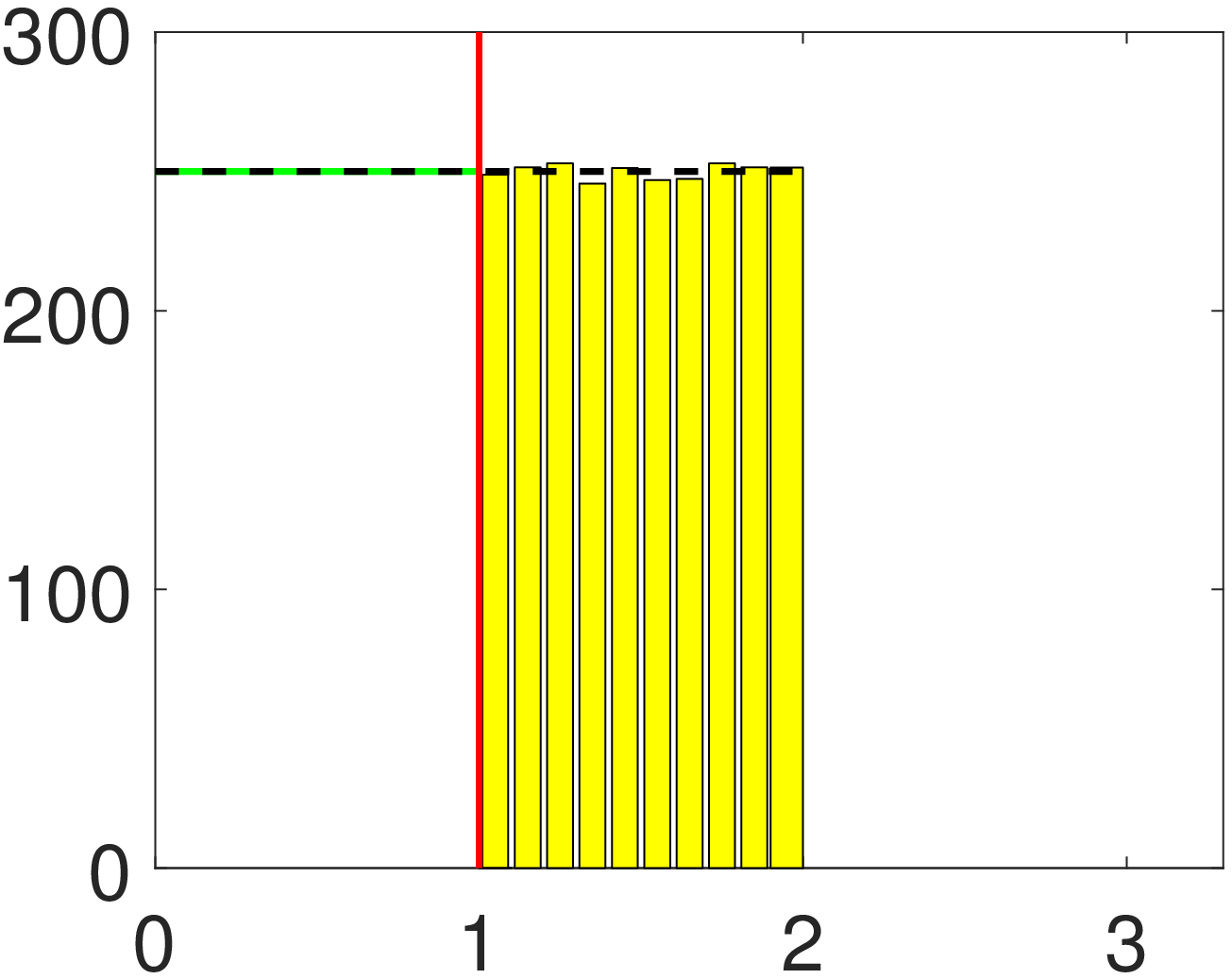}
\label{fig:Gradient_Results_ARM_0}}
\subfigure[][]{\includegraphics[width=0.31\textwidth]{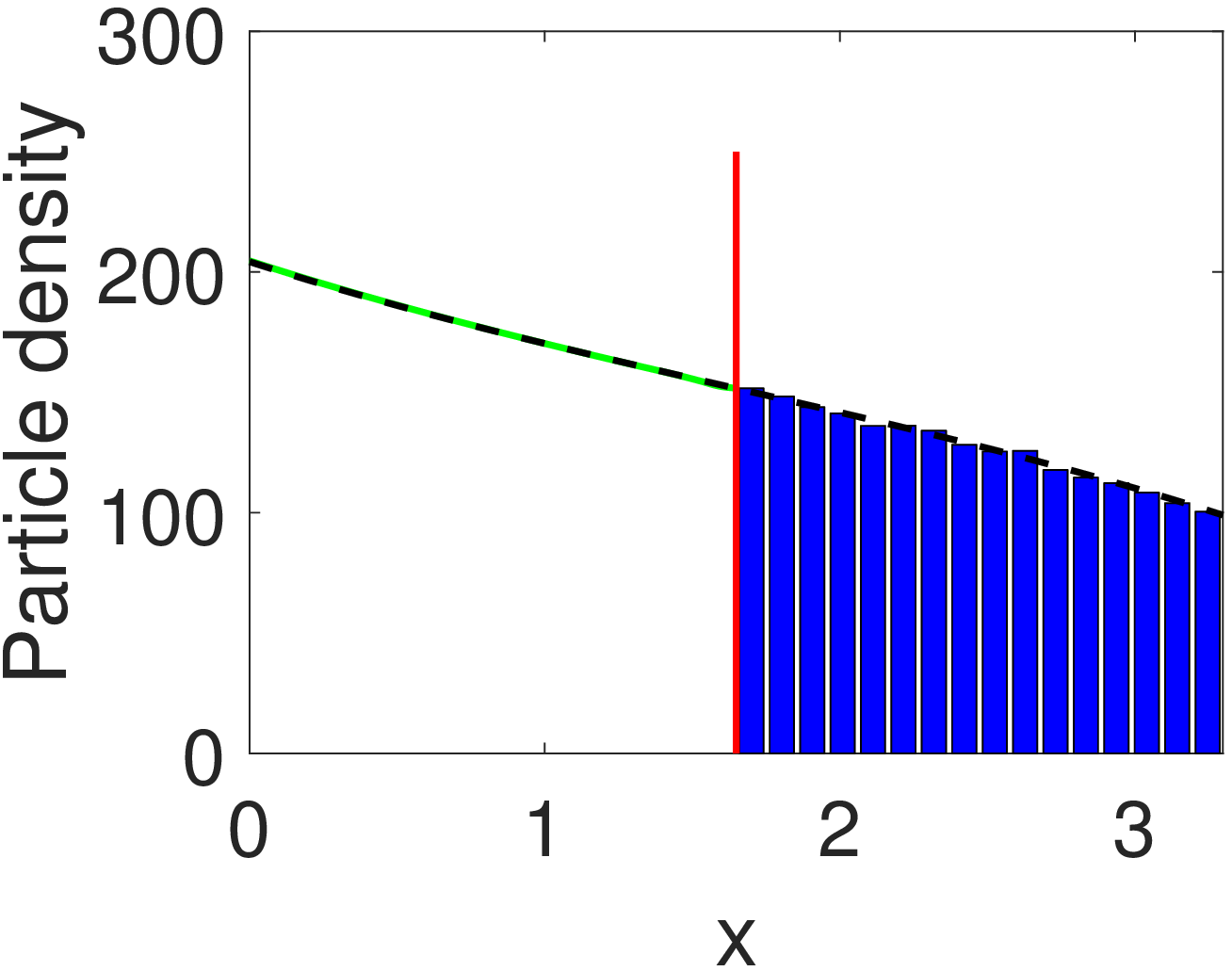}
\label{fig:Gradient_Results_PCM_T}}
\subfigure[][]{\includegraphics[width=0.31\textwidth]{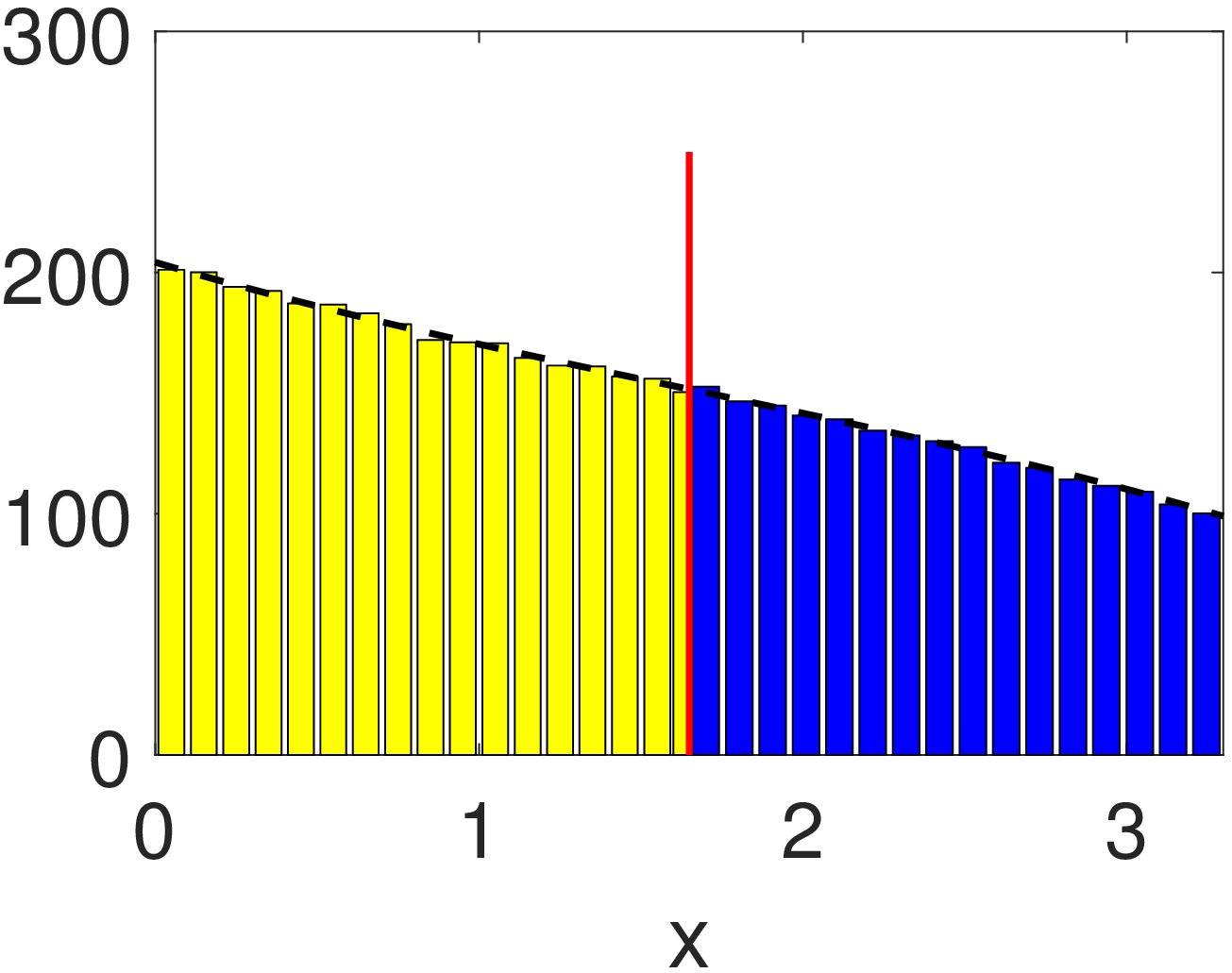}
\label{fig:Gradient_Results_GCM_T}}
\subfigure[][]{\includegraphics[width=0.31\textwidth]{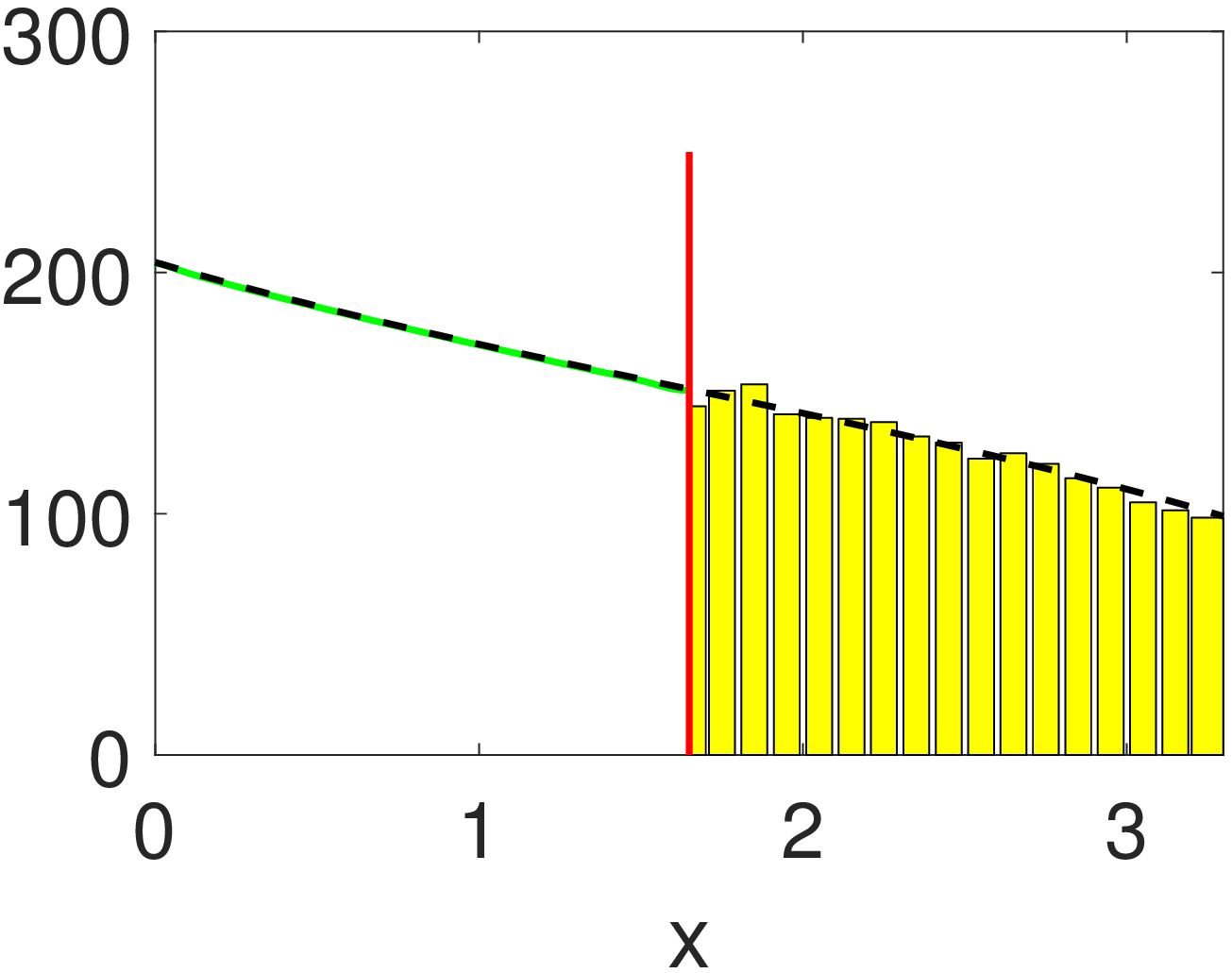}
\label{fig:Gradient_Results_ARM_T}}
\subfigure[][]{\includegraphics[width=0.31\textwidth]{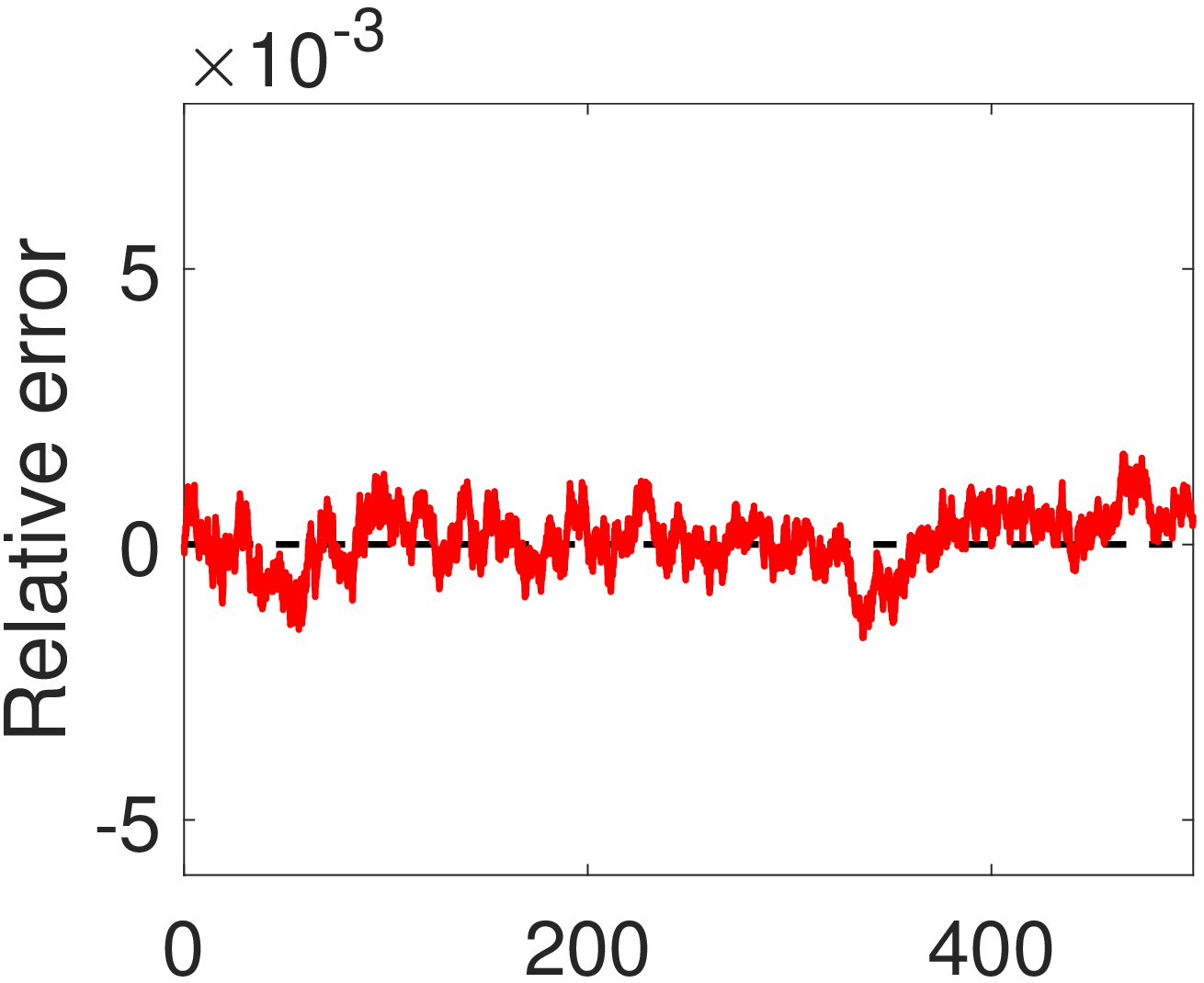}
\label{fig:Gradient_Results_PCM_L}}
\subfigure[][]{\includegraphics[width=0.31\textwidth]{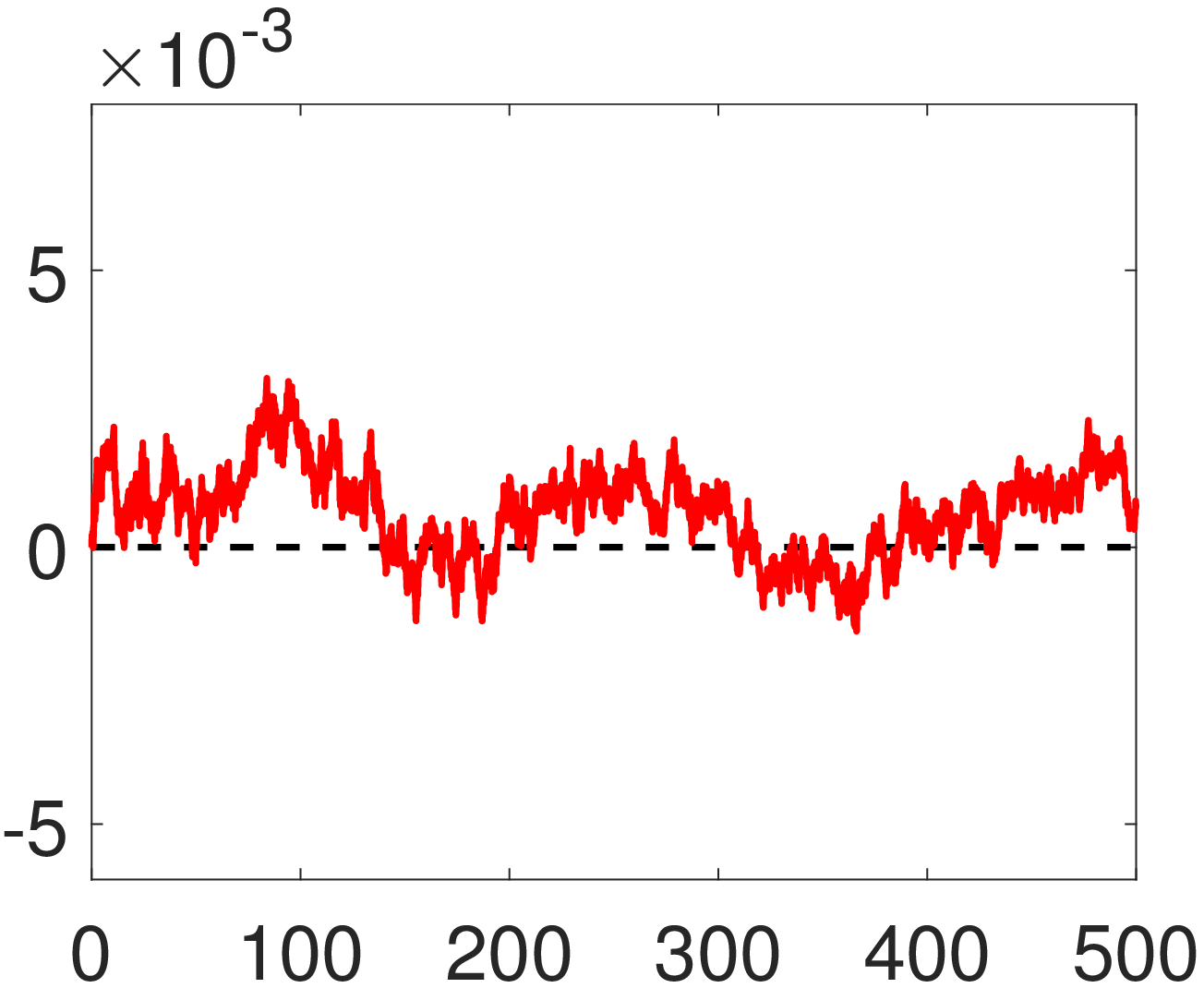}
\label{fig:Gradient_Results_GCM_L}}
\subfigure[][]{\includegraphics[width=0.31\textwidth]{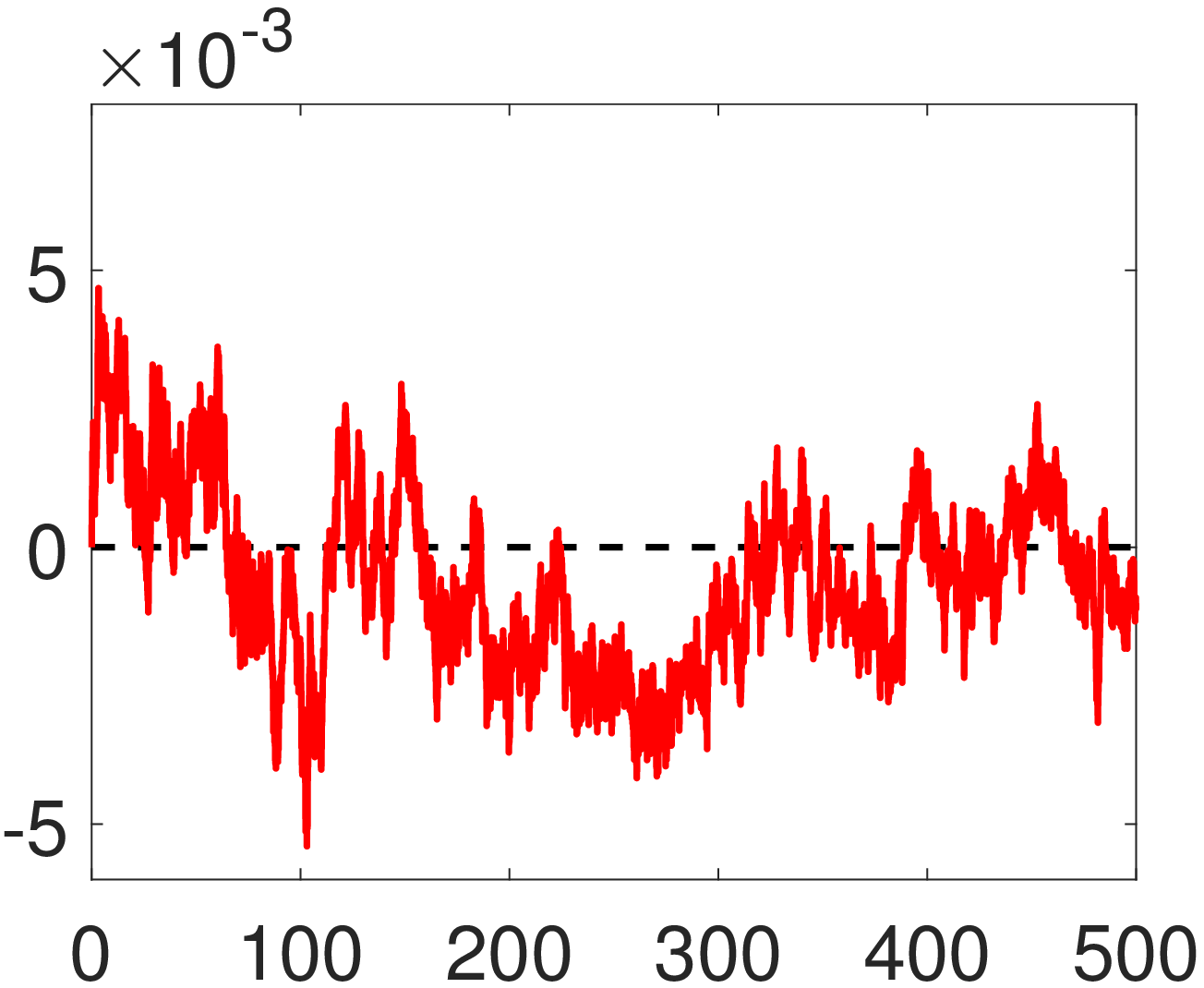}
\label{fig:Gradient_Results_ARM_L}}
\subfigure[][]{\includegraphics[width=0.31\textwidth]{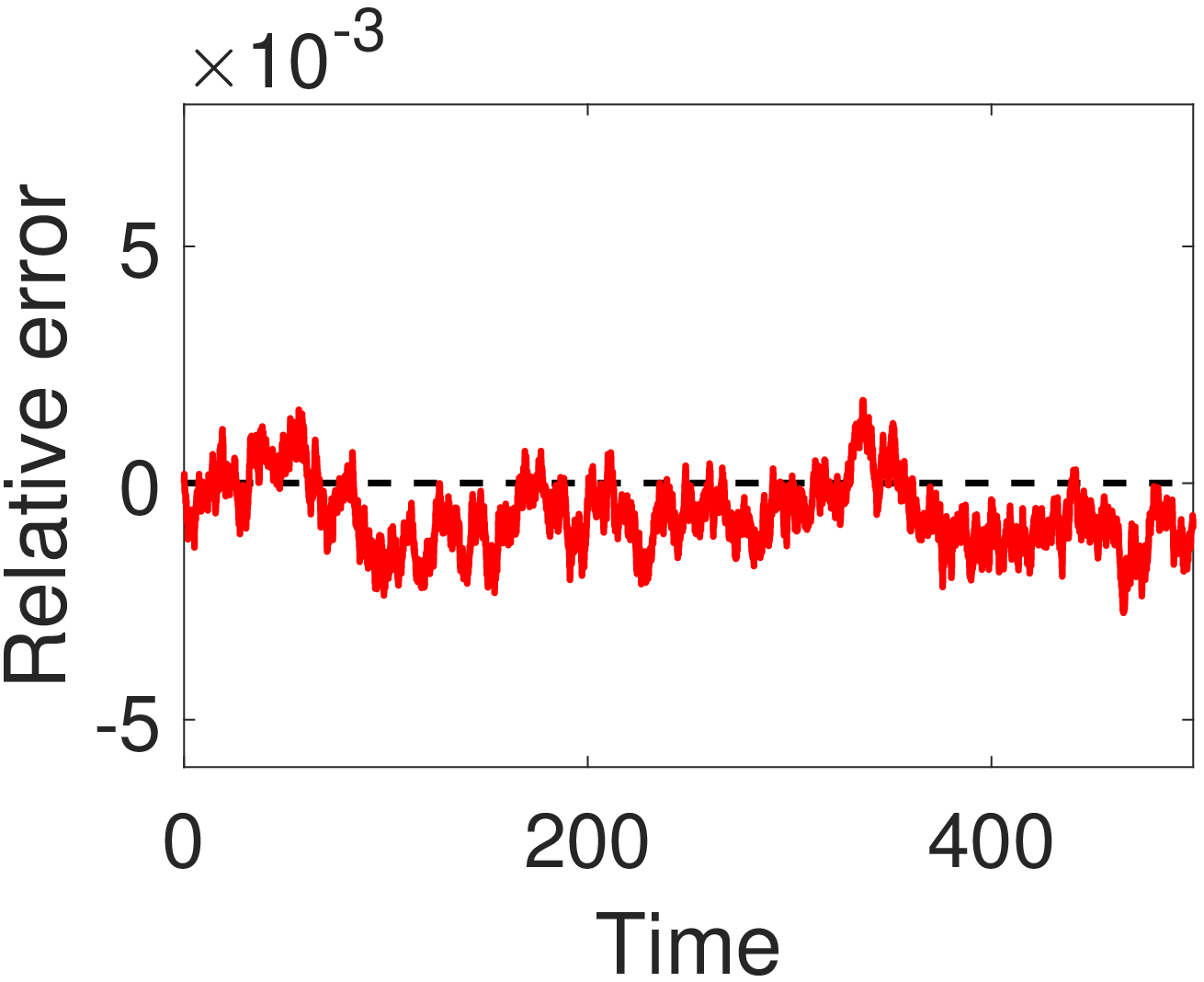}
\label{fig:Gradient_Results_PCM_R}}
\subfigure[][]{\includegraphics[width=0.31\textwidth]{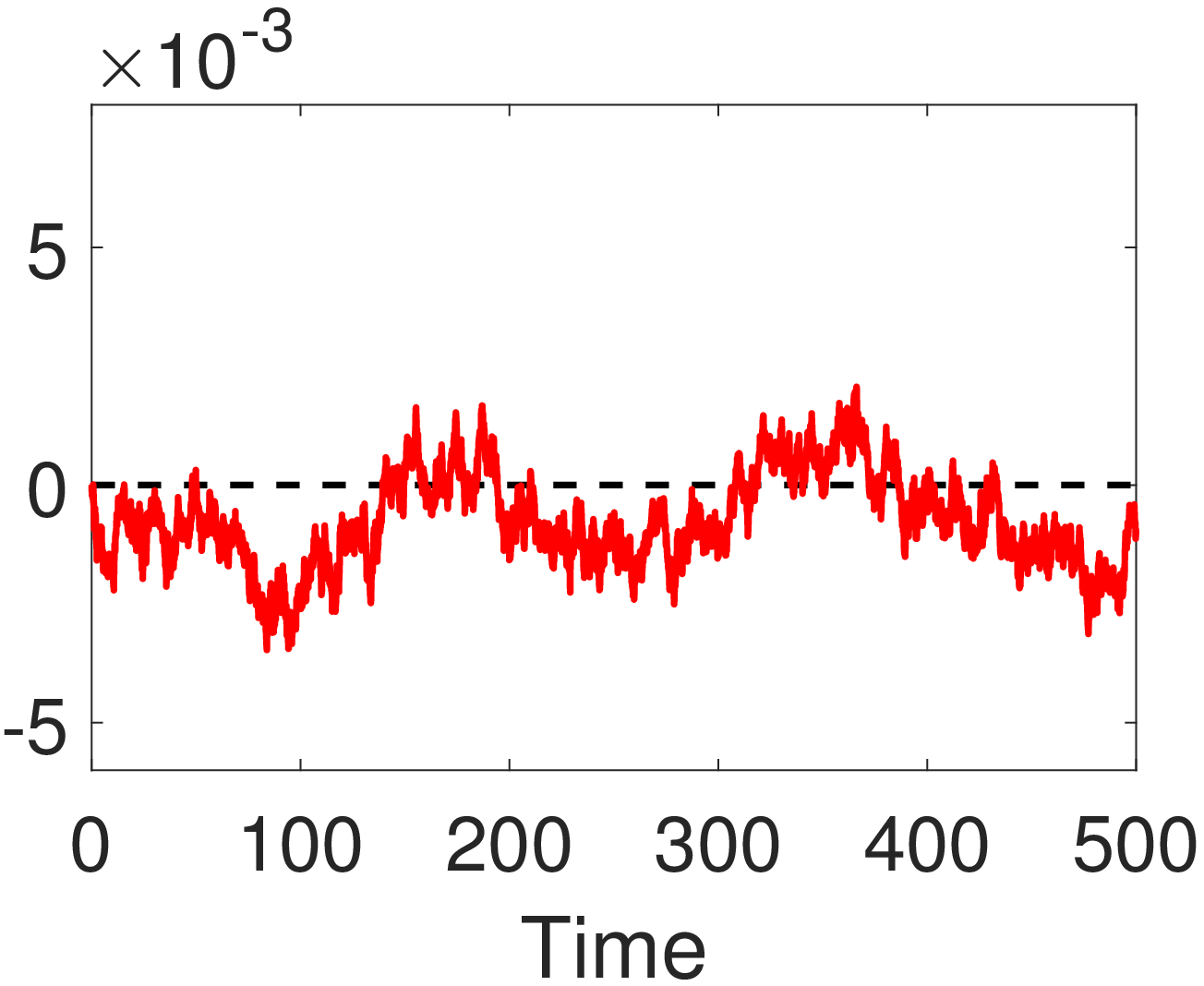}
\label{fig:Gradient_Results_GCM_R}}
\subfigure[][]{\includegraphics[width=0.31\textwidth]{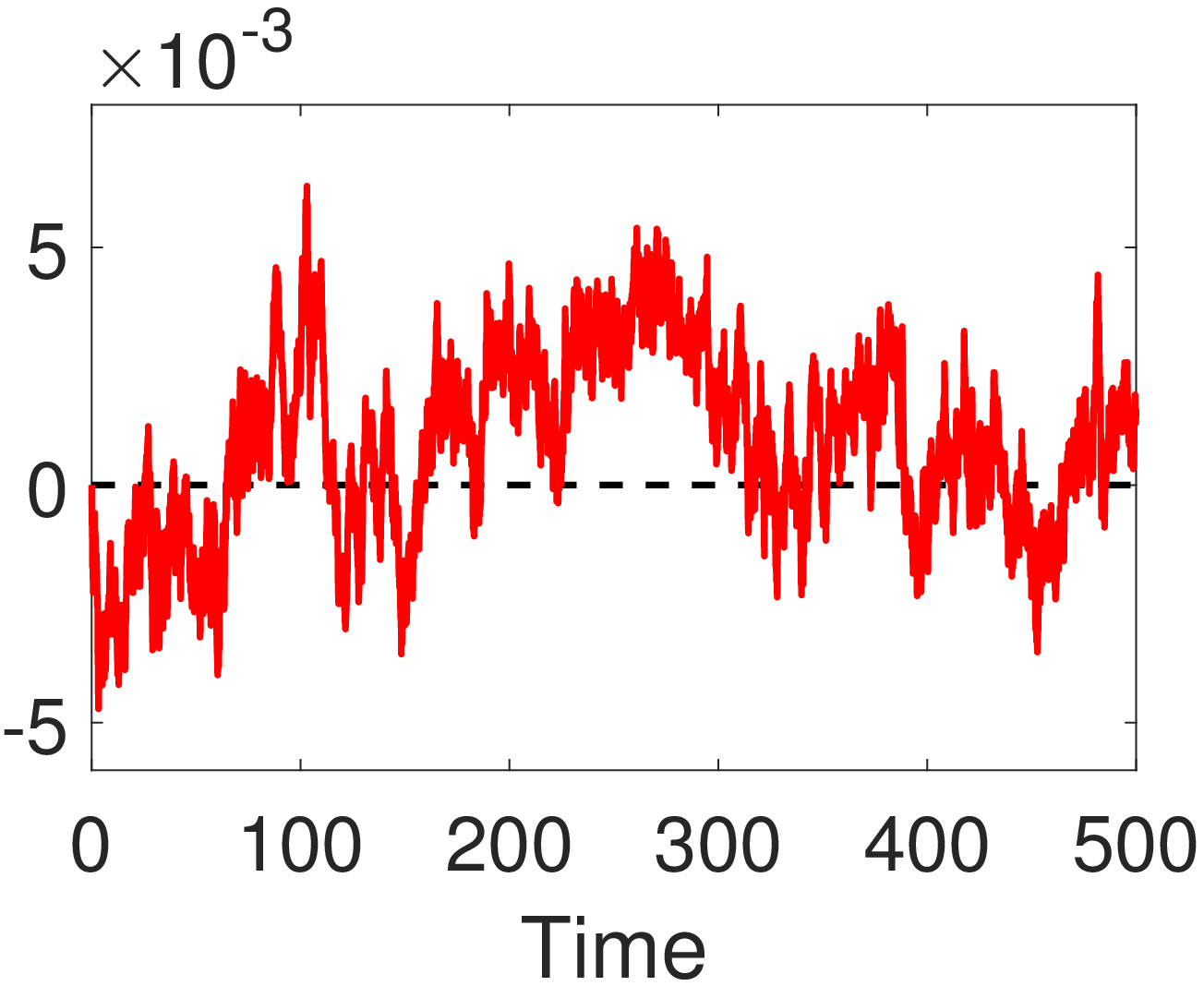}
\label{fig:Gradient_Results_ARM_R}}
\end{center}
\caption{Results for Test Problem 2. Figure descriptions are as in Figure \ref{fig:Uniform_Results}.}
\label{fig:Gradient_Results}
\end{figure}

We present our results for this test problem in Figure \ref{fig:Gradient_Results}. As in the case with the first test problem, we can see that all three of the hybrid methods perform accurately with no discernible bias in particle numbers on either side of the domain. This demonstrates that the hybrid methods are able to resolve gradients over the interface. 

\subsection{Test problem 3: Morphogen gradient formation}

For the final test problem, we investigate the formation of a morphogen gradient on a growing domain.
This example is designed to determine whether the hybrid methods are able to accurately perform when zeroth- and first-order reactions are incorporated\footnote{In this proof of principle paper we do not include examples of second- or higher-order interactions. We know that the mean-field PDE model does not correspond exactly to the mean of the stochastic models in these cases due to the necessity for moment closure when deriving the continuum equations from the individual-based models. To accurately determine whether our hybrd coupling introduces bias we consider only examples in which the expected behaviour of the individual-based method matches the behaviour of the equivalent continuum model.}. The PDE, boundary and initial conditions are below: \begin{align}
\partder{u}{t}(x,t) &= D\secpartder{u}{x}(x,t) - \rho\partder{(xu(x,t))}{x} - \mu u&& x\in(0,2\exp\{\rho t\}),\ t>0,\\
\partder{u}{x}(0,t) &= -\lambda && t > 0,\\
\partder{u}{x}(2\exp\{\rho t\},t) &= 0 && t>0,\\
u(x,0) &= \frac{M}{2} && x \in [0,2].
\end{align}
The regular diffusion and dilution in the PDE is augmented with a sink term, $-\mu u$. This degradation of mass manifests as a first-order reaction of the form: 
\begin{equation*}
A\xrightarrow{\mu}\emptyset
\end{equation*} 
in both of the mesoscopic and microscopic representations, where $A$ represents a particle whose density is given by $u$. The boundary condition at the left-hand end of the domain represents an influx of particles with rate $D\lambda$. This can be thought of as a zeroth-order reaction at the left-hand boundary, of the form: 
\begin{equation*}
\emptyset \xrightarrow{\kappa} A,
\end{equation*}
where $\kappa$ is the rate of introduction of new particles which is related to $\lambda$ via $\kappa = \lambda D$.

\begin{figure}
\begin{center}
\subfigure[][]{\includegraphics[width=0.31\textwidth]{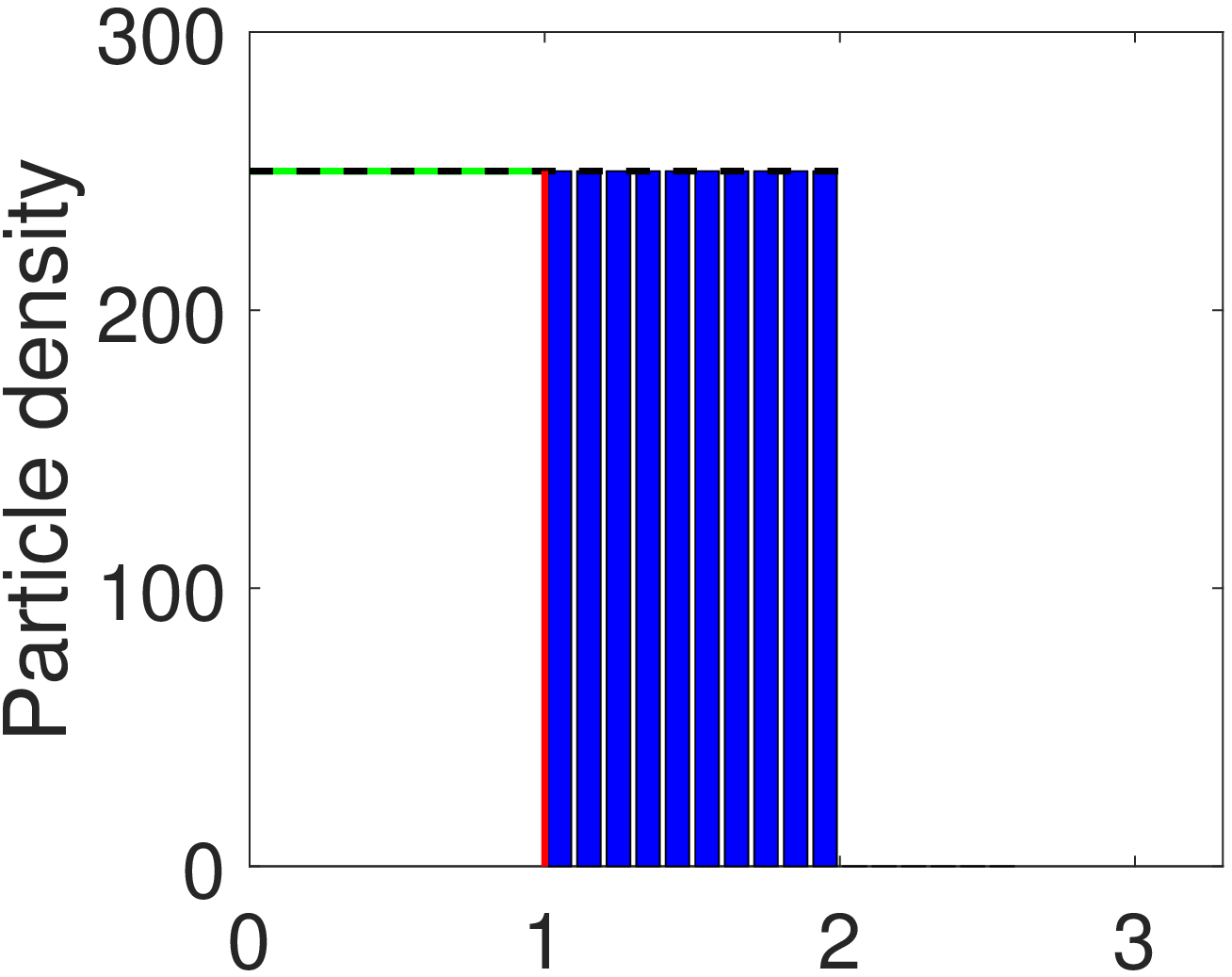}
\label{fig:Morphogen_Results_PCM_0}}
\subfigure[][]{\includegraphics[width=0.31\textwidth]{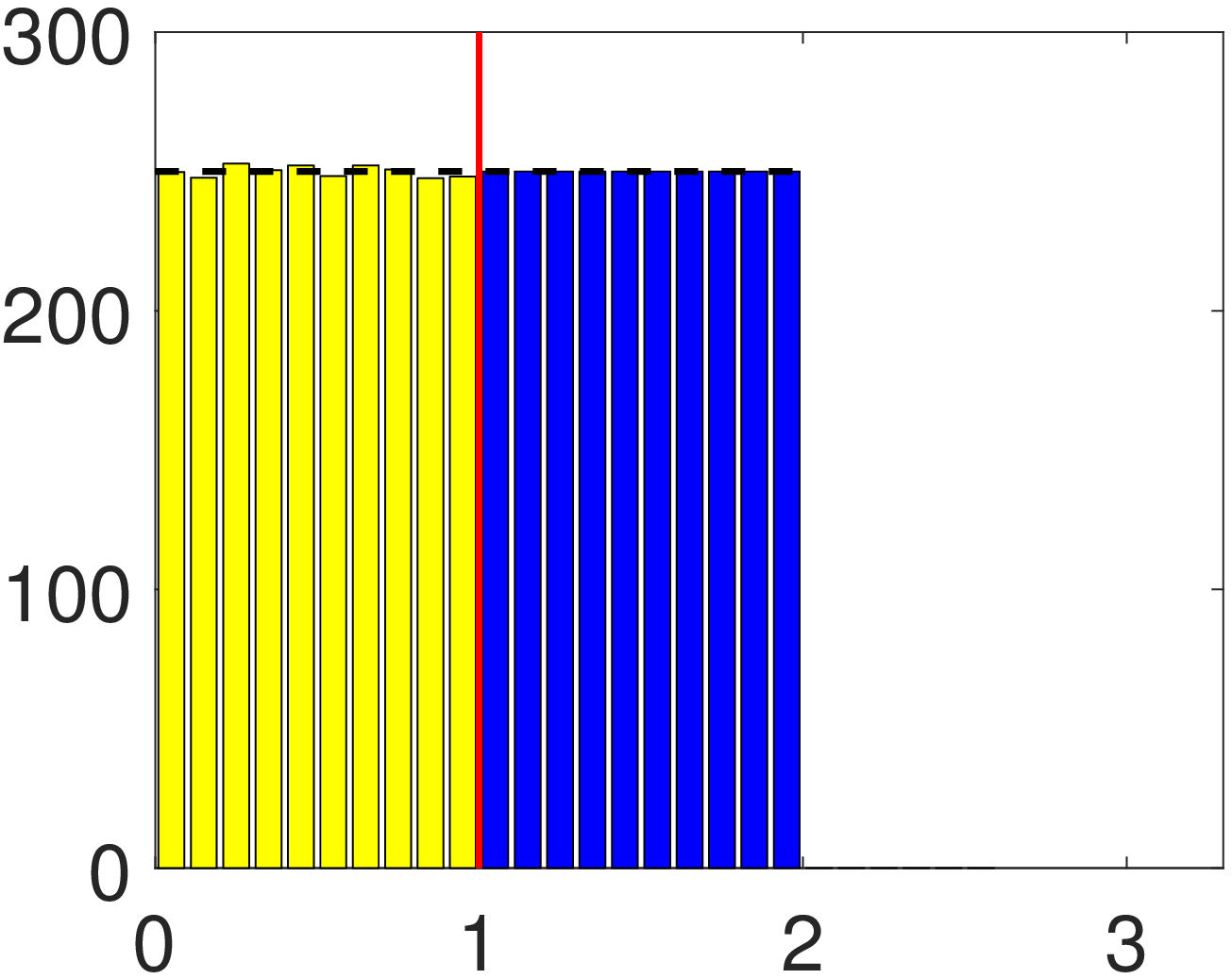}
\label{fig:Morphogen_Results_GCM_0}}
\subfigure[][]{\includegraphics[width=0.31\textwidth]{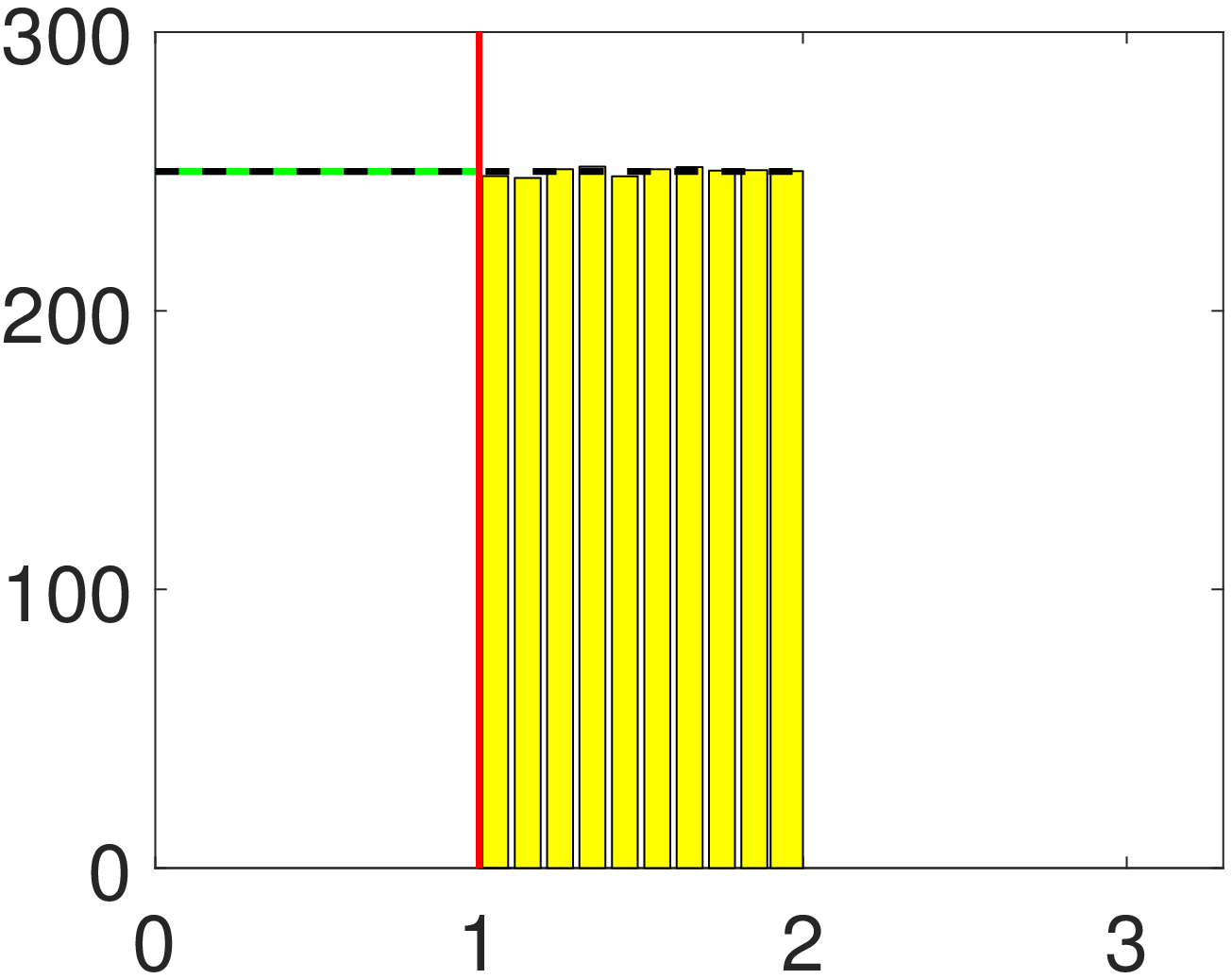}
\label{fig:Morphogen_Results_ARM_0}}
\subfigure[][]{\includegraphics[width=0.31\textwidth]{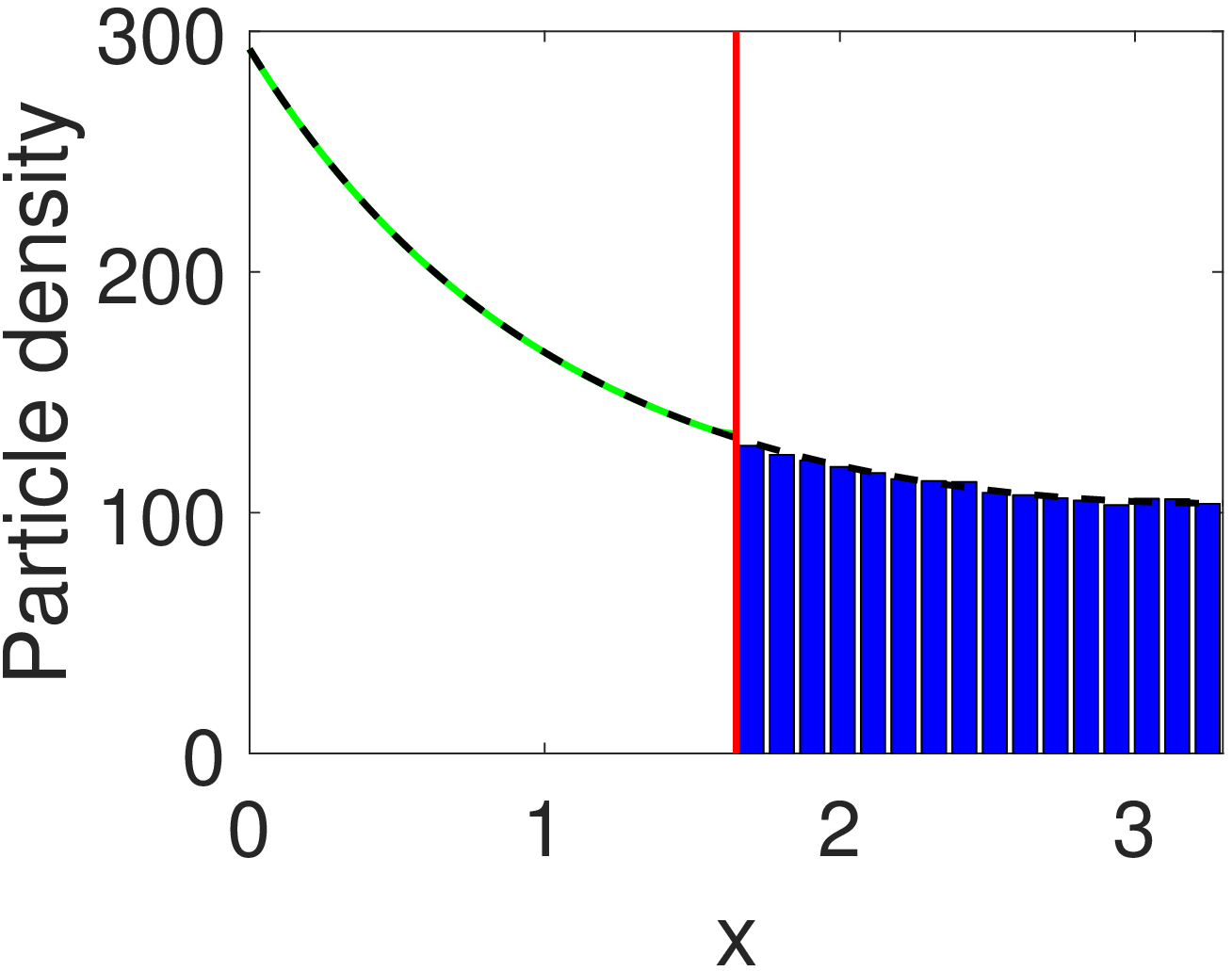}
\label{fig:Morphogen_Results_PCM_T}}
\subfigure[][]{\includegraphics[width=0.31\textwidth]{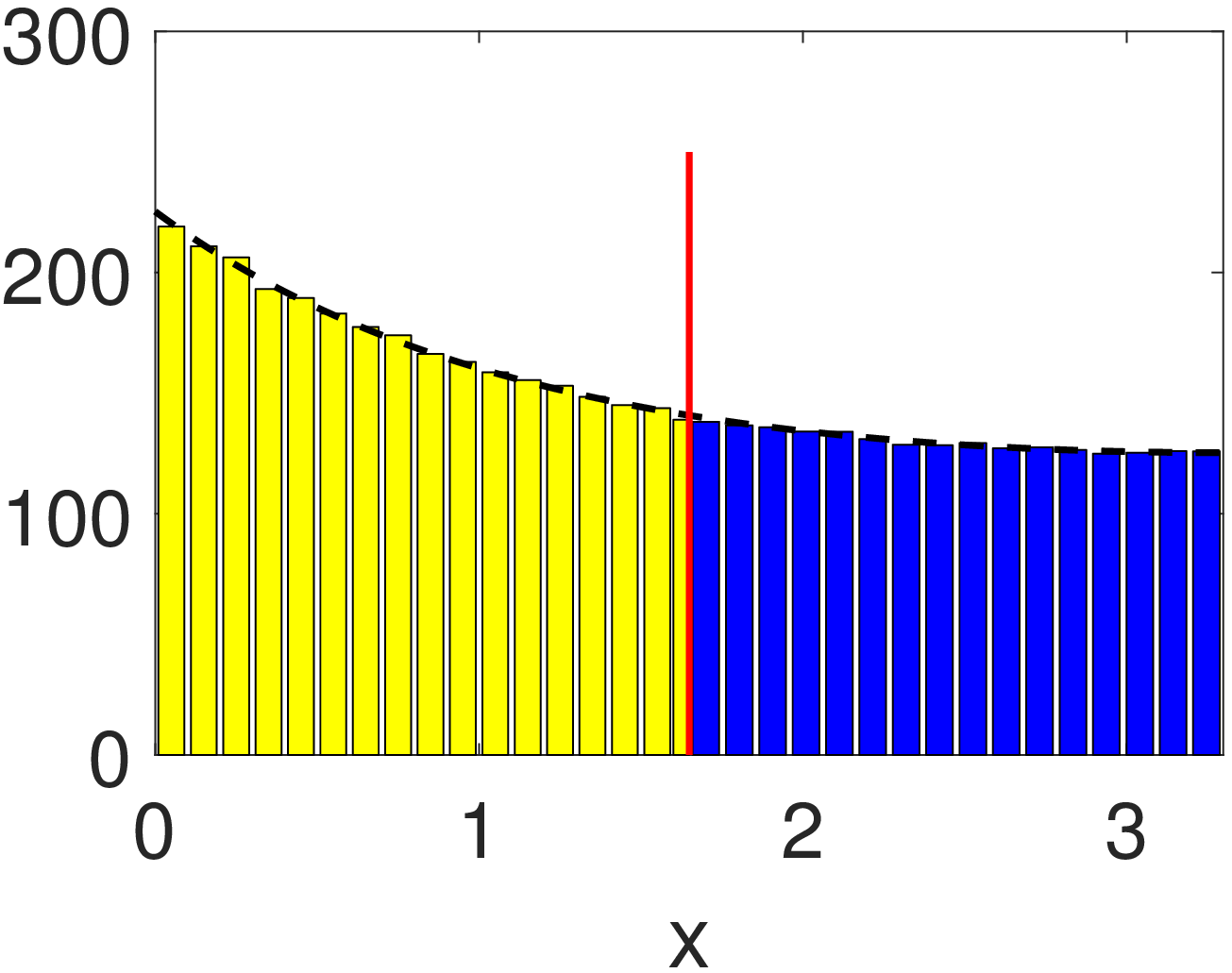}
\label{fig:Morphogen_Results_GCM_T}}
\subfigure[][]{\includegraphics[width=0.31\textwidth]{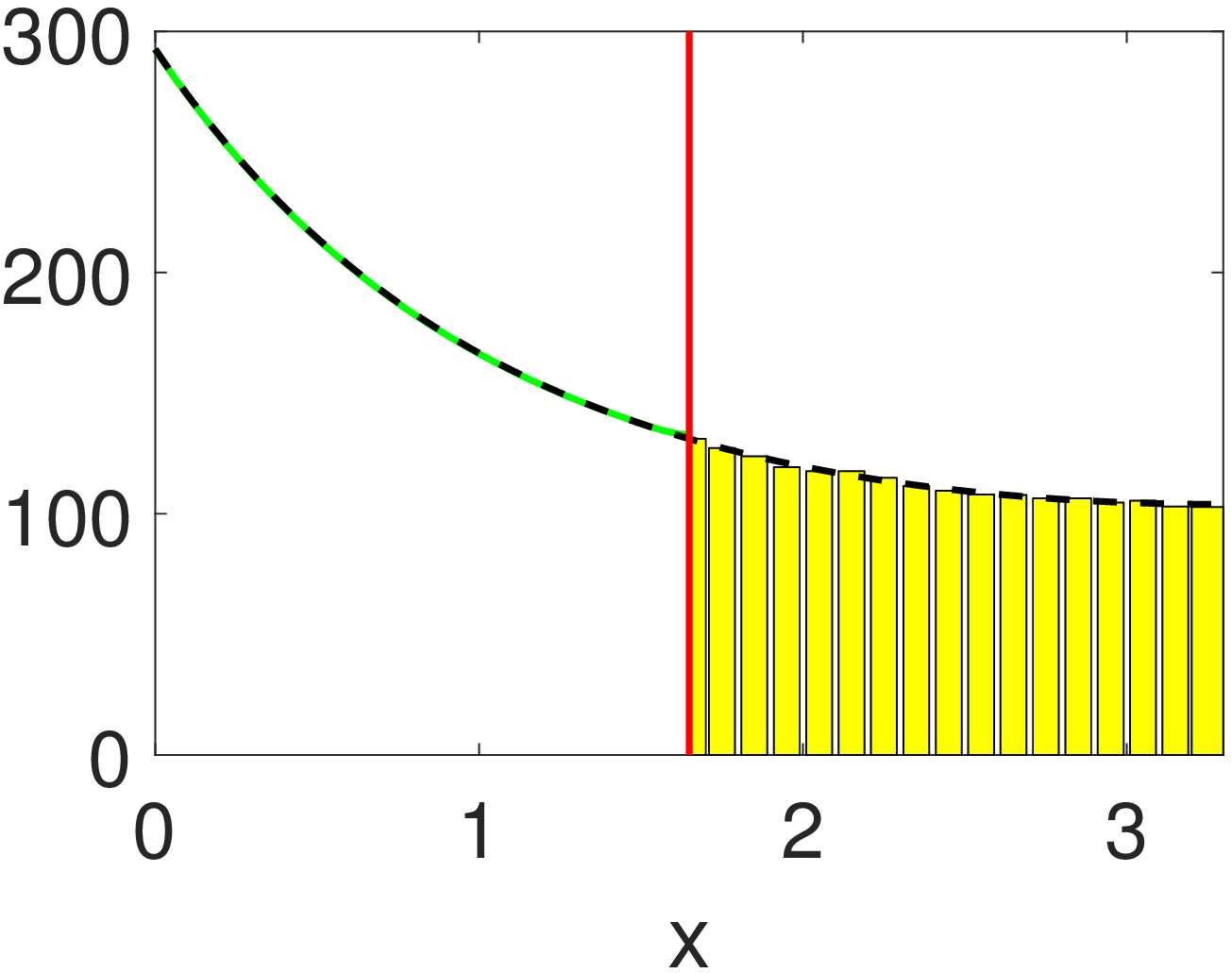}
\label{fig:Morphogen_Results_ARM_T}}
\subfigure[][]{\includegraphics[width=0.31\textwidth]{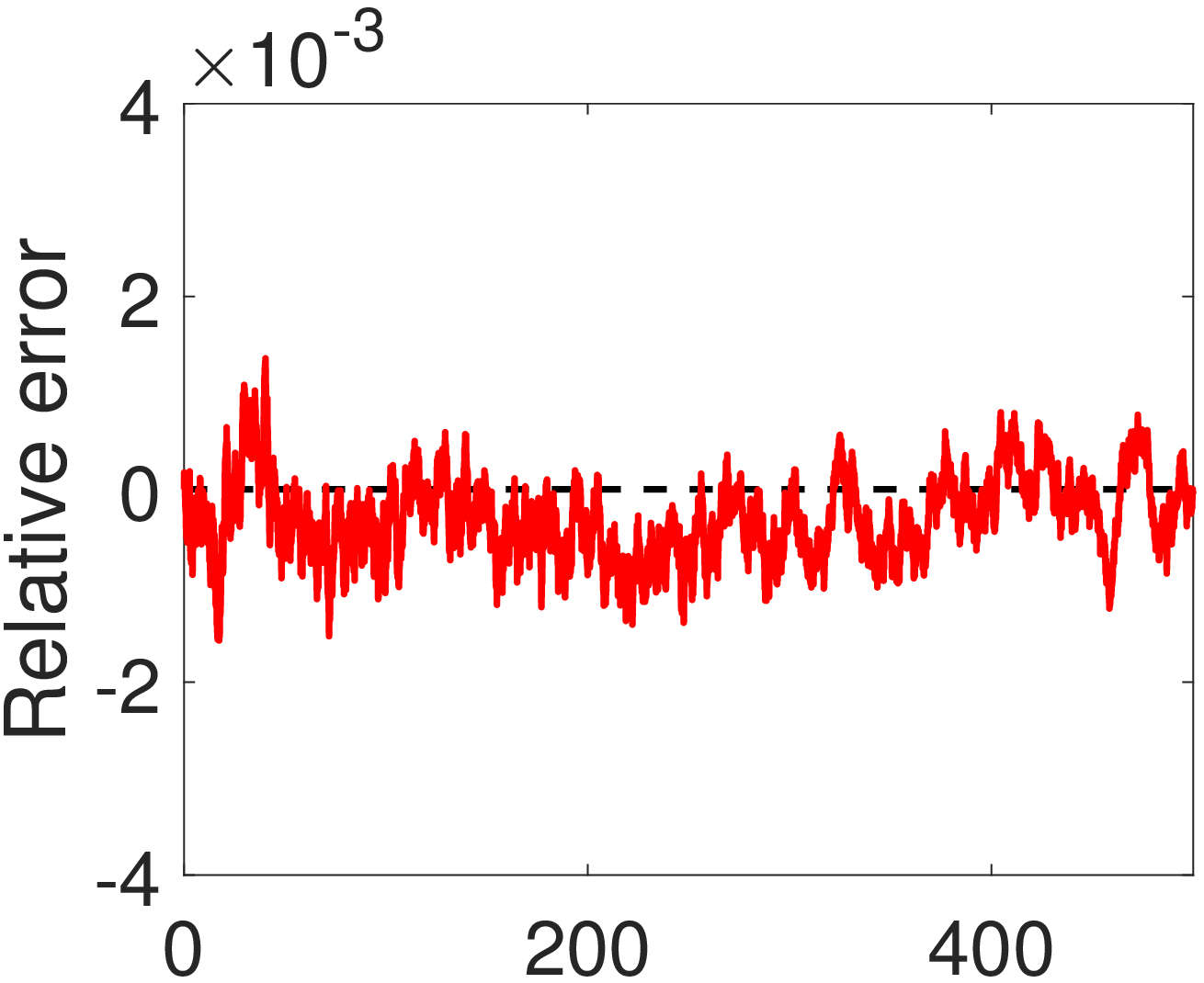}
\label{fig:Morphogen_Results_PCM_L}}
\subfigure[][]{\includegraphics[width=0.31\textwidth]{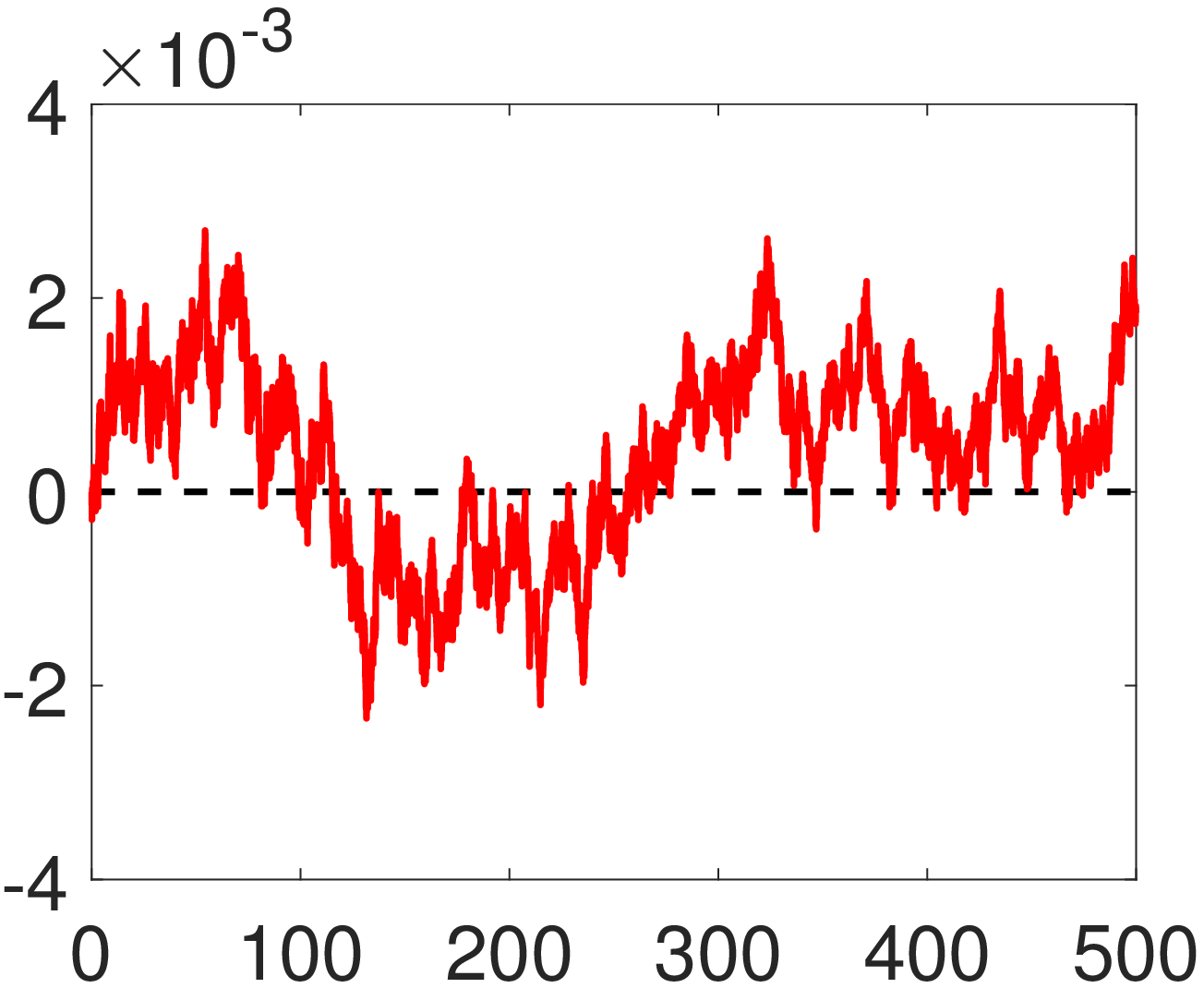}
\label{fig:Morphogen_Results_GCM_L}}
\subfigure[][]{\includegraphics[width=0.31\textwidth]{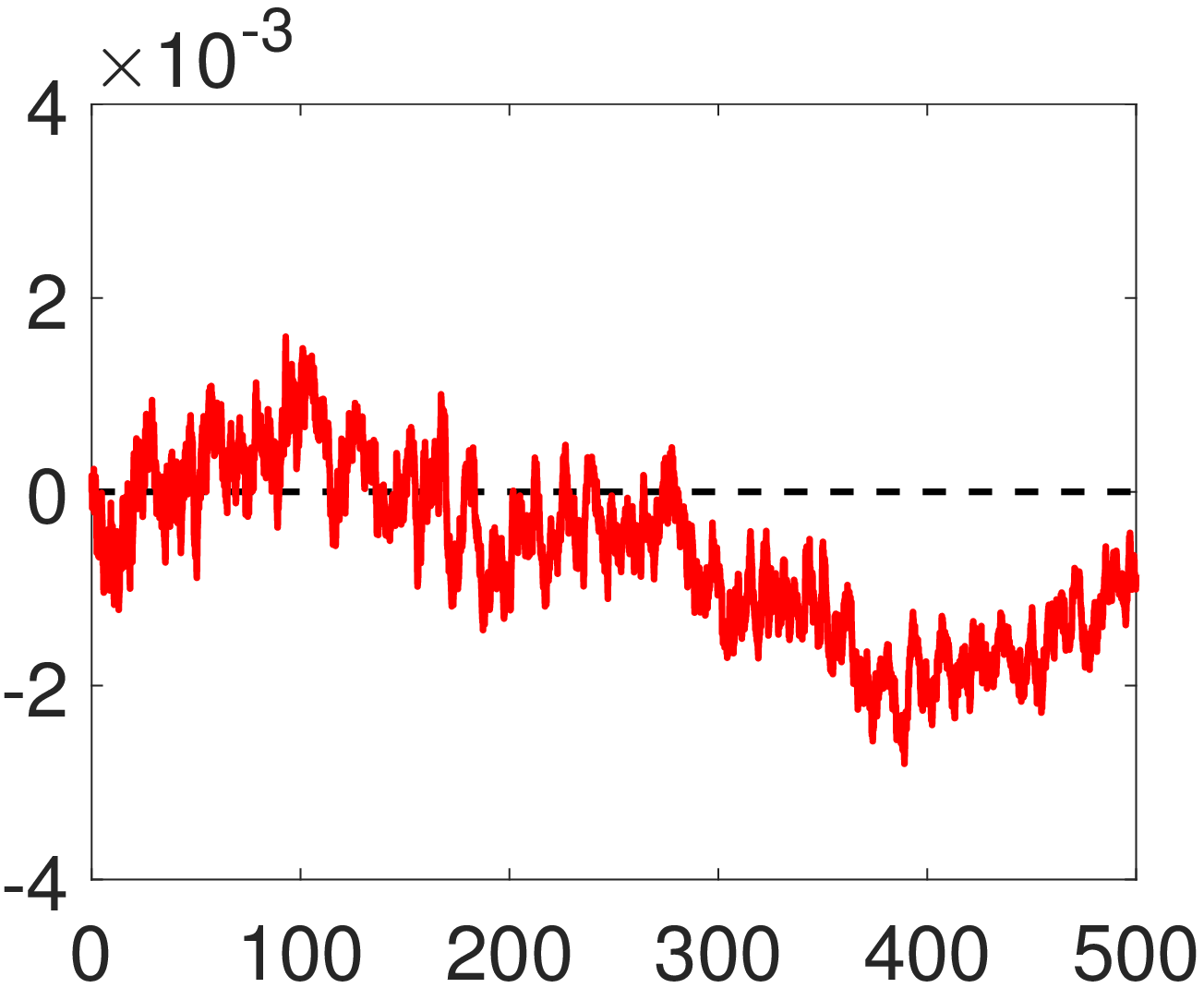}
\label{fig:Morphogen_Results_ARM_L}}
\subfigure[][]{\includegraphics[width=0.31\textwidth]{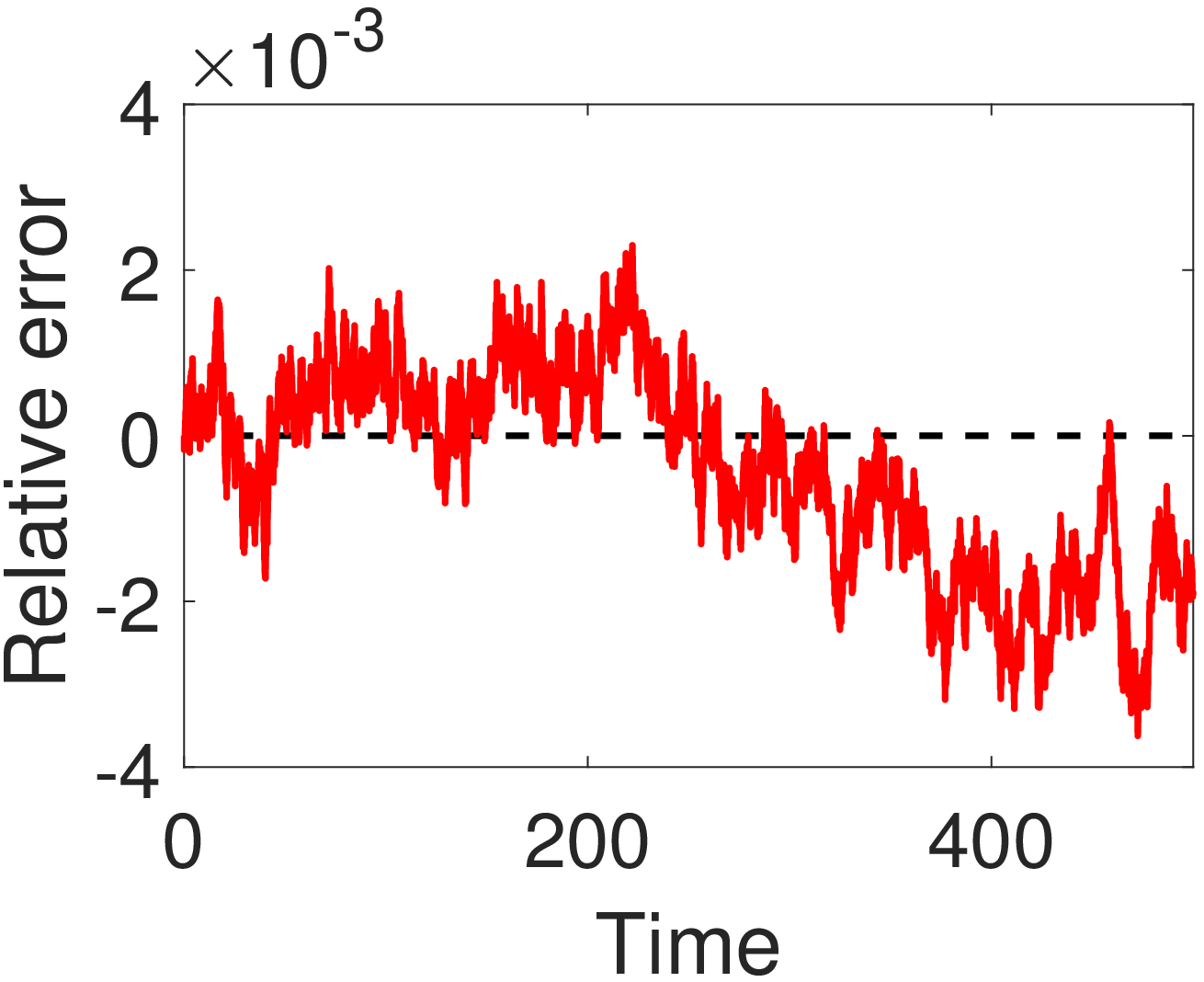}
\label{fig:Morphogen_Results_PCM_R}}
\subfigure[][]{\includegraphics[width=0.31\textwidth]{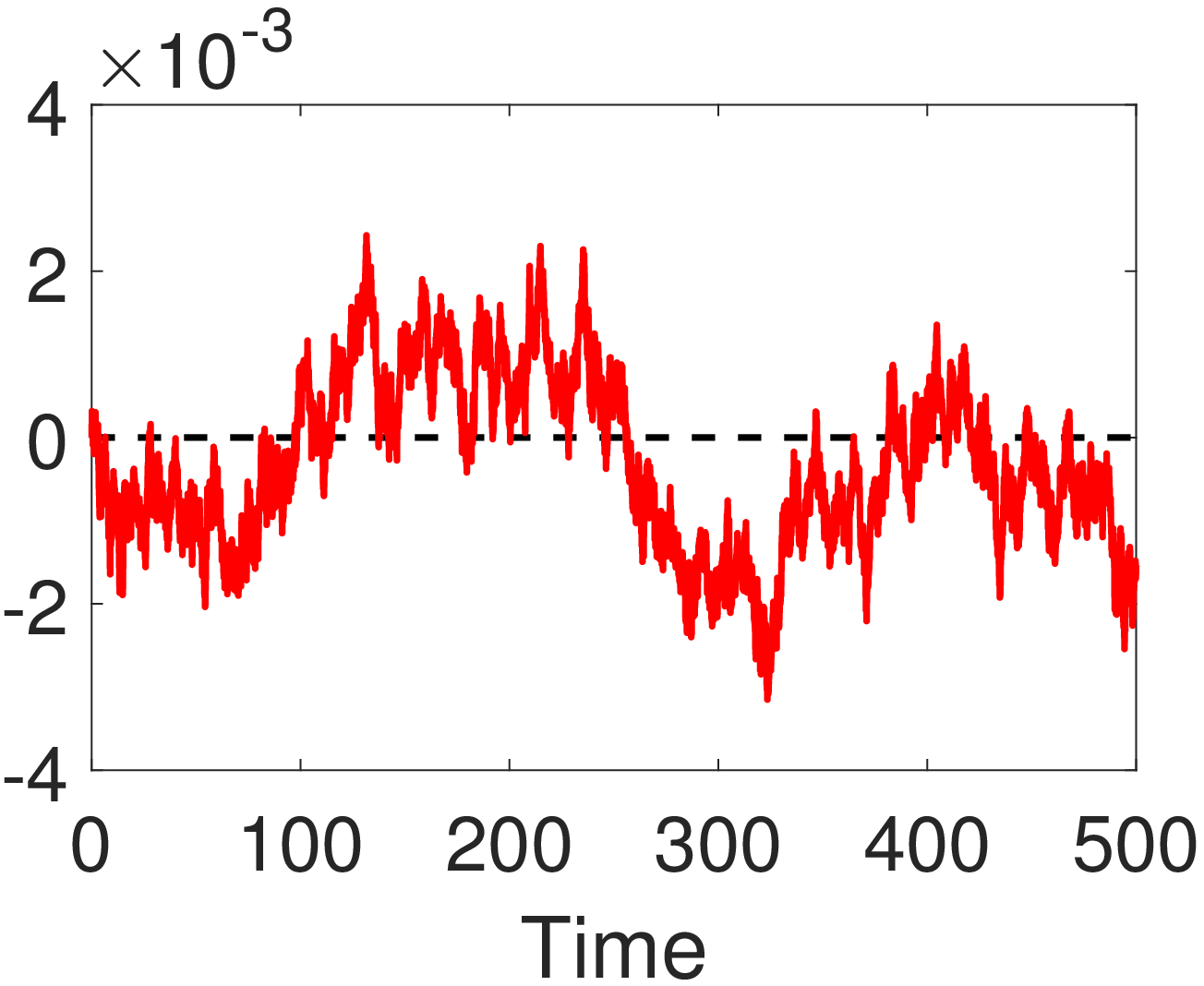}
\label{fig:Morphogen_Results_GCM_R}}
\subfigure[][]{\includegraphics[width=0.31\textwidth]{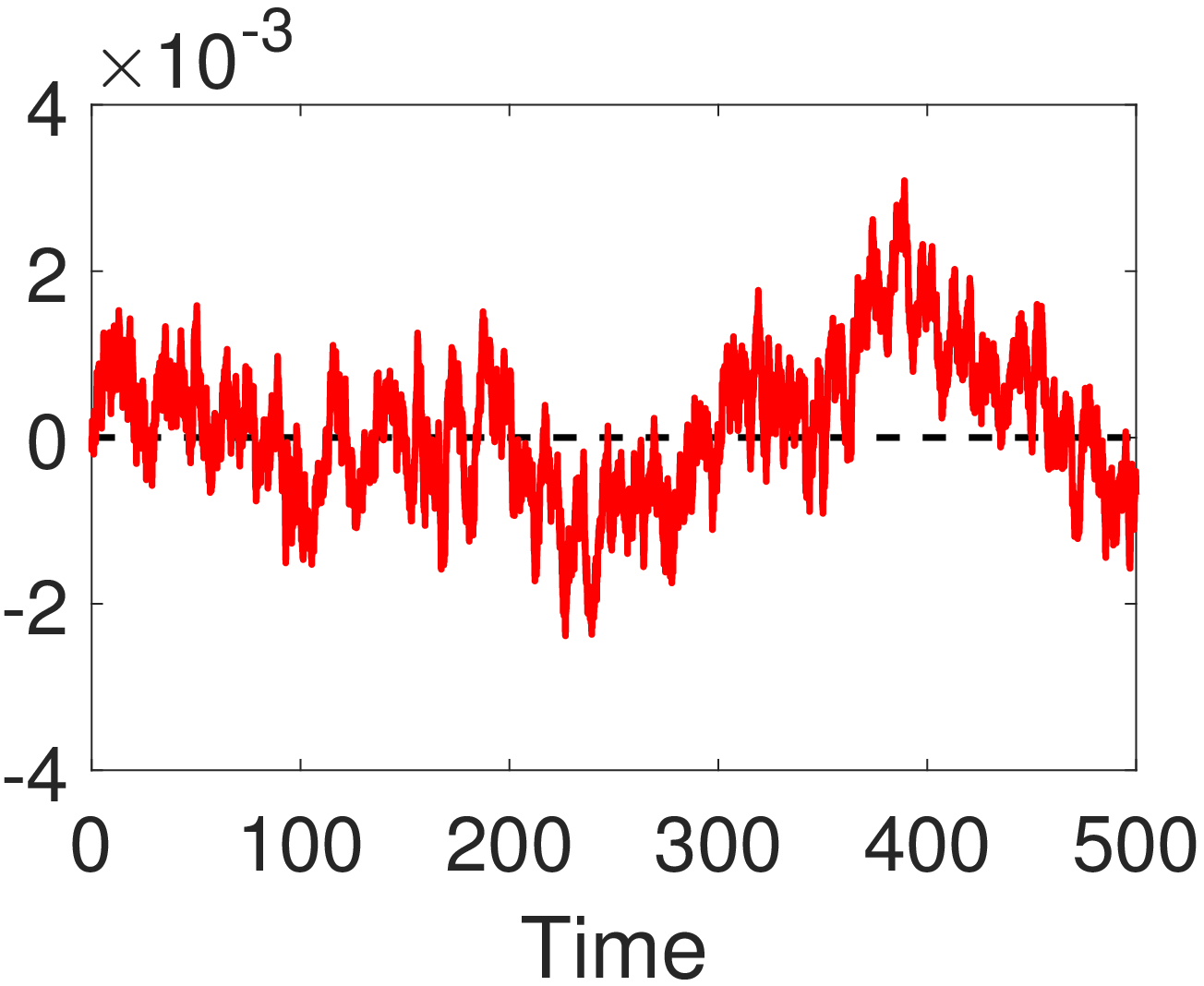}
\label{fig:Morphogen_Results_ARM_R}}
\end{center}
\caption{Results for Test Problem 3. Figure descriptions are as in Figure \ref{fig:Uniform_Results}.}
\label{fig:Morphogen_Results}
\end{figure}

We present the results for test problem 3 in Figure \ref{fig:Morphogen_Results}. As with each of the previous test problems, we see that there is a very good qualitative agreement with all three methods and the ground truth numerical PDE solution (\ref{fig:Morphogen_Results_PCM_0}-\ref{fig:Morphogen_Results_ARM_T}) and this is further confirmed by the relative error plots, which show no discernible bias in either direction for any of the hybrid methods.

\section{Discussion} \label{sect:Discussion}

In this paper, we have developed three spatially extended hybrid methods which are capable of simulating multi-scale reaction diffusion processes on uniformly growing domains. Each method has been extended from previous methods originally specified on static domains \citep{yates2015pcm, flegg2015cmc, smith2018arm}. We have provided descriptions, schematics and detailed algorithms for the implementation of these methods. Furthermore, we have demonstrated that each of these methods are accurate and unbiased under three test problems: pure diffusion with zero-flux boundaries, diffusion with partially absorbing periodic boundaries, and the formation of a morphogen gradient with particle degradation and an influx boundary.

We have focussed on exponential growth, however it should be noted that the methods set out here will work for any type of uniform growth, provided that there is an equivalence framework between the three individual modelling paradigms. For example, if the domain were to grow linearly with rate $\rho$, the domain length at time $t$ (given it is of length $L_0$ to begin with) would be $L(t)=L_0(1+\rho t)$. The advection term in equation \eqref{eqn:PDE-EC} would become $\rho(ux)_x/(1+\rho t)$, where the subscript here denotes differentiation with respect to the $x$ variable. In order for the mesoscale to be equivalent, domain growth events must occur with rate $\rho/(h_cK)$ \citep{baker2010fmm}. We must also be careful to adapt the calculation of the internal times of the mesoscopic events in each of the algorithms for this altered domain growth scenario \citep{anderson2007mnr}.

We have also presented each of our examples here in one dimension for simplicity and clarity of explanation, however, it would be straightforward to extend the methods to higher dimensions which have (hyper-)planar interfaces. Care must be taken if domain growth is to be implemented in higher dimensions, that the interface remains coherent. For example, when implementing domain growth in the compartment-based method, it makes sense to employ deterministic growth as in \citep{smith2019uol} (rather than stochastic growth \citep{baker2010fmm}) so that the rate of domain growth along the interface matches in each of the coupled methodologies.

There are several extensions (many adapted from the static hybrid literature) that might be included in order to render these growing hybrid methods more versatile. The inclusion of an adaptive interface, for example, would allow the methods to more robustly and efficiently deal with examples in which the concentration profile changes significantly. Static interfaces (interfaces which remain in the same place when considered in Lagrangian coordinates) are useful when we either know something about what the solution looks like \textit{a priori}, and hence we can place the interface in an appropriate position, or if there is a region of the spatial domain that requires a more detailed representation (such as around ion channels when considering transmembrane transport  - see, for example \citep{dobramysl2015pbd}). However, if neither of these are true, an adaptive interface may be required. Implementing an adaptive interface requires the use of local densities around the interface in order to determine its position; if the density in the finer scale subdomain is too high, the interface can move into that subdomain - extending the coarser subdomain, and vice-versa if the local density in the coarser subdomain is too low \citep{robinson2014atr,harrison2016hac,smith2019uol}.

While domain growth is an important phenomena in many biological scenarios \citep{baker2010fmm}, domain shrinkage also has equally important applications. As an example, directed apoptosis is an important component of wound healing, and requires models that incorporate domain shrinkage \citep{yates2014dcm}. The hybrid methods presented in this paper would extend equally well to a domain that uniformly shrinks in size as they do to those that grow.

The hybrid methods presented here provide accurate ways of simulating reaction-diffusion systems on uniformly growing domains, and will be of interest to those who model such systems that either have a scale difference in particle numbers across their domain, or have regions of space which need to be modelled in more detail than others. The methods developed here will allow members of the modelling community to probe the important effects of stochasticity in their multiscale systems without the suffering potential simulation penalties that modelling the entire system using a fine-scale methodology might bring.

\bibliographystyle{plainnat}
\bibliography{Growing_Hybrid}

\end{document}


\begin{flushleft}
{\Large
\textbf{Supplementary Material\\Incorporating domain growth into hybrid methods for reaction-diffusion systems}
}
\\
Cameron A. Smith$^{1,\ast}$, 
Christian A. Yates$^{1}$
\\
$^1$\textbf{Centre for Mathematical Biology, Department of Mathematical Sciences, University of Bath, Claverton Down, Bath, BA2 7AY, United Kingdom}\\
\end{flushleft}

Key words: Reaction--diffusion, domain growth, hybrid methods.


In this document, we introduce any of the mathematics required to demonstrate equivalence between the three methods described in Section \ref{MAIN-sect:Equiv}, and give the numerical algorithms that we employ. It should be noted that other numerical schemes may be used.

\section{Macroscale} \label{sect:Macro}

\subsection{The Lagrangian PDE} \label{sect:Macro_PDE}

We wish to be able to simulate the solution of the PDE \eqref{MAIN-eqn:PDE-EC}, with appropriate boundary and initial conditions. However, conventional techniques are unable to do so when the PDE is written in this form due to the existence of the growing domain \citep{simpson2015esl}. Instead, we transform the PDE onto a fixed coordinate system, known as the Lagrangian coordinates. We apply the following change of variables: \begin{align}
x &= X\exp\{\rho \tau\}, \label{eqn:X-LC}\\
t &= \tau, \label{eqn:tau-LC}
\end{align} where $X\in(0,L_0)$ and $\tau>0$. Applying change of variables \eqref{eqn:X-LC}-\eqref{eqn:tau-LC} to the Eulerian PDE \eqref{MAIN-eqn:PDE-EC} yields the Lagrangian PDE
\begin{equation}
\partder{u}{\tau}(X,\tau) = \frac{D}{\exp\{2\rho\tau\}}\secpartder{u}{X}(X,\tau) - \rho u(X,\tau). \label{eqn:PDE-LC}
\end{equation} 
The domain growth manifests itself in equation \ref{eqn:PDE-LC} in two ways: firstly by creating a time-dependent diffusion coefficient, which comes from the conversion of the second order derivative; and secondly by converting the dilution from an advective term to a particle sink via a first-order degredation reaction. This second term is one of the two terms obtained when the dilution term in \eqref{MAIN-eqn:PDE-EC} is expanded using the product rule. The second of these product rule terms cancels with a term that originates from the conversion of the time derivative.

\subsection{Numerical scheme} \label{sect:Macro_Num}

In order to simulate the PDE \eqref{eqn:PDE-LC}, we will use the $\theta$-method (see, for example \citep{smith1985nsp}), however we note that any different simulation methodologies may be appropriate. The $\theta$-method involves writing the second derivative as a combination of implicit and explicit terms, and using a time-stepping algorithm in order to find the solution at a later time. a description can be found in Algorithm \ref{alg:PDE} (below) for the PDE \eqref{eqn:PDE-LC} together with zero-flux boundary conditions. Because the PDE contains time-varying coefficients, the matrices used in order to update the solution need to be calculated at every time-step.

\begin{Algorithm}{The $\theta$-method (diffusion only)} \label{alg:PDE}
\item[]\textbf{Input}: Current PDE solution vector --- $\bs{U}$; Time of previous update --- $t_n$; Time-step --- $\Delta t$; $\theta$-value --- $\theta$; Diffusion coefficient $D$; Lagrangian mesh size --- $h_p$; Number of mesh points --- $S+1$; Growth rate --- $\rho$; Initial domain length $L_0$.
\item Calculate the time-dependent matrices $A_n$ and $B_n$:
\begin{align*}
(A_n)_{ij}&=\left\{\begin{array}{ll}
1+\frac{D\Delta t\theta}{h_p^2}\exp\{-2\rho(t_n+\Delta t)\} & i=j=1,J+1 \\ 
1+2\frac{D\Delta t\theta}{h_p^2}\exp\{-2\rho(t_n+\Delta t)\} & i\in\{2,...,S\},j=i \\ 
-\frac{D\Delta t\theta}{h_p^2}\exp\{-2\rho(t_n+\Delta t)\} & i\in\{2,...,S+1\},j=i-1 \\ 
-\frac{D\Delta t\theta}{h_p^2}\exp\{-2\rho(t_n+\Delta t)\} & i\in\{1,...,S\},j=i+1 \\ 
0 & \text{otherwise}
\end{array} \right.\\
(B_n)_{ij}&=\left\{\begin{array}{ll}
1-(\rho+\mu)\Delta t - \frac{D\Delta t(1-\theta)}{h_p^2}\exp\{-2\rho t_n\} & i=j=1,S+1 \\ 
1-(\rho+\mu)\Delta t - 2\frac{D\Delta t(1-\theta)}{h_p^2}\exp\{-2\rho t_n\} & i\in\{2,...,S\},j=i \\ 
-\frac{D\Delta t(1-\theta)}{h_p^2}\exp\{-2\rho t_n\} & i\in\{2,...,S+1\},j=i-1 \\ 
-\frac{D\Delta t(1-\theta)}{h_p^2}\exp\{-2\rho t_n\} & i\in\{1,...,S\},j=i+1 \\ 
0 & \text{otherwise}
\end{array} \right.
\end{align*}
\item Update the PDE solution $\bs{U}$ according to the following:
\begin{equation*}
\bs{U} \leftarrow A_n^{-1}B_n\bs{U}.
\end{equation*}
\end{Algorithm}

\subsection{Remeshing} \label{sect:Macro_Remesh}

Due to the difference in coordinate systems described in Section \ref{MAIN-sect:PCM} of the main text, the solution vector is calculated on a static Lagrangian mesh. Considered in Eulerian coordinates, however, this mesh grows. As such, it is prudent to remesh the solution vector when the PDE mesh grows by a certain amount in Eulerian coordinates. We present the remeshing algorithm which requires the pre-calculation of the times that we remesh. A natural time to remesh is when the domain has doubled in length, and this is indeed how we calculate these remeshing times. However, alternative scale factors may be used.

We begin by calculating the number of times the domain doubles in length before the final time of the simulation, $t_f$, via \begin{equation}
s_{max} = \lfloor \exp\{\rho t_f\}/2\rfloor,
\label{eqn:Remesh_Number}
\end{equation}
where $\lfloor\cdot\rfloor$ denotes the floor function. Then for $i\in\{1,...,s_{max}\}$, the $i$th time of remeshing is given by solving 
\begin{equation}
2i L_0 = L_0\exp\{\rho t_i\}.
\label{eqn:Remesh_Times}
\end{equation} The full algorithm for remeshing can be found in Algorithm \ref{alg:Remesh}

\begin{Algorithm}{Remeshing the PDE} \label{alg:Remesh}
\item[] \textbf{Input}: Current PDE solution vector --- $\bs{U}$; Growth rate --- $\rho$; Current number of PDE mesh points --- $S+1$; Current remesh time --- $t_i$; Next remesh time --- $t_{i+1}$.
\item Create a temporary solution vector $\bs{V}$ of size $2S + 1$.
\item Set $V_{2j-1} = U_j$ for every $j\in\{1,...,S+1\}$.
\item For all $V_j$ with $j\in\{2,4,6,...,2S\}$, interpolate the solution. Any reasonable interpolation may be used, however for the purposes of this paper we use linear interpolation:
$$V_j = \frac{V_{j-1}+V_{j+1}}{2},\quad j\in\{2,4,6,...,2S\}$$
\item Set $\bs{U}\leftarrow \bs{V}$.
\item Update the number of mesh points to be $2S+1$.
\item Update the remesh time to be $t_{i+1}$.
\end{Algorithm}

\section{Mesoscale} \label{sect:Meso}

\subsection{Modified next reaction method} \label{sect:Meso_MNRM}

We will evolve the mesoscopic system according to the modified next reaction method \citep{anderson2007mnr}. We choose to use the modified next reaction method as opposed to the standard Gillespie algorithm \citep{gillespie1977ess} because the propensity functions in the hybrid methods in the main text depend explicitly on time, meaning that between two events taking place, the propensity function is changing. This violates the assumption of the Gillespie algorithm, that requires the propensity functions to remain constant between the times that events take place. 

In order to use the modified next reaction method, we require the definition of propensity functions $a_j(\bs{n},t)$, where $j\in\{1,...,2K(t)\}$ indexes the possible events that can occur, with the first $2K(t)$ being the diffusive jumps left and right from each compartment. The probability of a particular event $j$ occurring during a small time interval $(t,t+\delta t)$ is then defined to be $a_j(\bs{n}(t),t)\delta t$.

The modified next reaction method requires the calculation and tracking of internal clock times $T_i$ for every possible event, and next firing times $P_i$. These are initialised as $T_i=0$ and $P_i = \ln\left(1/r_i\right)$, where $r_i$ is a uniformly distributed random variable between zero and one. Propensity functions $a_i$ are initialised, and the absolute time until the next event of type $i$ fires can be calculated by solving: \begin{equation}
P_i-T_i = \int_t^{t+\Delta t_i}{a_i(\bs{n}(t),s)\ ds}, \label{eqn:MNRM_next_time}
\end{equation} for $\Delta t_i$ at $t = 0$.

The algorithm then locates the event which happens first, which is equivalent to finding the minimum of the $\Delta_i$, which we call $\Delta$. This event is enacted, all internal clock times are updated according to \begin{equation}
T_i \leftarrow T_i + \int_t^{t+\Delta}{a_i(\bs{n}(t),s)}\ ds, \label{eqn:Update-Internal}
\end{equation} and for the event that fires, say event $\beta$, a new next firing time is found by setting \begin{equation}
P_\beta \leftarrow P_\beta + \ln\left(\frac{1}{r_\beta}\right), \label{eqn:Update-Next_Firing}
\end{equation} where $r_\beta$ is a uniformly distributed random variable between zero and one. Time is updated to be $t + \Delta$ and then the algorithm repeats. The time-varying nature of the propensity functions is accounted for by using the internal clocks as opposed to absolute time. The algorithm for the modified next reaction method can be found in Algorithm \ref{alg:MNRM}.

\begin{Algorithm}{Modified next reaction method \citep{anderson2007mnr} (diffusion only)} \label{alg:MNRM}
\item[] \textbf{Input}: Current mesoscopic state --- $\bs{n}$, Current time --- $t$; Final time --- $t_f$; Current number of compartments --- $K$; Internal clock times --- $T_i$ for $i\in\{1,...,2K\}$; Next firing times --- $P_i$ for $i\in\{1,...,2K\}$; Propensity function --- $a_i$ for $i\in\{1,...,2K\}$; Times until next event --- $\Delta t_i$ for $i\in\{1,...,2K\}$ (see equation \eqref{eqn:MNRM_next_time}).
\item Find $\Delta = \min\set{i\in\{1,...,2K\}}{\Delta t_i}$ and $\beta = \argmin\set{i\in\{1,...,2K\}}{\Delta t_i}$. \label{alg_step:MNRM_return}
\item Enact the event $\beta$.
\item Update the internal clocks according to equation \eqref{eqn:Update-Internal}.
\item For event $\beta$, update the next firing time using equation \eqref{eqn:Update-Next_Firing}.
\item Recalculate the propensity functions at the new time.
\item Update absolute time $t = t + \Delta$.
\item If $t < t_f$, return to step \ref{alg_step:MNRM_return}, otherwise end. 
\end{Algorithm}

\subsection{Domain stretching} \label{sect:Meso_Stretch}

Next we address how to stretch the domain, which we do according to the stretching method \citep{smith2019uol}. Suppose we have a state $\bs{N}^{K-1}(t)$ which contains $K(t)-1$ compartments, and we wish to extend the domain by a single compartment. In order to do this, we will define the pre-growth state $\bs{N}^{K-1}(t)=\bs{r}\in\mathbb{R}^{K(t)-1}$ and the post-growth state to be $\bs{N}^{K}(t)\bs{n}\in \mathbb{R}^{K(t)}$. In order to determine how the particles in the pre-growth state should be redistributed to the post-growth state, we will calculate ``overlap regions''. Consider the compartments being of length $h_c$. When there are $K(t)$ compartments, we have a domain of length $K(t)h_c$. We now stretch our pre-growth state to be of the same length, meaning that each of the $K(t)-1$ compartments are of length $h_cK(t)/(K(t)-1)$, yielding a domain of length $K(t)h_c$. From these, we can calculate the length of the $i^\text{th}$ pre-growth compartment that overlaps the $i^\text{th}$ post-growth compartment. These are the overlap regions. Using this set up, it can be calculated that the $i^\text{th}$ overlap region, with $K(t)-1$ pre-growth compartments, is of size \begin{equation}
\delta_i^{K(t)-1} = \frac{K(t)-i}{K(t)}, \label{eqn:Overlap}
\end{equation}
which holds for $i\in\{1,...,K(t)-1\}$.

These overlap regions are treated as the probability of placing a particle from pre-growth compartment $i$, into post-growth compartment $i$, independent of all others. Therefore, for each pre-growth compartment, draw a Binomially distributed random variable $b_i$ with $r_i$ trials and probability of success $\delta_i^{K(t)-1}$. Then for each post-growth compartment $j\in\{1,...,K(t)\}$ we set: \begin{equation}
n_j = \left\{\begin{array}{ll}
b_1, & \text{if }j = 1, \\ 
b_j + (r_{j-1}-b_{j-1}), & \text{if }j\in\{2,...,K(t)-1\}, \\ 
r_{K(t)-1}-b_{K(t)-1}, & \text{if }j = K(t).
\end{array} 
\right.
\label{eqn:Meso-Growth-Update}
\end{equation} We then set the new current state $\bs{N}(t)$ to be $\bs{n}$. This algorithm can be found in Algorithm \ref{alg:Stretch}.

\begin{Algorithm}{The stretching method \citep{smith2019uol}} \label{alg:Stretch}
\item[] \textbf{Input}: Current mesoscopic state --- $\bs{n}$; Current number of compartmemts --- $K$. 
\item Define the pregrowth state to be $\bs{r}\leftarrow\bs{n}$.
\item Calculate the overlap proportions $\delta_i^K = (K+1-i)/(K+1)$.
\item Draw $K$ binomial random variables $b_i\sim\text{Bin}(r_i,\delta_i^K)$. \item Create an extra compartment at the right end of the postgrowth domain (increasing $K$ by 1 by setting $K \leftarrow K + 1$). 
\item For $j\in\{1,...,K\}$, set $n_j$ according to \eqref{eqn:Meso-Growth-Update}
\end{Algorithm}

We can incorporate the stretching into the overall mesoscopic simulation algorithm in two ways. The first is to add an extra propensity function to the list which represents the stretching event. Over multiple repetitions of the algorithm, this will yield domains of different lengths due to the stochasticity in the number of times this event will fire. However, in the limit as the number of repeats tends to infinity, the length of the domain will become the average length of the domain, which will be the length of the equivalent PDE domain, i.e. $L(t)=L_0\exp\{\rho t\}$. We will not take this approach, as we wish to assess the accuracy of the hybrid method, which is aided by the removal of as many sources of statistical error from other parts of the algorithm as possible. Alternatively, we use the deterministic length $L(t) = L_0\exp\{\rho t\}$, to calculate the times at which, on average, the number of compartments should increase by 1, and will always enact a stretching event at these times.

\subsection{Equivalence} \label{sect:Meso_Equiv}

To demonstrate equivalence of this mesoscopic scheme to the Eulerian PDE, we need to formulate the equations for the evolution of the mean number of particles in the mesoscopic description, which we do by firstly writing down the master equation for the system. Define $p(\bs{n},k,t)$ to be the probability that the state variable $\bs{N}^K(t)$ is $\bs{n}$ and the number of compartments $K(t)$ is $k$ at time $t$. The master equation describes the evolution of this probability. Then the rate of change of this probability is \begin{equation}
\begin{split}
\partder{p}{t}(\bs{n},k,t) &= d\sum_{i=1}^{k-1}{\left[(n_i+1)p(J_i^+\bs{n},k,t) - n_ip(\bs{n},k,t)\right]} + d\sum_{i=2}^{k}{\left[(n_i+1)p(J_i^-\bs{n},k,t) - n_ip(\bs{n},k,t)\right]}\\
&+\rho(k-1)\sum_{\bs{r}\in\mathcal{M}_{k-1}^M}{\left[p(\bs{r},k-1,t)\pi(\bs{n}|\bs{r})\right]} - \rho kp(\bs{n},k,t).
\end{split}
\label{eqn:Master_Equation}
\end{equation} 
Here, we have that $d$ is the rate for any particle to jump from its compartment to one of its neighbours, $J_i^\pm$ are operators which move a single particle from compartment $i$ to compartment $i\pm 1$, then set $\mathcal{M}_k^M = \set{\bs{m}\in\mathbb{N}^k}{\sum_{i=1}^k{m_i}=M}$ is the set of all state variables with $k$ compartments and $M$ total particles, and $\pi(\bs{n}|\bs{r})$ is the transition probability from state $\bs{N}^{k-1}=\bs{r}$ to $\bs{N}^k=\bs{n}$. The first two sums on the right-hand side capture the impact of the diffusive jumps right and left respectively, and the final two terms correspond to the effects of domain growth.

In order to obtain the mean equations, we define 
\begin{equation}
\bar{M}_i^k(t) = \sum_{\bs{n}\in\mathcal{M}_k^M}{n_ip(\bs{n},k,t)}.
\label{eqn:Mean_Sum}
\end{equation}
to be the average number of particles in compartment $i$ when there are $k$ compartments in total, at time $t$. We can then calculate the mean equations by multiplying equation \eqref{eqn:Master_Equation} by $n_i$ and summing over the state space $\bs{n}$ to give: 
\begin{equation}
\ordder{\bar{M}_i^k}{t} = d\bar{M}_{i-1}^k - (2d+\rho k)\bar{M}_i^k + d\bar{M}_{i+1}^k + \rho(k-1)\left[\left(1-\frac{k-i+1}{k}\right)\bar{M}_{i-1}^{k-1}+\frac{k-i}{k}\bar{M}_i^{k-1}\right].
\label{eqn:Mean_Equations}
\end{equation}
%
In order to write this as a continuous system, we  let $x=ih_c$, and suppose that $\bar{M}_i^k(t) \approx u(x,t)$ and $\bar{M}_{i\pm1}^k(t) \approx u(x\pm h_c,t)$. We will finally assume that mass spreads uniformly when it grows, so that $(k-1)\bar{M}_i^{k-1} = k\bar{M}_i^k$. Substituting this into the mean equations \eqref{eqn:Mean_Equations} yields \begin{equation*}
\begin{split}
\partder{u}{t}(x,t) \approx du(x-h_c,t) &- 2du(x,t) + du(x+h_c,t))\\
&+ \rho\frac{k(i-1)}{k}u(x-h_c,t) + \rho\frac{k(k-i)}{k}u(x,t)-\rho ku(x,t).
\end{split}
\end{equation*}We apply a second order Taylor expansion about $x$ and rearrange in derivatives of $u$. We also omit all arguments of the function $u$ as all will be evaluated at $(x,t)$.  This gives us the following: \begin{align*}
\partder{u}{t} &\approx d\left[u - h_c\partder{u}{x} + \frac{h_c^2}{2}\secpartder{u}{x}\right] - 2du + d\left[u + h_c\partder{u}{x} + \frac{h_c^2}{2}\secpartder{u}{x}\right]\\
&\qquad\qquad\qquad\qquad\qquad+ \rho(i-1)\left[u - h_c\partder{u}{x} + \frac{h_c^2}{2}\secpartder{u}{x}\right] + \rho(k-i)i - \rho ku,\\
&= \left[d-2d+d+\rho(i-1)+\rho(k-i)-\rho k\right]u + h_c\left[-d+d-\rho(i-1)\right]\partder{u}{x}\\
&\qquad\qquad\qquad\qquad\qquad+\frac{h_c^2}{2}\left[d+d+\rho(i-1)\right]\secpartder{u}{x},\\
&=-\rho u - \rho h_c(i-1)\partder{u}{x} +dh_c^2\secpartder{u}{x} + \frac{\rho}{2}h_c^2(i-1)\secpartder{u}{x}.
\end{align*} We utilise the fact that $ih_c = x$ in order to simplify this further: \begin{equation*}
\partder{u}{t} = -\rho u - \rho x\partder{u}{x} + dh_c^2\secpartder{u}{x} + h_c\left[\rho\partder{u}{x} + \frac{\rho}{2}(x-h_c)\secpartder{u}{x}\right].
\end{equation*} 
We now take the diffusive limit, by taking compartment size, $h_c$ to 0 while making the jump rate, $d$, infinite and keeping $dh_c^2$ constant. We also recognise the first two terms as being an expansion of the derivative of $-\rho u x$ to obtain \begin{equation}
\partder{u}{t} = -\rho\partder{(ux)}{x} + dh_c^2\secpartder{u}{x}.
\end{equation} 
Comparing this with equation \eqref{MAIN-eqn:PDE-EC}, we note that equivalence requires the Fickian diffusion coefficient $D$ to be related to the mesoscopic jump rate $d$ via $D = dh_c^2$.

\section{Microscale} \label{sect:Micro}

\subsection{Numerical scheme} \label{sect:Micro_Num}

In order to simulate the SDE \eqref{MAIN-eqn:SDE-EC}, we utilise the Euler-Maruyama method. Suppose we have a system containing $N$ particles, whose positions at time $t$ are given by $y_i(t)$. Then the Euler-Maruyama method allows us to update the positions in the time interval $(t,t+\Delta t)$ according to\begin{equation}
y_i(t+\Delta t) = y_i(t) + \rho y_i(t) \Delta t + \sqrt{2D\Delta t}\ \xi_i, \label{eqn:SDE-Comp}
\end{equation} where the $\xi_i$ are normally distributed random variables with zero mean and unit variance. The algorithm for the movement of particles can be found in Algorithm \ref{alg:Ind-Update}

\begin{Algorithm}{Individual-based update (diffusion only)} \label{alg:Ind-Update}
\item[] \textbf{Input}: Current positions of particles --- $\bs{y}$; Current time --- $t$; Time-step to evolve particles --- $\Delta t$; Diffusion coefficient --- $D$; Growth rate --- $\rho$;
\item Update the particles' positions according to the computational SDE \eqref{eqn:SDE-Comp}.
\item For every $y_i$ such that $y_i < X_0\exp\{\rho(t+\Delta t)\}$, set $$y_i \leftarrow y_i + 2(X_0\exp\{\rho(t+\Delta t)\}-y_i).$$
\item For every $y_i > X_1\exp\{\rho(t+\Delta t)\}$, set $$y_i\leftarrow y_i-2(y_i-X1\exp\{\rho(t+\Delta t)\}).$$
\end{Algorithm}

We now briefly discuss how reactions are to be enacted. Zeroth-order reactions, where particles appear without the influence of any other particle, will occur with a probability $\kappa_0\Delta t$ over a time interval $(t,t+\Delta t)$, where $\kappa_0$ is the rate per unit time of the reaction. First-order reactions depend on the particles already present in the system. For every particle, the reaction is enacted with a probability $\kappa_1\Delta t$, where again, the $\kappa_1$ is the rate for each particle to react. Second-order reactions involve the interaction of two particles that are ``close enough'' to react. There are several methods that are able to resolve these interactions, such as the $\lambda$-$\rho$ method \citep{erban2009smr}. We refer to those interested in such methods to the following references \citep{smoluchowski1917vem, andrews2004ssc, erban2009smr, lipkova2011abd, van2005gfr}.

\subsection{Equivalence} \label{sect:Micro_Equiv}

Finally, to establish equivalence with the macroscale, we define the probability of finding a particle located at a position $x$ at time $t$, given that it was at a position $y$ at time $s<t$ to be $q(x,t|y,s)$. Then, through the use of an intermediate point, we can write the Chapman-Kolmogorov equation \begin{equation*}
q(z,t+\Delta t|y,s) = \int_{-\infty}^\infty{q(z,t+\Delta t|x,t)q(x,t|y,s)\ dx}.
\end{equation*} This equation holds for any $\Delta t$, not necessarily small. We can multiply this equation by a smooth test function $\varphi(z)$, integrate over $z$ and relabel the integral on the LHS to be an integral with respect to $x$ instead of $z$ to yield 
\begin{equation}
\int_{-\infty}^\infty{q(x,t+\Delta t|y,s)\varphi(x)\ dx} = \int_{-\infty}^\infty{\left[\int_{-\infty}^\infty{q(z,t+\Delta t|x,t)\varphi(z)\ dz}\right]q(x,t|y,s)\ dx}.\label{eqn:CK_test}
\end{equation}
Taylor expanding $\varphi(z)$ on the RHS about $x$ to second order, we obtain three integrals which can be interpreted as the zeroth- first- and second-order moments of the distribution $q(\cdot, t+\Delta t|z,t)$. These quantities are found straightforwardly using the SDE \eqref{MAIN-eqn:SDE-EC}. Once substituted into equation \eqref{eqn:CK_test}, all derivatives on the $\varphi$ function are transferred to the $q$ function, and a rearrangement yields the Eulerian PDE \eqref{MAIN-eqn:PDE-EC}, known in this context as the Fokker-Planck equation.

\bibliographystyle{plainnat}
\bibliography{./SM_Growing_Hybrid}

\makeatletter\@input{xx.tex}\makeatother
\makeatletter\@input{xx_SM.tex}\makeatother